\documentclass[11pt,letterpaper]{article}
\pdfoutput=1
\usepackage{jheppub}

\usepackage[T1]{fontenc} 
\usepackage{color}
\usepackage{graphicx}
\usepackage{tabularx}
\usepackage{xspace}
\usepackage{empheq}
\usepackage{amsmath}
\usepackage{amssymb}
\usepackage[caption=false]{subfig}
\usepackage{url}
\usepackage{bbold}
\usepackage{array}
\usepackage{multirow}
\usepackage{paralist}
\usepackage{makecell}
\usepackage[normalem]{ulem}
\usepackage{pdflscape}
\usepackage{afterpage}
\usepackage[framemethod=TikZ]{mdframed}

\allowdisplaybreaks 
\def \i {\mathrm{i}\mkern1mu} 
\newcommand{\gagg}{g_{a \gamma\gamma}}
\newcommand{\gp}{g_{\varphi \gamma\gamma}}
\newcommand{\solenoid}{{(s)}} 
\newcommand{\toroid}{{(t)}}

\makeatletter
\gdef\@fpheader{}
\makeatother

\preprint{\hbox{CERN-TH-2023-093}}

\title{Symmetries and Selection Rules: Optimising Axion Haloscopes for Gravitational Wave Searches}

\author[1]{\small Valerie Domcke,}
\author[2]{\small Camilo Garcia-Cely,}
\author[1,3]{\small Sung Mook Lee,}
\author[1]{\small Nicholas L. Rodd}

\affiliation[1]{\footnotesize Theoretical Physics Department, CERN, 1 Esplanade des Particules, CH-1211 Geneva 23, Switzerland}
\affiliation[2]{Instituto de F\'isica Corpuscular (IFIC), Universitat de Val\`{e}ncia-CSIC, Parc Cient\'ific UV, C/ Cate\-dr\'{a}tico Jos\'{e} Beltr\'{a}n 2,
E-46980 Paterna, Spain}
\affiliation[3]{\footnotesize Department of Physics \& IPAP \& Lab for Dark Universe, Yonsei University, Seoul 03722, Korea}

\emailAdd{valerie.domcke@cern.ch}
\emailAdd{camilo.garcia@ific.uv.es}
\emailAdd{sungmook.lee@yonsei.ac.kr}
\emailAdd{nrodd@cern.ch}

\abstract{In the presence of electromagnetic fields, both axions and gravitational waves (GWs) induce oscillating magnetic fields: a potentially detectable fingerprint of their presence.
We demonstrate that  the response is largely dictated by the symmetries of the instruments used to search for it.
Focussing on low mass axion haloscopes, we derive selection rules that determine the parametric sensitivity of different detector geometries to axions and GWs, and which further reveal how to optimise the experimental geometry to maximise both signals.
The formalism allows us to forecast the optimal sensitivity to GWs in the range of 100~kHz to 100~MHz for instruments such as ABRACADABRA, BASE, ADMX SLIC, SHAFT, WISPLC, and DMRadio.
}

\begin{document} 

\maketitle
\flushbottom

\section{Introduction}

Gravitational wave (GW) experiments have begun to probe the GW spectrum over a vast range, from the Gigaparsec wavelengths probed by the CMB~\cite{Planck:2018jri} to thousands of kilometers, covered by current ground-based interferometers which operate in the 100~Hz range~\cite{aLIGO:2020wna},  yielding fundamental insights into cosmology, astrophysics, and particle physics.
Reaching even higher frequencies poses a significant experimental challenge, but would represent a unique opportunity to probe possible extensions of the Standard Models of particle physics and cosmology~\cite{Aggarwal:2020olq}.
A cosmological source of GWs produced at a temperature $T_*$ could generate a stochastic GW background at frequencies of $f \gtrsim 1\,{\rm kHz}~(T_*/10^{10}\,{\rm GeV})$, and leave a signature from modifications to the Standard Model at the highest temperatures.\footnote{An explicit example of such a source would be cosmological phase transitions; for recent reviews, see e.g. Refs.~\cite{Caprini:2019egz,Athron:2023xlk}.}
Unfortunately, probing relics of a possible high-temperature phase of the early Universe is fundamentally challenging.
Experimental sensitivity to GWs can be expressed in terms of the strain $h$. 
As the energy density in GWs scales as $\rho \sim h^2 f^2 M_{\rm Pl}^2$, at higher frequencies even greater reach in terms of $h$ is required to reach energy densities below the current bounds on the total energy in radiation in the early Universe derived from BBN and CMB observations.
Instead, exotic astrophysical events sourcing transient signals appear to be a more promising medium-term target~\cite{Franciolini:2022htd,Domcke:2022rgu}.
For example, the merger of two equal mass objects yields GWs at $f \sim 1\,{\rm kHz}~(M_{\odot}/m)$, so that sources such as primordial black holes with $m \ll M_{\odot}$ could populate the high frequency landscape.

In recent years, significant progress has been made in understanding the sensitivity of electromagnetic GW detectors in this frequency regime.
In a flat spacetime perturbed by a gravitational wave, $g_{\mu \nu} = \eta_{\mu \nu} + h_{\mu \nu}$, 
the usual expressions for electrodynamics in flat spacetime receive corrections of the schematic form $\sim h F^2$, yielding a graviton-two-photon vertex.
As long appreciated, this interaction can lead to photon-GW mixing~\cite{Gertsenshtein,Boccaletti, Raffelt:1987im}.
More generally, however, a GW in the presence of an electromagnetic background will induce an electromagnetic response, in close analogy to the signal from axion (scalar) dark-matter arising from the coupling $a F \tilde{F}$ ($\varphi F^2$).
Exploiting the considerable experimental efforts to search for an electromagnetic response from wave-like dark matter, it has been shown that these same instruments can be used as GW telescopes~\cite{Ejlli:2019bqj,Berlin:2021txa,Domcke:2022rgu,Herman:2020wao,Gao:2023gph}.
Largely motivated by the QCD axion, dark matter searches focus on signals of a MHz or above, and are therefore naturally suited to look for high-frequency GWs.

In this paper, we will continue the study of the sensitivity of axion haloscopes to GWs, with a particular focus on instruments operating in the ``low-mass'' magnetoquasistatic regime, and sensitivity in the MHz-GHz window; experiments already operating in this range include ABRACADABRA~\cite{Kahn:2016aff,Ouellet:2018beu,Salemi:2021gck}, ADMX SLIC~\cite{Crisosto:2019fcj}, BASE~\cite{Devlin:2021fpq}, SHAFT~\cite{Gramolin:2020ict}, and WISPLC~\cite{Zhang:2021bpa}.
These devices feature a strong static magnetic field which in the presence of an axion -- or a GW -- sources a small, oscillating induced magnetic field which is captured by a suitably placed pickup loop.
Many of the existing instruments are effectively prototypes, with a sensitivity that can be improved significantly by increasing the volume of the magnetic field, and by reading out the magnetic flux induced in the pickup loop resonantly.
By combining both of these improvements, the goal of the DMRadio collaboration is to reach the QCD axion prediction for ${\rm neV} \lesssim m_a \lesssim \mu{\rm eV}$~\cite{DMRadio:2022jfv,DMRadio:2022pkf,DMRadio:2023igr}.
In view of the expected progress, it is timely to consider how synergies in axion and GW searches can be optimally exploited, in particular in view of different detector geometries currently proposed for axion searches.

Reference~\cite{Domcke:2022rgu} first proposed the use of low-mass axion haloscopes as GW detectors, and demonstrated that a toroidal magnetic field -- as employed by ABRACADABRA, SHAFT, and the upcoming DMRadio-50L -- could detect a passing GW.
Here, we generalise that analysis to additional detector geometries, with a particular focus on the solenoidal magnetic field used by ADMX SLIC, BASE, WISPLC and which as been moreover proposed for DMRadio-m$^3$.
We provide analytical expressions for the effective current which the GW sources, the resulting induced magnetic field, as well as for resulting magnetic flux for various pickup loop geometries.
Armed with these results, we will bootstrap the expected sensitivities to GW signals from axion searches.
A further improvement over Ref.~\cite{Domcke:2022rgu} is a careful treatment of the different timescales involved, in particular the potentially short duration of the GW signal.

Whilst the GW sensitivity for a solenoidal magnetic field is a practical result, as for the toroidal magnetic field, the calculation remains involved, and ultimately it becomes inefficient to compute the GW interaction with all possible magnetic field geometries.
Motivated by this, we derive a series of symmetry based selection rules, which determine the parametric sensitivity to a GW signal depending upon the symmetries of the experimental magnetic field and the pickup loop used to read out the signal.
From these results, we will demonstrate that configurations with a high degree of symmetry can kill the leading order sensitivity to a GW, even though they may be desirable to maximise the axion sensitivity.
An analogue of this was already observed in Ref.~\cite{Domcke:2022rgu}, where it was shown that if the flux from a toroidal magnetic field is read out through a circular pickup loop, then the leading order GW sensitivity, expected at ${\cal O}[(\omega L)^2]$, vanishes, while sensitivity at ${\cal O}[(\omega L)^3]$ remains.
Here $\omega$ is the angular frequency of the GW, $L$ is a characteristic length scale for the experiment, and in the magnetoquasistatic regime of interest for low-mass axion haloscopes, $\omega L \ll 1$.
We show that if both the external magnetic field and the pickup loop have cylindrical symmetry, i.e.\ if they are invariant under azimuthal rotations and reflections in the $z$ coordinate, any orientation of the pickup loop which is sensitive to the axion suffers from a cancellation of the leading order term ($\propto (\omega L)^2$) for the GW signal.
This symmetry is commonly exhibited by axion haloscopes, which make use of solenoidal or toroidal magnetic fields.
To recover the dominant scaling, the cylindrical symmetry must be broken, for instance through the placement or geometry of the pickup loop.
The latter can be most easily achieved by modifying the pickup loop to span only a fraction of the azimuthal angle, with the optimal GW sensitivity obtained when the cylindrical symmetry for the pickup loop is maximally broken (for instance, with a figure-8 configuration as in Ref.~\cite{Domcke:2022rgu}).

For existing experiments, as the largest axion signal is obtained for detectors with full cylindrical symmetry, this explains the observation in Ref.~\cite{Domcke:2022rgu} that the optimal axion and GW sensitivities cannot be simultaneously obtained for a haloscope based on a toroidal magnetic field, and furthermore demonstrates that this conclusion is generic.
Nevertheless, we find that modifying the pickup loop geometry (or including several different pickup loops) allows one to obtain sensitivity to both the axion and GW signal, {in a manner that at worst reduces the axion sensitivity by an ${\cal O}(1)$ amount}.\footnote{This same approach would also allow for discrimination between a GW and axion signal.
Of course, we note that there are many ways to distinguish these signals, the most important being that in the accessible parameter space the GW signal will be transient, whereas that from dark matter is persistent.}
We illustrate the power of symmetry arguments by determining the leading power in $(\omega L)$ sensitivity for a range of different detector geometries without explicit computation, in view of determining the optimal geometries for GW searches.
For the most relevant cases, we provide the computation to confirm our results.

At the outset, we can already provide an intuitive argument as to why cancellations in highly symmetric detectors might be expected.
To do so, rather than contrasting GW and axion electrodynamics, as we will in the remainder of the paper, let us consider a simpler comparison: a scalar versus a pseudoscalar.
In particular, consider first the induced magnetic field arising from the interaction of a toroidal magnet, ${\bf B} = B_0 \hat{\bf e}_{\phi}$, with a pseudoscalar via the interaction, $g a F \tilde{F}$.
If we consider the induced magnetic field in the $z$ direction at the center of the toroid, as measured by the ABRACADABRA collaboration, we find $B^a_z \sim g (\partial a) B_0 L$.
The consistent transformation of this result under parity, which can be confirmed directly, is critically reliant on the pseudoscalar nature of the axion.
Indeed, if we ask what the induced field would be for a scalar interaction, $g \varphi F^2$, there is no expression we can write consistent with parity and the cylindrical symmetry of the instrument.
An explicit computation confirms that $B^{\varphi}_z = 0$.
This argument can be formalised into symmetry based selection rules which determine the geometries that are sensitive to scalar versus pseudoscalar coupling -- indeed, there are configurations where $B_a=0$ whilst $B_{\varphi} \neq 0$ -- and we undertake that exercise in App.~\ref{app:scalarEM}.
The general lesson, however, is that highly symmetric detectors impose symmetry constraints on the induced fields that can be so restrictive that the measurable signal vanishes.
This is true also for GWs, and we will determine an appropriate set of selection rules to determine the interplay between signals and geometry.

We can actually determine an additional general lesson by comparing the scalar and pseudoscalar interaction.
As is well known, the axion interaction generates an effective current proportional to $\partial_{\nu} (a \tilde{F}^{\nu \mu}) = (\partial_{\nu}a) \tilde{F}^{\nu \mu}$, so that the interactions depends only on a derivative of the axion, as expected for a pseudo-goldstone boson.
The equivalent expression for a scalar is  $\partial_{\nu} (\varphi F^{\nu \mu}) = (\partial_{\nu} \varphi) F^{\nu \mu} - \varphi j^{\mu}$, where $j^{\mu} = \partial_{\nu} F^{\mu \nu}$ is the current that generates the leading order fields in the laboratory. 
Accordingly, for the scalar there is an additional contribution to the effective current localised at the boundary of the magnetic volume, which turns out to be generic: it will be present also for the GW, although it has so far been overlooked in the literature.
This contribution can be interpreted as an effective current at the boundary of the magnetic volume, determined by the component of the effective magnetisation vector (introduced in Ref.~\cite{Domcke:2022rgu} for GWs) parallel to the boundary surface.

In the remainder of this paper we will flesh out these ideas for the GW signal, and we organise our discussion as follows.
Section~\ref{sec:framework} lays out the theoretical framework for our work, reviewing the relevant aspects of electrodynamics in a spacetime perturbed by a GW.
Several points, such as a discussion of the symmetry properties of the induced magnetic field and response matrix formalism are presented here for the first time in this context.
This sets the stage for deriving the GW sensitivity of axion haloscopes with solenoidal magnetic field configurations in Sec.~\ref{sec:GWforSolenoid}.
The results are then generalised in Sec.~\ref{sec:othergeom} where we derive symmetry principles which allow us to determine the parametric scaling of the GW sensitivity for various detector geometries without explicit computation.
The symmetry arguments will then enable us to draw general conclusions about the optimal strategy for axion and GW searches in axion haloscopes.

Many details of our analyses are deferred to appendices.
Appendix~\ref{app:Maxwell} reviews Maxwell's equation in curved space time.
Within it, we provide a careful derivation of the main equations governing the interaction of a GW with a background electromagnetism (EM) field, the derivation of the effective surface current, and an explanation of why the GW effects we consider scale at lowest order as $(\omega L)^2$.
In App.~\ref{app:scalarEM} we study scalar and axion electrodynamics, with a focus on sharpening an analogy to the GW case.
We will explain how our GW selection rules extend to these spin-0 waves, and the consequences for various detector geometries.
In App.~\ref{app:Bdecomposition} we summarise the symmetry properties of the cylindrical magnetic field configurations employed by axion haloscopes, and demonstrate that they can be decomposed into a solenoidal and toroidal component.
Appendix~\ref{app:responsematrix} expands our discussion of the response matrix formalism used to describe the detector response to a passing GW.
In App.~\ref{app:explicitcurrent} we summarise the explicit analytical expressions for all components of the effective current induced by a GW, up to order $(\omega L)^3$ and for both toroidal and solenoidal external magnetic field configurations.
These expressions may be used as input for full detector simulations, or for detailed numerical calculations of the relevant GW effects.
In App.~\ref{app:CoherenceRatio} we discuss in detail the bootstrapping of axion search results to establish GW sensitivity, carefully taking into account the different time scales involved in the possible signals and detectors.
The appendix further discusses details of several possible sources for high-frequency GWs.
Finally, App.~\ref{app:Efield} is dedicated to the possibility of using an external electric instead of magnetic field for GW detection, and demonstrates how our symmetry arguments extend to this case.

\section{Gravitational Wave Electrodynamics}
\label{sec:framework}

To begin with, we review the general formalism used to compute the magnetic flux induced by a GW passing through a lumped-element circuit axion haloscope.
We will review the discussion of Ref.~\cite{Domcke:2022rgu} (see also Ref.~\cite{Berlin:2021txa}), pointing out an additional contribution to the induced magnetic flux due to effective surface currents at the boundary of the magnetic volume, which was previously overlooked.
We then extend this approach to account for the transformation properties and symmetries of the various quantities, in particular the induced magnetic field, under rotations and reflections.
The axion haloscopes targeting the magneto-quasistatic regime ($m_a \lesssim \mu{\rm eV}$) generally have a high degree of cylindrical symmetry, and we will study the impact of this on the GW signal systematically.
Doing so will allow us to develop a systematic approach to the geometries of the external background magnetic field and pickup loop, and resolve fundamental questions such as determining the optimal geometry for GW and axion searches.

\subsection{Proper detector frame}

Throughout this paper we will work in the proper detector frame,\footnote{This is in contrast to the transverse traceless (TT) frame, in which coordinate distances are set by the geodesics of free-falling test masses, and a rigid instrument and experimental magnetic field no longer have a simple description.} in which coordinate distances to the origin match the proper distance, and thus coincide with those measured by ideal rigid rulers.
As a consequence of this, up to non-inertial forces such as those associated with the rotation of the Earth (which can be neglected at high frequencies~\cite{Ni:1978zz,Maggiore:2007ulw}), the effect of GWs is simply given by a small Newtonian force proportional to their amplitude.
Throughout this paper, we will assume that these GW forces do not mechanically deform the experimental setup, in particular the static electromagnetic fields applied in the experiment remain static in the presence of a GW.
Critically, this implies that in the proper detector frame the experimentally generated magnetic field coincides with that of flat spacetime, see App.~\ref{app:Maxwell} for details.
This assumption is in particular valid for GW frequencies below the mechanical resonance frequencies of the setup.
At frequencies around and above the lowest mechanical resonance $\omega_0 \sim v_s/L$, with $v_s$ denoting the speed of sound in the material, the Newtonian GW force is no longer negligible~\cite{Berlin:2023grv,Bringmann:2023gba}.
We expect this to impact part of the parameter space relevant for the experimental setups discussed here, and we leave a quantitative analysis to future work.
Interestingly, in the case of microwave cavities, it was demonstrated that this effect can enhance the GW sensitivity~\cite{Berlin:2023grv}.

Expanding the metric as $ g_{\mu\nu} = \eta_{\mu\nu} + h_{\mu\nu}$ with $\eta_{\mu \nu}$ denoting the flat metric with sign convention $(- + + +)$,
in the proper detector frame the GW at the position~${\bf r}$ can be expressed as~\cite{Berlin:2021txa,Domcke:2022rgu},\footnote{
This expression is equivalent to Eq.~(S5) in Ref.~\cite{Domcke:2022rgu}, as can be shown using the completeness relation $\hat{k}_i\hat{k}_j + \hat{U}_i\hat{U}_j +\hat{V}_i\hat{V}_j = \delta_{ij}$ 
in the third line of Eq.~\eqref{eq:metricperturbation}.}
\begin{equation}\begin{aligned}
h_{00} & = \omega^2 e^{-\i \omega t} F({\bf k} \cdot {\bf r}) \, r_m r_n \sum_{A = +, \times} h^A e^A_{mn} (\hat{\bf k}), \\
h_{0i} & = \frac{1}{2} \omega^2 e^{-\i \omega t} 
[ F({\bf k} \cdot {\bf r})  - \i F^{\prime}({\bf k} \cdot {\bf r}) ] 
[\hat{\bf k} \cdot {\bf r} ~ r_m \delta_{ni} - r_m r_n \hat{k}_i] \sum_{A = +, \times} h^A e^A_{mn} (\hat{\bf k}), \\
h_{ij} & = - \i \omega^2 e^{-\i \omega t} F^{\prime}({\bf k} \cdot {\bf r}) 
[ \vert {\bf r} \vert^2 \delta_{im} \delta_{jn} + r_m r_n \delta_{ij} - r_n r_j \delta_{im} -  r_m r_i \delta_{jn}  ]
\sum_{A = +, \times} h^A e^A_{mn} (\hat{\bf k}),
\label{eq:metricperturbation}
\end{aligned}\end{equation}
where $F(\xi) = [e^{\i\xi} - 1 - \i \xi]/\xi^2$, $h^{+, \times}$ denotes the amplitude of the two GW polarisations in the TT frame, and the polarisation tensor $e_{ij}^{+, \times} (\hat{\bf k})$ for a given direction of GW propagation $\hat{\bf k} = \sin \theta_h \, \hat{\bf e}^{\phi_h}_{\rho} + \cos \theta_h \, \hat{\bf e}_z$ can be defined as
\begin{align}
e_{ij}^+ = \frac{1}{\sqrt{2}} \left[	\hat{U}_i \hat{U}_j - \hat{V}_i \hat{V}_j	\right]\!, &&
e_{ij}^{\times} = \frac{1}{\sqrt{2}} \left[	\hat{U}_i \hat{V}_j + \hat{V}_i \hat{U}_j	\right]\!, &&
\hat{\bf V} = \hat{\bf e}_{\phi}^{\phi_h}, && 
\hat{\bf U} = \hat{\bf V} \times \hat{\bf k}. 
\label{eq:polarisation}
\end{align}
Here $\theta_h$ and $\phi_h$ denote the azimuthal and polar angle of the GW, and $\hat{\bf e}^{\phi_h}_{\rho}$, $\hat{\bf e}_{\phi}^{\phi_h}$ and $\hat{\bf e}_z$ denote unit vectors in the radial, angular and vertical direction for a polar angle $\phi_h$, with both coordinate systems defined with origin at the center of the experiment.
In particular, note that $h_{\mu i}r_i =0$, and consequently $ds^2=g_{\mu\nu} dx^\mu dx^\nu=\eta_{\mu\nu} dx^\mu dx^\nu $ for $dx^\mu = (0, dr\, \hat{{\bf r}} )$.
From this we see that coordinate distances to the origin coincide with the corresponding proper distance, a defining characteristic of the proper detector frame~\cite{Manasse:1963zz}.

In this work, we will limit ourselves to the regime of $\omega L \ll 1$, as appropriate over most of the range covered by lumped-element circuit instruments.\footnote{For the result when taking $\omega L \sim 1$ for the axion induced signal in these instruments, see Ref.~\cite{Benabou:2022qpv}.}
{We can therefore treat $\omega L$ as a perturbative parameter, and will do so often, for instance it will be implicit in our use of the Biot-Savart law and used throughout our discussion of the implications of the symmetry transformations.
Further, as $F(\xi) = - \tfrac{1}{2} + {\cal O}(\xi)$, it follows from Eq.~\eqref{eq:metricperturbation} that $h_{\mu \nu}$ in the proper detector frame has a leading order contribution at ${\cal O}[(\omega L)^2]$.}
The absence of any contribution at ${\cal O}[\omega L]$ is a consequence of working in a freely falling reference frame assumed to be rigid, as we demonstrate in 
App.~\ref{app:PDframe}.
An immediate implication of this scaling is that for a GW incident on an electromagnetic field that is static in the proper detector frame, the leading order electromagnetic response induced will scale as ${\cal O}[(\omega L)^2]$.\footnote{By gauge invariance, the same must also be true in the TT frame, where $h \propto e^{-\i \omega t}$ and therefore has contributions at all orders in $\omega L$.
This implies there must be a detailed cancellation of the linear frequency contribution, and the need to keep track of this highlights the advantage of working in the proper detector frame.
For additional discussion, see Ref.~\cite{Berlin:2021txa}.}
This demonstrates that the optimal observables for the GW one can construct will also be at ${\cal O}[(\omega L)^2]$.

As outlined in the introduction, one of the primary goals of the present work is to understand the role symmetry plays in the GW interactions.
In particular, we will be studying the interaction between a GW and detectors with a high degree of symmetry.
Existing axion instruments tend to have full cylindrical symmetry, that is, invariance under rotations about $\hat{\bf e}_z$ with an angle $\varphi$, $R_z(\varphi)$, and arbitrary reflections, $P_{\alpha}$, with $\alpha = x,y,z$.
Therefore, it is worthwhile already to characterise the transformation of the GW polarisations and proper detector frame components when these transformations are applied to the position and incident direction at which we evaluate these quantities.
To begin with,
\begin{equation}
e_{ij}^A (P_{\alpha} \hat{\bf k}) = \sigma \, [P_{\alpha}]_{ik}\, e_{kl}^A (\hat{\bf k})\, [P_{\alpha}]_{lj}, \hspace{0.5cm}
e_{ij}^A (R_z \hat{\bf k}) = [R_z]_{ik}\, e_{kl}^A (\hat{\bf k}) \,[R_z]_{lj}.
\label{eq:polunderP}
\end{equation}
Here, $[P_{\alpha}]_{ij}$ is a $3 \times 3$ matrix corresponding to the reflection of the $\alpha$-component, and $[R_z]_{ij}$ is similarly the matrix describing the rotation about $\hat{\bf e}_z$.\footnote{We emphasise that $e_{ij}^A$ transforms under general rotations as a tensor only up to gauge transformations~\cite{Weinberg:1964ew}.
From the definitions in Eq.~\eqref{eq:polarisation}, however, the polarisation tensors are true tensors under rotations about $\hat{\bf e}_z$.}
We have further introduced $\sigma= +1$  ($\sigma = -1$) for the  $A=+$ ($A = \times$) polarisation, which keeps track of their different transformations under the reflections.
From these results, we conclude
\begin{equation}\begin{aligned}
&h_{00}( P_{\alpha} {\bf r}, P_{\alpha}{\bf k}) = \sigma \, h_{00}({\bf r,k}),
\hspace{0.5cm} 
h_{0i}(P_{\alpha} {\bf r}, P_{\alpha}{\bf k} ) = \sigma \, [P_{\alpha}]_{ij}\, h_{0j} ({\bf r,k}), \\
&\hspace{2.cm}h_{ij}(P_{\alpha} {\bf r}, P_{\alpha}{\bf k}) = \sigma \, [P_{\alpha}]_{ik}\, h_{kl}({\bf r,k})\, [P_{\alpha}^T]_{lj},
\label{eq:h_parity}
\end{aligned}\end{equation}
and $h_{\mu \nu}$ transforms as a regular tensor under rotations about the $z$-axis.

\subsection{Effective current induced by GWs}

The interaction of GWs with electromagnetic fields can be effectively described as an additional current augmenting Maxwell's equations in a flat spacetime.
Specifically,
\begin{equation}
\partial_\nu  F^{\mu\nu}  =  j^\mu +j_{\rm eff}^{\mu},
\hspace{0.5cm}
\partial_{\nu} F_{\alpha \beta}+\partial_{\alpha} F_{\beta \nu}+\partial_{\beta} F_{\nu \alpha} =0,
\label{eq:MaxwellGW}
\end{equation}
where $j^\mu  $ is the electromagnetic current in the absence of the GW (i.e. the ordinary currents in flat spacetime) whereas the effective current can be written as
\begin{equation}
j_{\rm eff}^{\mu}  \equiv  \partial_{\nu} \left(	- \frac{1}{2} h F^{\mu\nu} 
+ F^{\mu \alpha} {h^{\nu}}_{\alpha}
- F^{\nu \alpha} {h^{\mu}}_{\alpha}
\right)\!,
\label{eq:effectivecurrent}
\end{equation}
with $h \equiv {h^{\mu}}_{\mu}$.
This is derived in the App.~\ref{app:Maxwell} (see also Refs.~\cite{Herman:2020wao,Berlin:2021txa,Domcke:2022rgu}), where we also discuss why the second equation in Eq.~\eqref{eq:MaxwellGW} -- the homogeneous Maxwell's equations -- are unaffected by the presence of the GW.
Throughout this paper we will be working to linear order in $h$, so that $F^{\mu \nu}$ as it appears on the right-hand side of this equation contains only the background fields.
In further analogy to EM, one can define~\cite{Domcke:2022rgu}
\begin{equation}\begin{aligned}
P_i & \equiv - h_{ij} E_j + \frac{1}{2} h E_i + h_{00} E_i - \epsilon_{ijk} h_{0 j} B_{k}, \\ 
M_i & \equiv - h_{ij} B_j - \frac{1}{2} h B_i + h_{jj} B_i + \epsilon_{ijk} h_{0j} E_{k},
\label{eq:PMdef}
\end{aligned}\end{equation}
so that
\begin{equation}
j_{\rm eff}^{\mu} = \left(	-\nabla \cdot {\bf P},\, \nabla \times {\bf M} + \partial_{t} {\bf P} \right)\!.
\end{equation}

This final formulation is reminiscent of polarisation and magnetisation vectors for EM in a medium.
Hence, the task of calculating the electromagnetic fields induced by a GW is equivalent to performing standard EM calculations in such media.
If we consider the leading order effect in ${\cal O}[(\omega L)^2]$, we can already see a simplification when the external field is purely magnetic: the spatial part of the current will be generated only by ${\bf M}$, and be sourced by $h_{00}$ and $h_{ij}$ but not $h_{0i}$, as the time derivative acting on ${\bf P}$ ensures it will be higher order.

Up to this point, we have not specified the geometry of the background fields or the pickup loop that will be used to measure the induced fields.
The majority of the axion haloscopes in consideration exploit solely an external magnetic field with full cylindrical symmetry.
For such magnetic field configurations it is possible to decompose the background magnetic field into a solenoidal and toroidal piece, ${\bf B}({\bf r}) = {\bf B}^{\solenoid}({\bf r}) + {\bf B}^{\toroid}({\bf r})$.\footnote{We have employed a slight abuse of notation here.
The solenoidal and toroidal components of the field do not correspond to the Helmholtz decomposition, where, for instance, the solenoidal component is divergence free.
Rather, in a sense that we make clear in App.~\ref{app:Bdecomposition}, it is possible to decompose the fields into two components that resemble those of a solenoidal magnet and a toroidal magnet.}
We will further assume the field depends only on $\rho$, namely $|{\bf B}({\bf r})| = B(\rho)$ for an unspecified function $B$.
The benefit of this decomposition is that  ${\bf B}^{\solenoid}({\bf r}) \propto \hat{\bf e}_z$ and ${\bf B}^{\toroid}({\bf r}) \propto \hat{\bf e}_{\phi}$, and each component has a well-defined set of transformations under reflections, which we keep track of through a parameter $\eta_{\alpha}$, defined through ${\bf B} (P_\alpha {\bf r}) = \eta_{\alpha} P_\alpha {\bf B} ({\bf r})$.
In addition to the partial reflections, we will track the transformation under the full parity transformation $P = P_x P_y P_z$, ${\bf B} (P {\bf r}) = \eta P {\bf B} ({\bf r})$, where consistency requires $\eta = \eta_x \eta_y \eta_z$.

For each pair of magnetic field configuration and spatial reflection, explicit values of $\eta_{\alpha}$ are summarised in Tab.~\ref{table:eta}.
Each detector configuration is usually associated uniquely with either ${\bf B}^{\solenoid}$ or ${\bf B}^{\toroid}$.
This is certainly true for the existing and planned axion haloscopes we consider.
Accordingly, we will suppress the superscripts $(s)$ or $(t)$ moving forward.
Combining the transformation properties of the magnetic field with Eq.~\eqref{eq:h_parity}, the spatial part of the induced current -- which fully determines the induced magnetic field -- then obeys
\begin{equation}
{\bf j}_{\rm eff}\left( P_{\alpha} {\bf r},\,P_{\alpha} {\bf k} \right)
= - \sigma  \eta_{\alpha} P_{\alpha}\, {\bf j}_{\rm eff} ({\bf r}, {\bf k}),
\label{eq:j_parity}
\end{equation}
and ${\bf j}_{\rm eff}\left( R_z {\bf r}, R_z {\bf k} \right) = R_z \, {\bf j}_{\rm eff}$.
We emphasise this transformation holds for both the ${\bf M}$ and ${\bf P}$ contributions to ${\bf j}_{\rm eff}$ separately.

\begin{table}[t] \centering \large
\renewcommand{\arraystretch}{1.2}
\begin{tabular}{c|c|c|c|c|c|}
\cline{2-6}
& ${\bf B}^{\solenoid} \propto \hat{\bf e}_z$ (solenoid) &
${\bf B}^{\toroid} \propto \hat{\bf e}_{\phi}$ (toroid) & 
$\hat{\bf n}' \propto \hat{\bf e}_z$ &
$\hat{\bf n}' \propto \hat{\bf e}_{\phi}$ &
$\hat{\bf n}' \propto \hat{\bf e}_{\rho}$
\\ \hline
\multicolumn{1}{ |c| }{ $P_x$ } & $+1$ & $-1$ & $+1$ & $-1$ & $+1$ \\ \hline
\multicolumn{1}{ |c| }{ $P_y$ } & $+1$ & $-1$ & $+1$ & $-1$ & $+1$ \\ \hline
\multicolumn{1}{ |c| }{ $P_z$ } & $-1$ & $+1$ & $-1$ & $+1$ & $+1$ \\ \hline \hline
\multicolumn{1}{ |c| }{ $P$ } & $-1$ & $+1$ & $-1$ & $+1$ & $+1$ \\ \hline
\end{tabular}
\caption{The transformation of a solenoidal and toroidal magnetic field configuration as well as various pickup loop orientations under reflections.
For the magnetic field, the values in the table determine $\eta_{\alpha}$ as defined in ${\bf B}^{(s,t)} (P_\alpha {\bf r}) = \eta_{\alpha} P_\alpha {\bf B}^{(s,t)}({\bf r})$, or $\eta$ in the case of the complete parity transformation $P = P_x P_y P_z$.
The transformation of the normal vector to the pickup loop, $\hat{\bf n}'$, are collected in the final three columns, with $\hat{\bf n}'(P_{\alpha} {\bf r}') = \kappa_{\alpha} P_{\alpha}\hat{\bf n}'({\bf r}')$, and $\kappa = \kappa_x \kappa_y \kappa_z$.
See text for details.}
\label{table:eta}
\end{table}

Before we move on to consider the fields generated by ${\bf j}_{\rm eff}$, we note that from the above discussion we can see the presence of a boundary contribution that has previously been overlooked (for instance, in Ref.~\cite{Domcke:2022rgu}).
Recall that at the interface of two bodies with different values of the magnetisation vector ${\bf M}$, Maxwell's equations predict a surface current proportional to $\hat{\bf n} \times \Delta {\bf M}$,  where $\hat{\bf n}$ is the unit vector normal to the surface.
For the external magnetic fields considered in this work, the GW effective magnetisation ${\bf M}$ in Eq.~\eqref{eq:PMdef} sharply drops to zero at cylindrical surfaces, due to the drop in the external magnetic fields.\footnote{
In the setups of interest here, this sharp drop is due to the configuration of the external electromagnetic current generating the magnetic field (a cylindrical or toroidal spool), and does not require the presence of a conducting shield.
}
If this occurs at a radius $\rho=R$, the GW generates an effective surface current given by ${\bf j}_{\rm eff}^{(t,s)} = \pm \delta(\rho -R)\,\hat{\bf e}_\rho \times {\bf M}$, which must be accounted for.
As already noted, such a contribution does not occur for an axion, but does for a scalar coupled to EM.
(For the axion, such a contribution will not occur as $\Delta {\bf M} \propto \Delta {\bf E} \propto \hat{\bf n}$ at any surface.
Explicit calculations are provided in App.~\ref{app:scalarEM}.)

\subsection{The induced magnetic field}

The effective current induced by the GW will source an induced magnetic field, ${\bf B}_h$, which is determined by the Biot-Savart law,\footnote{This approach is valid only to ${\cal O}[(\omega L)^3]$, see Ref.~\cite{Domcke:2022rgu} for a discussion.
\label{footnote:BiotSavart}} 
\begin{equation}
{\bf B}_h({\bf r}^{\prime},{\bf k}) = \frac{1}{4\pi} \int_{V_B} \hspace{-0.25cm}d^3{\bf r} ~ \frac{{\bf j}_{\rm eff}({\bf r},{\bf k}) \times ( {\bf r}^{\prime} - {\bf r} )}{\vert {\bf r}^{\prime} - {\bf r} \vert^3},
\label{eq:BiotSavart}
\end{equation}
where $V_B$ denotes the detector volume filled by the external magnetic field.
Here and throughout, we use ${\bf r}^{\prime}$ to indicate the position where we evaluate the induced field, and the pickup loop inserted to measure that field will integrate over this variable, whereas ${\bf r}$ is where we evaluate the effective current.
Under the assumption that the integration region $V_{B}$ is invariant under $P_{\alpha}$ -- which it is for the cylindrically symmetric detectors we consider -- the Biot-Savart law Eq.~\eqref{eq:BiotSavart} together with Eq.~\eqref{eq:j_parity} implies the transformation of the induced magnetic field as,
\begin{equation}
{\bf B}_h (P_{\alpha} {\bf r}^{\prime},P_{\alpha} {\bf k}) = \sigma \eta_{\alpha} P_{\alpha}\, {\bf B}_h({\bf r}^{\prime}, {\bf k}),
\label{eq:B_parity}
\end{equation}
and correspondingly ${\bf B}_h (R_z {\bf r}^{\prime}, R_z {\bf k}) = R_z \,  {\bf B}_h({\bf r}^{\prime}, {\bf k})$.
The result in Eq.~\eqref{eq:B_parity} will be a key tool in studying the implications of detector symmetry for the associated GW signal, which we consider in detail in Sec.~\ref{sec:othergeom}.

We now have all the ingredients to compute the effect of GWs on the observable used in low-mass axion haloscopes, namely the induced magnetic flux through a suitably placed pickup loop,
\begin{equation}
\Phi_h =  \frac{1}{2} \int_{A_\ell} F_{\mu \nu}\,  dx^\mu \wedge dx^\nu = \int_{A_{\ell}}\hspace{-0.2cm} d^2{\bf r}^{\prime}\, {\bf B}_h({\bf r}^{\prime}) \cdot \hat{\bf n}^{\prime}({\bf r}'),
\label{eq:flux}
\end{equation}
with $A_{\ell}$ the surface enclosed by the pickup loop, and $\hat{\bf n}^{\prime}({\bf r}')$ the unit normal vector to that surface.
Our symmetry arguments will also depend on the transformation of the pickup loop under reflections, which in direct analogy to the transformation of the magnetic field we will trace using 
\begin{equation}
\hat{\bf n}'(P_{\alpha} {\bf r}') = \kappa_{\alpha} P_{\alpha}\hat{\bf n}'({\bf r}'),
\label{eq:n_parity}
\end{equation}
with the different possible values collected in Tab.~\ref{table:eta}.
In Sec.~\ref{sec:GWforSolenoid}, we will provide explicit expression of $\Phi_h$ for solenoidal geometries and review how to set constraints on GW signals by recasting the results of axion experiments.

Rather than employing Eqs.~\eqref{eq:BiotSavart} and \eqref{eq:flux}, one can also work directly with the vector potential in Coulomb gauge,
\begin{equation}
{\bf A}_h( {\bf r}^{\prime} ) = \frac{1}{4\pi} \int d^3 {\bf r}\, \frac{{\bf j}_{\rm eff}({\bf r})}{\vert {\bf r}^{\prime} - {\bf r} \vert },\hspace{0.5cm}
\Phi_h = \int_{\ell} d {\bf r}^{\prime} \cdot {\bf A}_h ({\bf r}^{\prime}),
\label{eq:BioSavartA}
\end{equation}
where $\ell$ is the closed curve describing the pickup loop.
This approach reduces the dimension of the flux integration by one, thereby simplifying several analytical calculations.
Nonetheless, the symmetry based arguments are often more intuitive when expressed in terms of the magnetic field.
We will use both formalisms as needed.

Moreover, we introduce the {\it response matrix}, $D^{mn}({\bf k})$, to study the dependence of the flux on the polarisation.
The response matrix exploits the observation that the flux in Eq.~\eqref{eq:flux} is a linear functional of the GW, which in turn is linear in the polarisation tensors, $e^A_{mn}$.
Hence, there must exist a matrix $D^{mn}({\bf k})$ such that 
\begin{equation}
\Phi_h =e^{- \i \omega t } D^{m n}({\bf k}) \sum_A h^A e^A_{m n}(\hat{{\bf k}}).
\label{eq:flux_Dmn}
\end{equation}
A more detailed discussion of the response matrix and explicit expressions for $D^{mn}({\bf k})$ are provided in App.~\ref{app:responsematrix}.

Using the response matrix, for a given GW wave vector ${\bf k}$ and polarisation $A$, we can construct pattern functions $D^{mn}({\bf k}) e^A_{mn}$ which encode the angular response of a detector, i.e.\ describe the antenna pattern relevant to determine the magnitude of the induced magnetic flux.
This is analogous to the formalism introduced for the response of interferometers to GWs, see Ref.~\cite{Maggiore:2007ulw}. 
We emphasise that Eq.~\eqref{eq:flux_Dmn} results from the fact that the magnetic flux is linear in the metric perturbation associated with GWs.

Although more details can be found in App.~\ref{app:responsematrix}, let us here provide two remarks on the general properties of $D^{ij}$.
First, as we see from the form of Eq.~\eqref{eq:metricperturbation}, the lowest order frequency contribution to $D^{ij}({\bf k})$ occurs at ${\cal O}[(\omega L)^{2}]$.\footnote{For interferometers, the observable is the GW strain and the calculation can be performed in the TT frame, where the antenna pattern function  at leading order in frequency depends only on the direction of the GW, ${\hat{\bf k}}$.}
This is a generic consequence of the use of proper detector frame.
Secondly, the matrices $D^{ij}$ are not unique, but $D^{ij}({\bf k}) \rightarrow D^{ij}({\bf k}) + c^i k^j  + c^j k^i$ with constants $c^{i,j}$ gives rise to the same magnetic flux since the polarisation tensors are transverse with respect to ${\bf k}$.

\section{The GW Sensitivity of Solenoidal Detector Geometries}
\label{sec:GWforSolenoid}

\begin{figure}[!t]
\centering
\vspace{-0.7cm}
\includegraphics[width=0.6\textwidth]{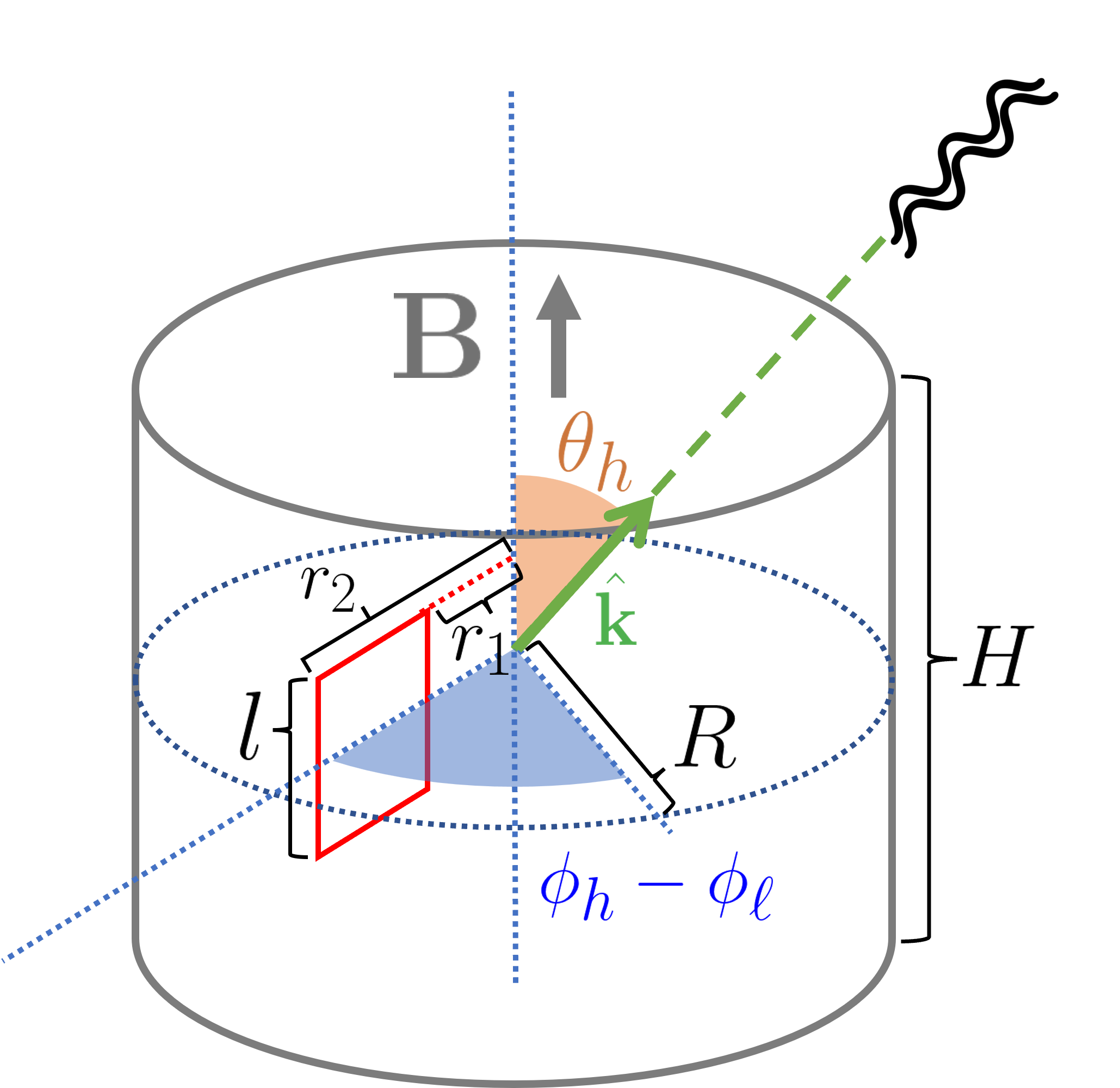}
\caption{A cartoon depiction of the geometry of a solenoidal detector that can be used to search for axions or a passing GW.
We show a solenoidal magnetic field (gray) with a rectangular vertical pickup loop (red).
The solenoid has a height $H$ and radius $R$, and throughout we will generally assume that $H$ is parametrically larger than all other scales.
The pickup loop is located at an angle $\phi_{\ell}$, and spans over coordinates $\rho \in [r_1,r_2]$ and $z \in [-l/2,l/2]$, giving it an area $l(r_2-r_1)$.
Finally, in green we depict the direction of the incident GW, which has a wave vector ${\bf k}$, and comes in at an angle in spherical coordinates of $\theta_h$ and $\phi_h$.}
\label{fig:solenoid}
\end{figure}

Having established the general framework of how a GW interacts with an experimental magnetic field, we now put it to use for the explicit case of solenoidal instrument.
Previous work, see Ref.~\cite{Domcke:2022rgu}, focused on a toroidal geometry for the external magnetic field, which is used for the axion searches performed by ABRACADABRA~\cite{Salemi:2021gck} and SHAFT~\cite{Gramolin:2020ict}.
This is modelled as
\begin{equation}
{\bf B} = B_{\max} \frac{R}{\rho} \left[ \Theta(R+a - \rho) - \Theta(R - \rho)  \right] \hat{\bf e}_{\phi},
\label{eq:toro_mag_field}
\end{equation}
where $\Theta(x)$ is the Heaviside step function, which ensures the external magnetic field only exists on $R < \rho < R+a$.
However, another magnetic field geometry that is being pursued by axion haloscopes is a {\it soleneoidal} field,
\begin{equation}
{\bf B} = B_0  \Theta(R - \rho) \hat{\bf e}_z.
\label{eq:sol_mag_field}
\end{equation}
Among the detectors making use of a solenoidal field, we will first focus on instruments where the induced flux is read out through a vertical pickup loop as depicted in Fig.~\ref{fig:solenoid} and implemented in the ADMX SLIC~\cite{Crisosto:2019fcj} and BASE~\cite{Devlin:2021fpq} experiments.
Moreover, the planned WISPLC~\cite{Zhang:2021bpa} and DMRadio-$ \text{m}^3$~\cite{DMRadio:2022pkf,DMRadio:2023igr} experiments are also planning to implement a related configuration.\footnote{Let us briefly comment on several of the differences between these experiments.
ADMX SLIC experiment has a single rectangular pickup loop at a fixed polar angle $\phi_\ell$.
The BASE experiment relies instead on many such pickup loops placed symmetrically in the horizontal plane, whereas DMRadio-m$^3$ will use a full toroidal sheath.
For the WISPLC experiment, the current design features a pickup loop at fixed $\phi_\ell$, but which is located outside the region of the external magnetic fields.}
(Although other future instruments will use a toroidal magnetic field, for instance DMRadio-50L.)
Therefore, as a straightforward generalisation of Ref.~\cite{Domcke:2022rgu}, we first calculate the expected magnetic flux from the incoming GW for a solenoidal magnetic field and different locations of the pickup loop.
Armed with an understanding of how the GW interacts with a detector for two explicit cases, in the next section we will then generalise our discussion for general geometries.

Before proceeding, we note that in the main text we will only consider the interaction between GW and laboratory magnetic fields.
The rationale for this is that axion haloscopes exclusively make use of magnetic fields, as larger energy densities can be built up in magnetic than electric fields.
Further, the axion interaction with a magnetic field is controlled by $\partial_t a$, which for dark matter is much larger than $\nabla a$, which the electric field couples to (for a review, see App.~\ref{app:scalarEM}).
As the GW is both relativistic and couples differently than the axion, this final consideration does not apply, and therefore for completeness we briefly discuss the interaction with an electric field in App.~\ref{app:Efield}.

\subsection{The GW signal for a solenoidal magnetic field}

Consider first the flux $\Phi_h(r)$ caught by a rectangular pickup loop at fixed polar angle $\phi_\ell$, radially ranging from $[0,r]$ with height $l$, and positioned symmetrically about $z=0$.
This scenario is equivalent to that depicted in Fig.~\ref{fig:solenoid} with $r_1=0$ and $r_2=r$.
We can then derive the equivalent result for an arbitrary width with $r_1 < r_2 \leq R$ from $\Phi_h(r_2) - \Phi_h(r_1)$.
Further, we will consider the case where $r \leq R$, with $R$ the radius of the detector, and $r \geq R$ separately.

For the solenoidal magnetic field in Eq.~\eqref{eq:sol_mag_field}, we can calculate the effective current, induced magnetic fields, and flux using the formalism of Sec.~\ref{sec:framework}.
As already discussed, we will study the problem perturbatively in $\omega L \ll 1$.
Complete expressions for all components of the current to order ${\cal O}[(\omega L)^3]$ are provided in App.~\ref{app:explicitcurrent}.
Here, we state the results for the flux, which we write as a series in $\omega$ as $\Phi_h = \Phi_h^{(2)} +  \Phi_h^{(3)} + \cdots$, with $\Phi_h^{(n)} $ denoting the flux at ${\cal O}[(\omega L)^n]$.
Explicitly, when $r \leq R$ we have
\begin{equation}
\Phi_h^{(2)} = \frac{e^{-\i \omega t}}{144 \sqrt{2} }\omega^2 B_0 l r \left( 30R^2 - 13 r^2 \right)
s_{\theta_h} \left(	h^+ c_{\theta_h} s_{\phi_h - \phi_{\ell}} + h^{\times} c_{\phi_h - \phi_{\ell}} \right)\!,
\label{eq:Fluxomega2}
\end{equation}
and
\begin{equation}\begin{aligned}
\Phi_h^{(3)}  =  & - \frac{\i e^{-\i\omega t}}{2304 \sqrt{2}} \omega^3 B_0 l r^2 \Big[
h^+ c_{\theta_h} s_{2(\phi_h - \phi_{\ell})} \{ 3 l^2 \!-\! 2 r^2 \!+\! 57 R^2 \!+\! ( l^2 \!-\! 22 r^2 \!+\! 27 R^2) c_{2\theta_h} \} \\
&+ 2 h^{\times} \{  (l^2 \!+\! 2 r^2 \!+\! 18R^2 \!+\! [l^2 \!-\! 14 r^2 \!+\! 24 R^2] c_{2\theta_h} )c_{2(\phi_h - \phi_{\ell})} \!+\! 6(5r^2 \!-\! 12R^2) s_{\theta_h}^2 \} \Big],
\label{eq:Fluxomega3}
\end{aligned}\end{equation}
where we employ the shorthands $c_x \equiv \cos x$ and $s_x \equiv \sin x$.
The factor of $\i$ in $\Phi_h^{(3)}$ indicates that this contribution enters with a $\pi/2$ phase shift in time as compared to $\Phi_h^{(2)}$.
If the loop instead is placed outside the solenoidal magnetic field, extending over $\rho^{\prime } \in [r_1,r_2]$ with $r_2> r_1 > R$, then the leading order flux is
\begin{equation}
\Phi_h^{(2)} = \frac{5 e^{-\i \omega t}}{48 \sqrt{2}}\omega^2 B_0  l R^4\, \frac{r_2 - r_1}{r_1 r_2} s_{\theta_h} \left(	h^+ c_{\theta_h}  s_{\phi_h - \phi_{\ell}} + h^{\times} c_{\phi_h - \phi_{\ell}}	\right)\!, 
\label{eq:Fluxomega2WISPLC}
\end{equation}
and
\begin{equation}\begin{aligned}
\Phi_{h}^{(3)} = &\frac{\i e^{- \i \omega t}}{2304\sqrt{2}} \omega^3 B_0 \frac{l R^4}{(r_1 r_2)^2}
\Bigg[ 2 h^{\times} \Bigg\{ - 72 (r_1 r_2)^2 \ln \frac{r_1}{r_2}  s_{\theta_h}^2 \\
&\hspace{1.2cm} + c_{2(\phi_h - \phi_{\ell})} \left( (r_2^2 - r_1^2) (l^{2} + 2R^{2} + (l^{2} - 8R^{2})c_{2\theta_{h}} ) + 24 (r_{1} r_{2})^{2} \ln \frac{r_{1}}{r_{2}} \right) \Bigg\} \\
&\hspace{1.2cm} + h^+ c_{\theta_h} s_{2(\phi_h - \phi_{\ell})} 
\Bigg\{ (3 l ^2 + R^2) (r_2^2 - r_1^2) + 60 (r_1 r_2)^2 \ln \frac{r_1}{r_2} \\
&\hspace{4.2cm}- \left((13R^2 - l^2 )(r_2^2 - r_1^2) + 12 (r_1 r_2)^2 \ln \frac{r_1}{r_2} \right) c_{2\theta_{h}} \Bigg\}\Bigg].
\label{eq:Fluxomega3WISPLC}
\end{aligned}\end{equation}
These are the appropriate results one would use for WISPLC.

Consider the symmetry of the above expressions.
Firstly, observe that $\Phi_h^{(2)} \neq 0$, $\Phi_h^{(3)} \neq 0$, and in fact both polarisations contribute at each order.
However, with a single pickup loop at $\phi=\phi_{\ell}$, the considered configuration manifestly breaks azimuthal symmetry.
If we were to restore this symmetry -- through an array of pickup loops arranged symmetrically in $\phi$ as for BASE, or with a coaxial arrangement as for DMRadio-m$^3$ -- then all trigonometric functions with the argument $\phi_h-\phi_{\ell}$ will vanish, and we would conclude $\Phi_h^{(2)} = 0$ and $\Phi_h^{(3)} \propto h^{\times}$, the $h^+$ contribution also vanishing.
Cylindrical symmetry hence induces a cancellation of the leading order sensitivity.
This same conclusion was reached for a toroidal magnetic field in Ref.~\cite{Domcke:2022rgu}.
There, it was shown that breaking the azimuthal symmetry by using a non-circular pickup loop restored sensitivity at ${\cal O}[(\omega L)^2]$, exactly as we find in Eqs.~\eqref{eq:Fluxomega2} and \eqref{eq:Fluxomega2WISPLC}.
In particular, Ref.~\cite{Domcke:2022rgu} showed that by using a magnetometer or figure-8 configuration of the pickup loop, the GW flux was given by
\begin{equation}
\Phi_8^{\toroid} = \frac{e^{-\i \omega t}}{3\sqrt{2}} \omega^2 B_{\max} r^3 R \ln(1+a/R) s_{\theta_h} \left( h^{\times} s_{\phi_h} - h^+ c_{\theta_h} c_{\phi_h} \right)\!.
\label{eq:toroid-8}
\end{equation}
We can imagine implementing a similar pickup loop configuration in the solenoidal case.
For instance, we can take Eq.~\eqref{eq:Fluxomega2} and combine the contribution from $\phi_{\ell} \in [0,\pi)$ with a $\pi$ phase shift to that from $\phi_{\ell} \in [\pi,2\pi)$.
(For BASE, we can conceive of implementing this by changing the orientation of the winding of the loops on, for instance, $0\leq\phi_\ell<\pi$.)
Doing so, we obtain,\footnote{This is an oversimplification.
One cannot simply add the contribution from a set of differentially spaced loops, as this neglects the mutual inductances between the loops.
Instead, one would need to model the full sheath and compute the induced current density across across it.
We thank Joshua Foster for discussions on this point.
\label{footnote:solenoid-8}}
\begin{equation}
\Phi_8^{\solenoid} = \frac{e^{-\i \omega t}}{36 \sqrt{2} }\omega^2 B_0 l r \left( 30R^2 - 13 r^2 \right)
s_{\theta_h} \left(	h^{\times} s_{\phi_h} - h^+ c_{\theta_h} c_{\phi_h} \right)\!.
\label{eq:Fluxomega2-8}
\end{equation}
The results in Eqs.~\eqref{eq:toroid-8} and \eqref{eq:Fluxomega2-8} are very similar -- if we take $B_0=B_{\max}$ and set all spatial scales to $L$, the two only differ by a factor of $\simeq 2$ -- suggesting that the change in geometry has only a minor impact.
In Sec.~\ref{sec:othergeom}, we will explain the origin of these cancellations in the instruments with full cylindrical symmetry in terms of the symmetry transformations of the various quantities introduced in Sec.~\ref{sec:framework}, and in particular explain how the appearance or absence of different terms can be understood without an explicit calculation.
Such arguments will allow us to determine optimised detector geometries that maximise the GW sensitivity.

\begin{figure}[!t]
\centering
\includegraphics[width=0.45\textwidth]{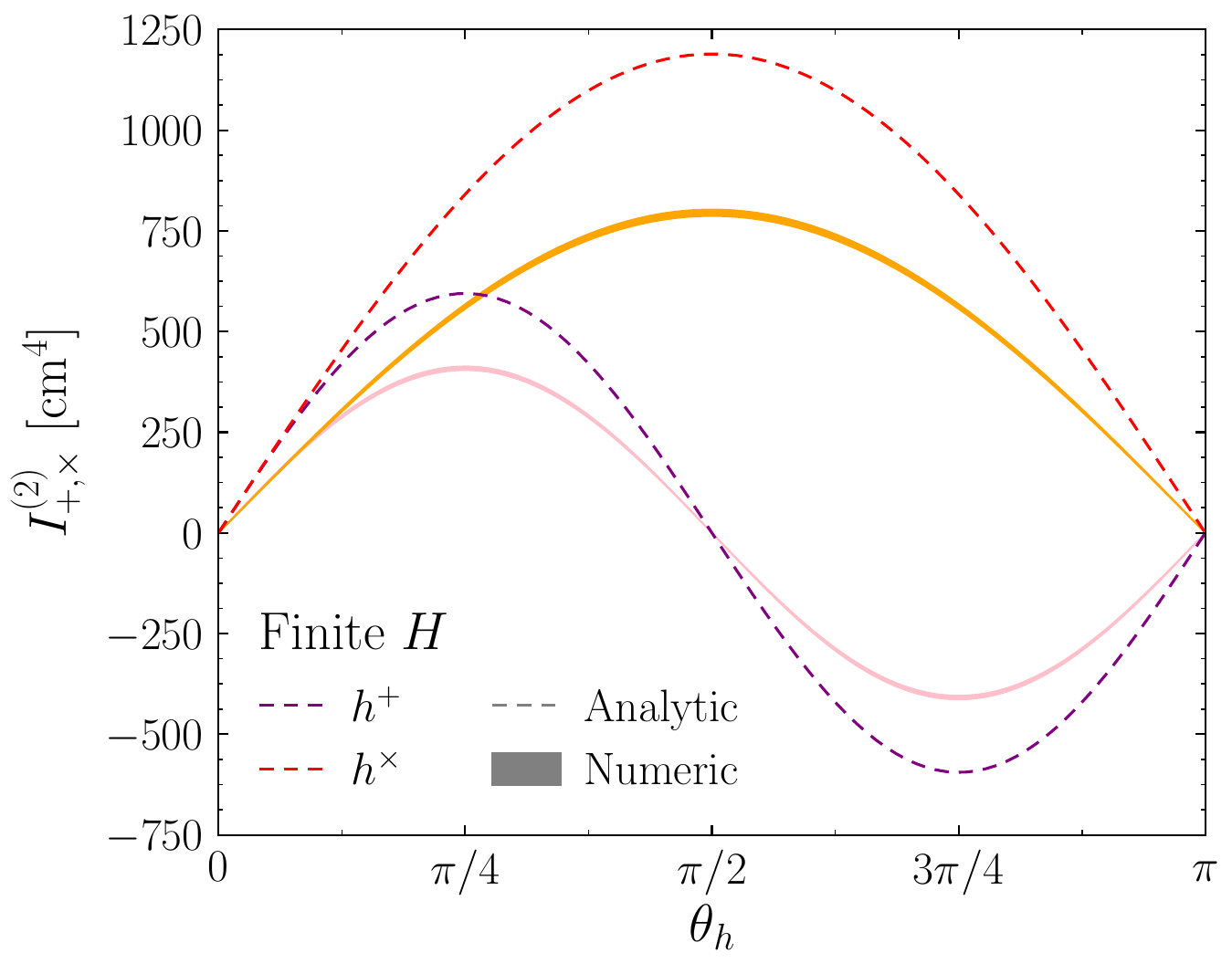}
\hspace{0.5cm}
\includegraphics[width=0.45\textwidth]{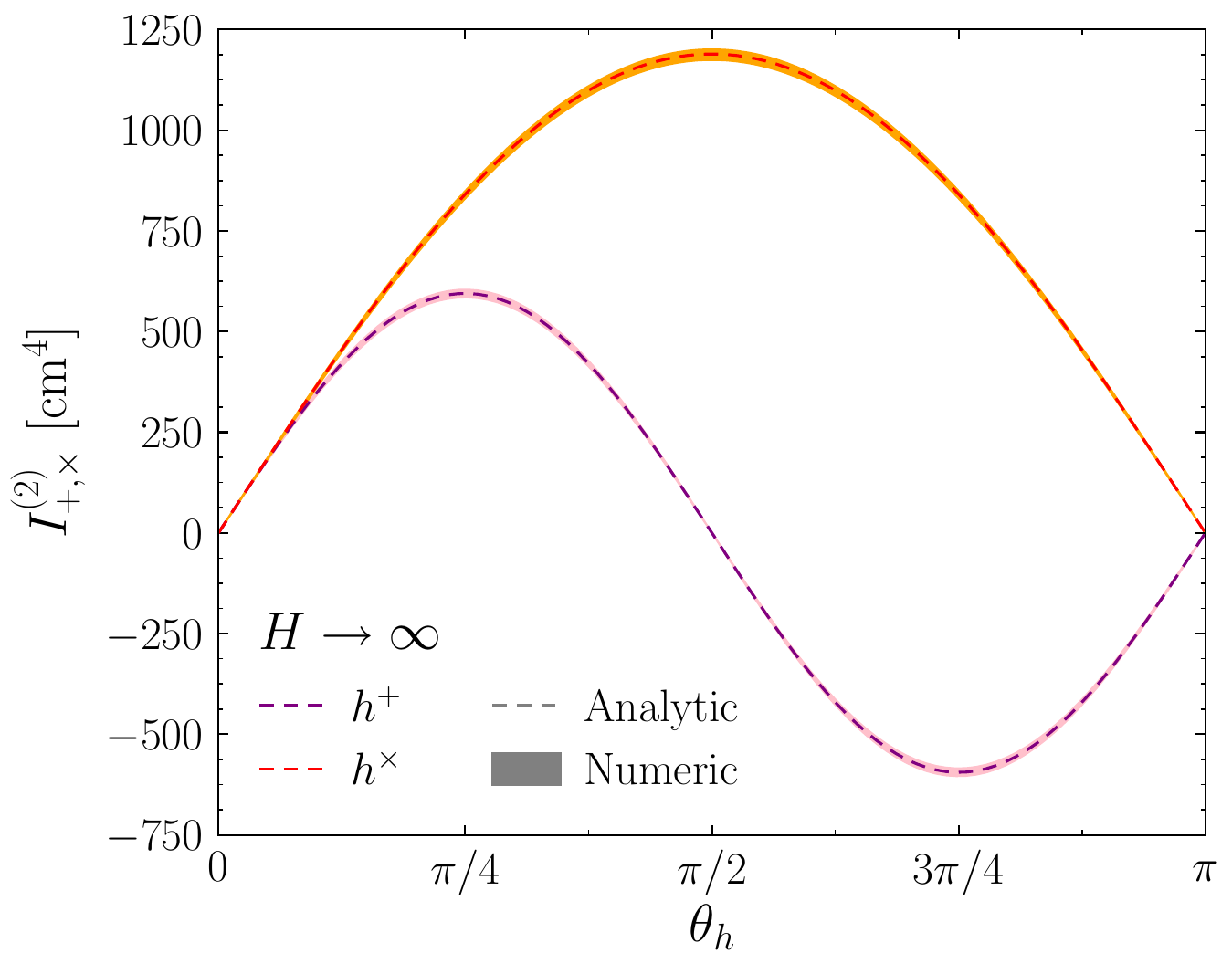}
\caption{A comparison between the analytic flux in Eq.~\eqref{eq:Fluxomega2} (dashed) to the exact result determined numerically (solid band), for the finite (left) and large $H$ (right) cases.
In detail, we show the non-trivial part of flux specified in Eq.~\eqref{eq:Idef}.
As expected, agreement is observed when $H \to \infty$, whereas an offset is seen for finite $H$, see text for a discussion.
For the geometry (see Fig.~\ref{fig:solenoid}), we adopt the various parameters used by ADMX SLIC~\cite{Crisosto:2019fcj}: $r_1=0$, $r_2 = 7.62~{\rm cm}$, $l=31.25~{\rm cm}$, $R=8.55~{\rm cm}$, and when finite $H=40~{\rm cm}$.
We take $\phi_{\ell}=0$, $\phi_h=\pi/4$, and show results as a function of $\theta_h$.}
\label{fig:NumericComparison}
\end{figure}

We emphasise that the analytic results above all assume the height of the solenoidal magnet, $H$, is parametrically larger than all other scales.
That we had to assume this is not unique to the GW signal, the equivalent axion flux (discussed in the next subsection) also only has an analytic form for parametrically large $H$.
In Fig.~\ref{fig:NumericComparison} we compare the leading order flux from the analytic result in Eq.~\eqref{eq:Fluxomega2}, to the exact result determined numerically, where we use the currents in App.~\ref{app:explicitcurrent}, and with the bands denoting the uncertainty in the integration.\footnote{As a point of caution, we note that the analytic results must be treated carefully as there can be branch cuts in the integration.}
In particular, we compute the non-trivial part of the flux, captured by $I^{(2)}_{+,\times}$ in
\begin{equation}
\Phi_h^{(2)} = e^{-\i \omega t} \omega^2 B_0 \left(h^+ I^{(2)}_+ + h^{\times} I^{(2)}_{\times} \right)\!.
\label{eq:Idef}
\end{equation}
We take all dimensions to match those of ADMX SLIC, and show results for the physical $H$ and $H \to \infty$.
Good agreement is observed in the large $H$ limit, whereas for finite $H$ the analytic results overestimate the amplitude of the flux.
This can be understood from considering the integral over $z$ that appears in Eq.~\eqref{eq:BiotSavart}: schematically, we have $\int_{-H/2}^{H/2} dz\, [A + (z-z')^2]^{-3/2} = 2/A - 4/H^2 + {\cal O}(H^{-4})$, so that the first term neglected will suppress the flux amplitude.

\subsection{Strain sensitivity from recasting axion limits and projections}
\label{subsec:GWSensitivity}

Combining the above results, we now determine the GW sensitivity of axion haloscopes making use of a solenoidal external magnetic field.
In detail, we recast the constraints on the axion photon coupling $\gagg$ from BASE~\cite{Devlin:2021fpq} and ADMX SLIC~\cite{Crisosto:2019fcj}, as well as the future instruments WISPLC~\cite{Zhang:2021bpa} and DMRadio-m$^3$~\cite{DMRadio:2022jfv}, as expected sensitivities to the amplitude of GWs.
All of these instruments are designed with the goal of detecting a coherently oscillating axion dark matter background field, which takes the schematic form
\begin{equation}
a(t) = \frac{\sqrt{2 \rho_{\scriptscriptstyle  {\rm DM}}}}{m_a} \sin (m_a t),
\end{equation}
with $\rho_{\scriptscriptstyle {\rm DM}}$ the local dark-matter density, and $m_a$ the unknown dark-matter mass.
In the presence of a magnetic field, the axion background generates an effective current ${\bf j}_{\rm eff} = \gagg (\partial_t a) {\bf B}$.
For the configuration in Fig.~\ref{fig:solenoid} the current induces the following magnetic flux,
\begin{equation}
\Phi_a = \frac{1}{4} \gagg (\partial_t a) B_0 l (r_2^2 - r_1^2) + {\cal O}(H^{-2}),
\label{eq:Phia}
\end{equation}
where the result for finite $H$ can again be determined numerically.

Our goal is to derive sensitivity to the GW strain, $h$, by reinterpreting results on $\gagg$, established assuming a signal flux as in Eq.~\eqref{eq:Phia} (modified for the specific experimental configuration).
To do so, we compare $\Phi_a$ and $\Phi_h$, but further we account for the fact that in all expressions derived so far the axion and GW are treated as persistent monochromatic waves, when this is not the case for either one.
The dark-matter axion is indeed persistent, but has a coherence time of $\tau_a = 2 \pi Q_a/m_a$, with a quality factor of $Q_a \sim 10^6$, indicating a highly coherent signal.
The GW signal on the other hand is model dependent (specific examples of superradiance and primordial black hole (PBH) mergers are discussed in App.~\ref{app:CoherenceRatio}), but can be described as lasting for a duration $T_h$ with coherence time scale $\tau_h$, centered at a frequency $f$, so that the signal has a quality factor $Q_h = \tau_h f$.
For resonant instruments, one must also account for the quality factor and coherence time of the instrument, given by $Q_r$ and $\tau_r = Q_r/f$.
The ultimate strain sensitivity is determined from considering the interplay of each of these scales, together with the experimental run time, an analysis of which we provide in App.~\ref{app:CoherenceRatio}.
The end result is that rather than simply matching the GW and axion fluxes, we instead arrive at
\begin{equation}
\Phi_h( h^+, h^{\times};\,\phi_h, \theta_h) = {\cal R}_c \, \Phi_a(\gagg),
\label{eq:fluxmatch}
\end{equation}
where the \textit{coherence ratio} ${\cal R}_c$ accounts for the difference in coherence between the signals.
As defined, ${\cal R}_c > 1$ implies the GW signal is harder to detect than a naive matching of the flux would suggest, as for a fixed $\Phi_a$ we want to probe the smallest $h \propto \Phi_h$ values possible.

\begin{figure}[!t]
\centering
\includegraphics[trim=0.2cm 0.2cm 0.2cm 0.15cm,clip,width=0.485\textwidth]{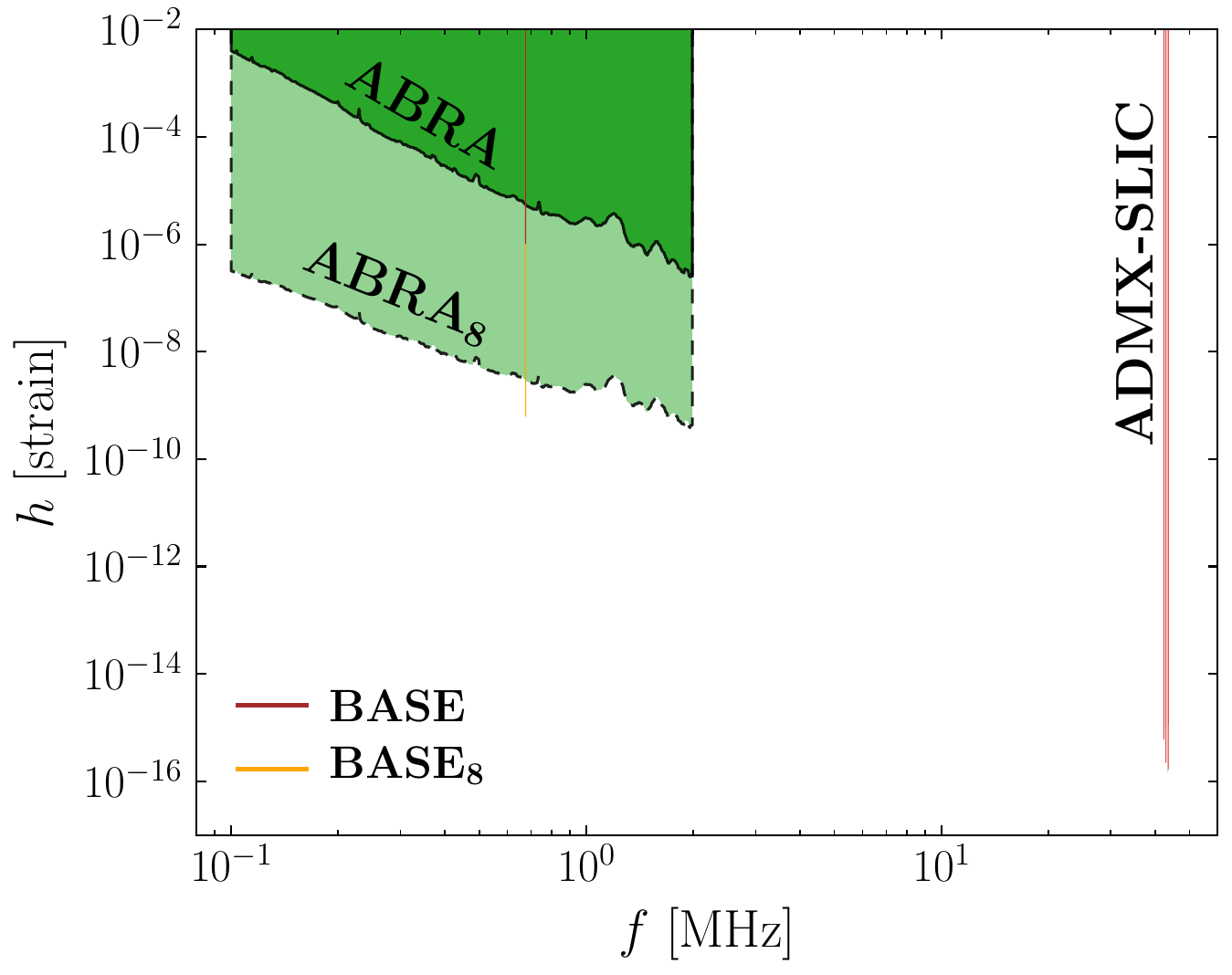}
\hspace{0.1cm}
\includegraphics[trim=0.2cm 0.2cm 0.2cm 0.15cm,clip,width=0.485\textwidth]{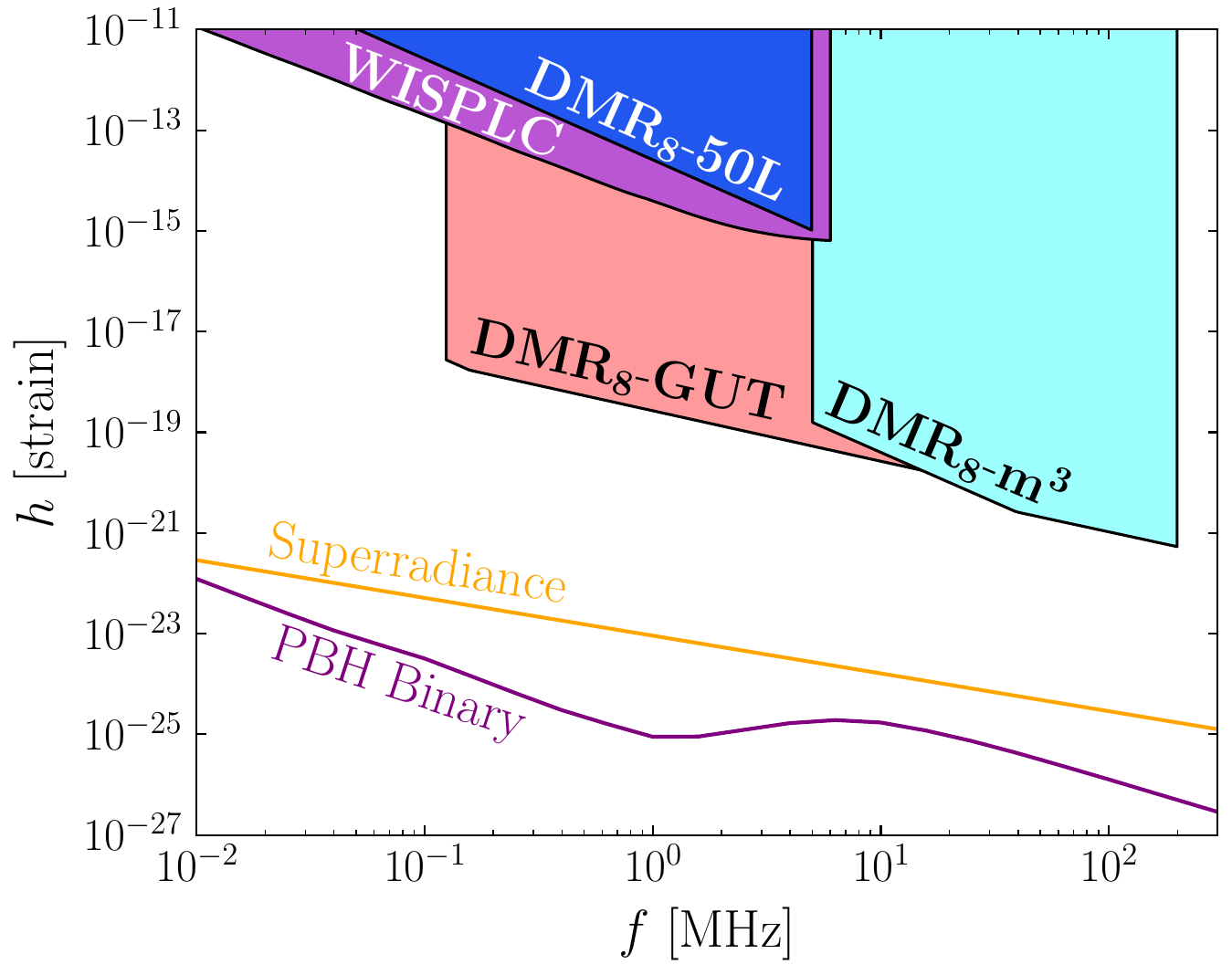}
\caption{Recast GW sensitivity for low-mass axion haloscopes.
On the left, we show the existing results from ABRA~\cite{Salemi:2021gck} (the sensitivity to SHAFT~\cite{Gramolin:2020ict} is similar), BASE \cite{Devlin:2021fpq}, and ADMX SLIC \cite{Crisosto:2019fcj}.
ADMX SLIC already achieves sensitivity to the leading ${\cal O}[(\omega L)^2]$ GW contribution, whereas the cylindrical symmetry of ABRA and BASE lead that to cancel.
Accordingly, we also show those instruments sensitivity to a figure-8 pickup loop configuration.
On the right we show the future reach to WISPLC~\cite{Zhang:2021bpa} and the three components of the DMRadio program~\cite{DMRadio:2022pkf,DMRadio:2022jfv}: 50L, m$^3$, and GUT.
We assume a figure-8 readout for all three DMRadio results, the toroidal result in Eq.~\eqref{eq:toroid-8} for 50L and GUT, and the solenoidal analogue in Eq.~\eqref{eq:Fluxomega2-8} for m$^3$.
In all cases, the sensitivity is determined from Eq.~\eqref{eq:fluxmatch} with ${\cal R}_c=1$.\protect\footnotemark~
Accordingly, to determine the detectability of any signal ${\cal R}_c$ must be included in the signal prediction, which we do for a putative PBH inspiral or axion superradiance signal.
See text for details.}
\label{fig:result}
\end{figure}

Here, we will restrict our attention to a single case, where we take $T_h = \tau_h$, and imagine a resonant instrument that spends a time $T_m \gg \tau_a,\, \tau_h,\, \tau_r$ scanning each axion mass.
In this case, the coherence ratio can be computed to be
\begin{equation}
{\cal R}_c =
\left( \frac{T_m}{\tau_h} \right)^{1/4}
\left( \frac{Q_a}{Q_h} \right)^{1/4} 
\left\{ \begin{array}{lcc}
1 & & Q_r <  Q_a,\,Q_h,  \\
(Q_a/Q_r)^{1/4} & & Q_a < Q_r < Q_h, \\
Q_r/Q_h& & Q_h < Q_r < Q_a, \\
(Q_a/Q_r)^{1/4}Q_r/Q_h  & & \textrm{otherwise}.
\end{array}\right.
\label{eq:Rexplicit}
\end{equation}
This expression is derived in App.~\ref{app:CoherenceRatio}, however, let us briefly describe the physical origin of each term.
The first factor encodes the suppression that arises as the GW signal does not persist for the full time the instrument scans this frequency.
The quarter scaling follows our assumption that $T_m$ is the largest scale, implying that the sensitivities have entered the asymptotic scaling regime consistent with the Dicke radiometer equation~\cite{Dicke:1946glx} (see also Ref.~\cite{Dror:2022xpi}).
The $(Q_a/Q_h)^{1/4}$ factor arises as signals that are more coherent are easier to discover, and was used in Refs.~\cite{Dror:2021nyr,Domcke:2022rgu}.
More coherent signals are narrower in the frequency domain, and therefore can generally be teased out over a smaller amount of background.
In addition, for $Q_a < Q_r$ there is a slight penalty to the axion signal of $(Q_a/Q_r)^{1/4}$ as the full axion signal is not resolved by the instrument.
Finally, there is a strong penalty of $Q_r/Q_h$ that applies to the GW signal whenever $Q_r > Q_h$.
When this occurs, the GW fails to fully ring up the resonance response of the instrument, which strongly decreases the power it deposits, explaining the linear scaling of this factor, as opposed to the quarter scaling of all others.
Using Eq.~\eqref{eq:Rexplicit} -- and its generalisations in App.~\ref{app:CoherenceRatio} -- for any given instrument, we can translate GW signals defined by their amplitude $h$, duration $T_h$, and coherence time $\tau_h$, to an effective signal strength $h/{\cal R}_c$.
We can then compare $h/{\cal R}_c$ to the equivalent plane wave sensitivity shown in Fig.~\ref{fig:result}, which is derived simply from matching the flux; explicitly, Eq.~\eqref{eq:fluxmatch} assuming ${\cal R}_c=1$.

\footnotetext{For comparison, Fig.~1 in Ref.~\cite{Domcke:2022rgu} employed ${\cal R}_c=\left(Q_a/Q_h\right)^{1/4} = 10^{3/4}$ for their adopted $Q_a=10^6$ and $Q_h=10^3$.
As explained in App.~\ref{app:CoherenceRatio}, this is only appropriate when comparing persistent signals (which PBHs are not) and in the case where $Q_a$ is not smaller than both $Q_r$ and $Q_h$.}

For the results in Fig.~\ref{fig:result}, the frequency range is fixed by the corresponding axion mass range, and therefore falls into the MHz band.
The left figure demonstrates the estimated sensitivity of three existing instruments: ABRA, BASE, and ADMX SLIC (the reach for SHAFT is comparable to ABRA, see Ref.~\cite{Domcke:2022rgu}).
Note that BASE and ADMX SLIC perform a resonant search strategy for the axion, and therefore at present have a deeper sensitivity, but narrower frequency coverage than ABRA which completed a broadband search.
For ABRA and BASE, the cylindrical symmetry of these instruments suppresses the leading order flux, for reasons we demonstrate in the next section.
For this reason we also show the sensitivity the instruments could obtain if they implemented a figure-8 style geometry, as given in Eqs.~\eqref{eq:toroid-8} and \eqref{eq:Fluxomega2-8}.
On the right we show the sensitivity for future instruments, in particular DMRadio and WISPLC.
For DMRadio, we show the projected reach of the 50L, m$^3$, and GUT variants of these instruments, in each case assuming they have adopted a figure-8 style readout.
We assumed a solenoidal magnet for m$^3$, but toroidal for 50L and GUT.
Beyond this, the only difference to these DMRadio projections and those derived in Ref.~\cite{Domcke:2022rgu} is the use of the $H \to \infty$ results and the treatment of the coherence ratio, which here we take as ${\cal R}_c=1$, leaving it for the signal predictions.\footnote{In detail, for 50L we took $r=a=R = 0.11\,{\rm m}$, for GUT $r=a=R = 0.64\,{\rm m}$, and for m$^3$, $R=0.8\,{\rm m}$, $r_1 = 0.34\,{\rm m}$, $r_2 = 0.64\,{\rm m}$, $l=1.4\,{\rm m}$.
For WISPLC, we used $R = r_1 = 0.063\,{\rm m}$, $r_2 = 0.19\,{\rm m}$, $l = 0.57\,{\rm m}$.
}
For WISPLC, we have repurposed their anticipated axion sensitivity with a resonant detection strategy, and assumed a single pickup loop employing Eq.~\eqref{eq:Fluxomega2} (more precisely, WISPLC has two individual pickup loops~\cite{Zhang:2021bpa}).
For these cases, we chose the angular direction of GW and relative pickup loop direction which maximise the GW sensitivity.
In computing all fluxes, we have adopted the $H \to \infty$ results; cf. Fig.~\ref{fig:NumericComparison}.
Further, as the GW signal depends on the incident direction, in each case we have simply taken one of the two polarisations, and chosen the incident direction to maximise the signal, although performing an angular average instead would only minimally impact the results.

For comparison, on the right we also draw the expected effective signals $h/{\cal R}_c$ coming from superradiance or PBH binaries as benchmarks for various parameter choices.
Details are given in App.~\ref{app:CoherenceRatio} and effective signals with different parameters for superradiance and PBHs are given in Figs.~\ref{fig:superradiance} and \ref{fig:pbh signal} respectively.
For drawing the signal curves in the right panel of Fig.~\ref{fig:result} we adopt the following simplified  experimental setups: $Q_r = 10^4$ and $T_m = 10^3$~s for superradiance while $Q_r = 10^4$ and $T_m = 1\,{\rm ms}$ for PBH.
Note that in the latter case, the signal depends on the PBH masses and the curve presented in the figure corresponds to the maximal signal for a given frequency. Further
example values of ${\cal R}_c$ for different instruments and two different benchmark signals are provided in Tab.~\ref{tab:benchmarks}, {with $f_*$ denoting a suitable detector reference frequency. 
The first benchmark (B1) is a GW signal of duration $1\,{\rm s}$, with a coherence limited only by its finite duration.
The second (B2) is an example of a persistent highly coherent signal.
In both cases, we assume the frequency spectrum of the signal to be centered around the detectors resonant frequency.
These two benchmarks are loosely inspired by the properties of GWs from primordial black hole mergers and from superradiance, respectively.
Further discussion of each is provided in App.~\ref{app:CoherenceRatio}.
The advantage of these simplified signals is to facilitate comparisons across different detector concepts for astrophysical signals. This is an alternative to the comparison of sensitivity curves in terms of noise spectral densities (see e.g. Ref.~\cite{Tobar:2022pie,Berlin:2023grv}) which requires detailed knowledge of the relevant noise sources.\footnote{This information is well known to the respective experimental collaborations.
If, in addition to publishing limits on $g_{a\gamma\gamma}$, experimental collaborations published a spectral density of axion-photon theta angle noise as proposed in Ref.~\cite{Tobar:2022pie}, the relevant noise spectral density for GW searches can be immediately obtained from the ratio of the induced magnetic flux sourced by the axion and by the GW, as derived here (corresponding to ${\cal R}_c = 1$).}
A further alternative is the use of the characteristic strain, $h_c\sim Q_h^{1/2} h$, as advocated e.g. in Ref.~\cite{Aggarwal:2020olq}.
However, in realistic cases the timescales mentioned above enter into the estimation of the characteristic strain, and hence again detailed knowledge of signal and detector properties becomes necessary for comparisons.
Of course, the method used here of defining simplified benchmarks for comparisons also has its drawbacks.

We can also use the formalism introduced above to estimate the sensitivity to stochastic gravitational backgrounds.
Their energy is constrained by BBN and CMB observations to $\rho_\text{gw}/\rho_c \lesssim 10^{-5} \Delta N_\text{eff}$ with $\rho_c$ denoting the critical energy density of the Universe today and $\Delta N_\text{eff} \lesssim 0.1$~\cite{Yeh:2020mgl}.
It can be expressed as $\rho = \rho_c \int d\ln \omega  \, \Omega_\text{gw} \sim \omega^2 M_\text{Pl}^2 h_c^2$ with $\Omega_\text{gw} = \omega^2 h_c^2 M_\text{Pl}^2 / 2 \rho_c$ denoting the GW spectrum expressed in terms of the characteristic strain $h_c$.
Comparing this with the energy in a plane gravitational wave, $\rho_\text{gw} \sim  \omega^2 M_\text{Pl}^2 h^2$, we identify $h \sim h_c$, so that the BBN and CMB bounds on the characteristic strain, $h_c \lesssim 10^{-29} (100~\text{MHz}/f) \Delta N_\text{eff}^{1/2}$ indicate the maximal GW strain achievable from cosmological stochastic backgrounds. 
The remaining task is to estimate the coherence ratio ${\cal R}_c$ in this case. Setting $\tau_h = T_h = T_m$ and $Q_h \sim 1 \ll Q_a, Q_r$ we obtain ${\cal R}_c  \gg 1$.
In other words the low coherence further suppresses the effective signal strength, which, together with the bound on $\Delta N_\text{eff}$ implies that these signals are unfortunately currently out of range by several orders of magnitude.
Of course, recasting axion searches for highly coherent signals is not the optimal strategy to search for stochastic backgrounds.
Nevertheless, this simple estimate illustrates the challenges that such a dedicated search will be facing.

Figure~\ref{fig:result} demonstrates that axion haloscopes can place competitive bounds on GWs in this frequency range. Due to the challenges mentioned above in comparing different sensitivity estimates, we refrain here from including other detector concepts in the this figure.
This is, however, a very active field and other concepts such as bulk accoustic wave devices~\cite{Goryachev:2014yra}, levitated sensors~\cite{Aggarwal:2020umq}, interferometers~\cite{Holometer:2016qoh}, other electromagnetic GW detectors~\cite{Ringwald:2020ist,Berlin:2021txa,Berlin:2023grv,Bringmann:2023gba} and indirect detection methods~\cite{Fujita:2020rdx,Domcke:2020yzq,Liu:2023mll} have reached, or are expected to reach, similar sensitivities.
As evident from Fig.~\ref{fig:result}, a further increase in sensitivity is needed to reach possible astrophysical (or cosmological) signals.
Our work should be seen as part of the quest of paving a possible path towards this.

\begin{table}[t]
\centering
\renewcommand{\arraystretch}{1.1}
\begin{tabular}{lccccc}
& $Q_r$ & $T_m$ & $f_*$ & ${\cal R}_c^{({\rm B}1)}$ & ${\cal R}_c^{({\rm B}2)}$  \\ \hline
ADMX SLIC \cite{Crisosto:2019fcj}   & $3 \times 10^3 $ & 320\,s\footnotemark & 50\,MHz & 1.6 & 0.1  \\
BASE \cite{Devlin:2021fpq}   &  $4 \times 10^4$  & 1\,min & 0.7\,MHz & 3.0 &  0.39 \\
WISPLC \cite{Zhang:2021bpa}  & $10^4$  & 1\,min & (30~kHz, 5~MHz)&  (6.7, 1.9) & (0.86, 0.24)   \\
DMRadio \cite{DMRadio:2022pkf}  & $2 \times 10^7$ &  (8\,mins, 60\,ns)  & (100\,kHz, 30\,MHz)   & (787, 1) &  (0.18, 1) 
\end{tabular}
\caption{Coherence ratio factor ${\cal R}_c$ for the different experiments considered in Fig.~\ref{fig:result} for two different benchmark signals: $T_h = \tau_h = 1$~s, $Q_h = f_* \tau_h $ (B1) and $T_h \gg T_m$, $Q_h = 10^{10}$ (B2).
For example, the best estimated sensitivity of ADMX SLIC when taking ${\cal R}_c = 1$ is $h \simeq 1.7 \times 10^{-16}$, so that the sensitivity to the signals B1 (B2) would be $\simeq 2.8 \times 10^{-16}$ ($\simeq 1.7 \times 10^{-17}$).
The $T_m$ values for DMRadio follow from our simple model for the instruments scan strategy in App.~\ref{app:CoherenceRatio}.
}
\label{tab:benchmarks}
\end{table}
\footnotetext{Technically the ADMX SLIC scan strategy involved averaging 10,000 individual 32\,ms scans at each frequency, which we here treat as a single 320\,s scan.}

\section{Selection Rules for General Detector Geometries}
\label{sec:othergeom}

In Sec.~\ref{sec:GWforSolenoid} we studied in detail the interaction of a GW with a solenoidal magnetic field, adding to the existing results where the wave interacts with a toroidal field~\cite{Domcke:2022rgu}.
In this section we seek to generalise these results with a symmetry based study of a broader class of magnetic fields and pickup loops.
We will consider detectors with both solenoidal and toroidal magnetic fields (using Eqs.~\eqref{eq:toro_mag_field} and \eqref{eq:sol_mag_field}), but in each case we will consider all possible directions for the pickup loops: designed to measure the induced magnetic field in the $\hat{\bf e}_z$, $\hat{\bf e}_{\phi}$, and $\hat{\bf e}_{\rho}$ directions.

A central goal of our analysis is to identify, by symmetry alone, what is the leading power in $(\omega L)$ of the GW flux, $\Phi_h$, for a given detector.
To do so, we will derive three selection rules that hold for the interaction of a GW with a cylindrical instrument, that allow us to study the more general case and identify promising detector geometries without explicit calculation.
The results are catalogued in Tab.~\ref{table:summary} supplemented by the outcome of explicit computations, but let us briefly summarise the key findings.
The leading contribution we expect to $\Phi_h$ is at ${\cal O}[(\omega L)^2]$, however, we have already seen for the BASE experiment and for ABRA that the flux at this order vanishes.
These are two examples of a general result: instruments with full cylindrical symmetry (of both the magnetic field and pickup loop) designed to search for axions have a leading power sensitivity of at most ${\cal O}[(\omega L)^3]$.
This is a consequence of two observations.
Firstly, as we will demonstrate, detectors with azimuthal symmetry are only sensitive to one of the two GW polarisations $h^+$ or $h^{\times}$.
They are also only sensitive to either a scalar or axion, as under parity the scalar transforms as $h^+$, whereas the axion transforms as $h^{\times}$ (see App.~\ref{app:scalarEM}).
Secondly, azimuthal symmetry enforces that only the $h^+$ contribution can enter at ${\cal O}[(\omega L)^2]$.
Consequently, instruments with full cylindrical symmetry which can detect scalars coupled to EM may also detect $h^+$ GW flux at ${\cal O}[(\omega L)^2]$, but no leading order contribution can appear for axion experiments which employ full cylindrical symmetry to enhance the axion signal.
If the cylindrical symmetry of the pickup loop is broken, the leading power sensitivity to the GW can be restored, at the cost of an ${\cal O}(1)$ factor to the axion signal.

\subsection{Three selection rules for the interaction of a GW with a cylindrical detector}

Let us now derive three general results regarding the form of $\Phi_h$ when we have a magnetic field and pickup loop with full cylindrical symmetry.

\begin{mdframed}[linewidth=1.5pt, roundcorner=6pt]
\vspace{-8pt}
\paragraph{Selection Rule 1:} For an instrument with azimuthal symmetry, $\Phi_h \propto h^+$ at ${\cal O}[(\omega L)^2]$.\footnotemark
\end{mdframed}
\vspace{-8pt}
\footnotetext{We emphasise that this statement is coordinate dependent, and holds for the definition of $h^+$ adopted in Eq.~\eqref{eq:polarisation}.
In general, one can convert $h^+$ into $h^{\times}$ by a $\pi/4$ rotation around the propagation direction of the GW.
The coordinate independent statement is that for geometries with azimuthal symmetry, only a single  polarisation appears at leading order.}

\textit{Proof:} At ${\cal O}[(\omega L)^2]$, the effective current ${\bf j}_{\rm eff}$ receives a contribution only from ${\bf M}$ (as $\partial_t {\bf P} \propto {\cal O}[(\omega L)^3]$), which itself depends on $h_{00}$ and $h_{ij}$, defined in Eq.~\eqref{eq:metricperturbation}.
Moreover, as both $F({\bf k} \cdot {\bf r})$ and $F^{\prime}({\bf k} \cdot {\bf r})$ are constant at leading order, $h_{00}$ and $h_{ij}$ depend only on the GW direction through the polarisation tensors in Eq.~\eqref{eq:polarisation}, to which the induced magnetic field and the flux are proportional.
Equivalently, the leading order response matrix in Eq.~\eqref{eq:flux_Dmn} is independent of the GW direction, explicitly $D^{mn}({\bf k}) = D^{mn}_{(2)} + {\cal O}[(\omega L)^3]$, where $D^{mn}_{(2)}$ depends on $\omega^2$ but not $\hat{\bf k}$.
These results hold in general.
We now invoke azimuthal symmetry, which implies that the result cannot depend on the azimuthal angle associated with the direction of the GW, so that $\Phi_h(\hat{\bf k}) = \Phi_h(R_z(\varphi)\hat{\bf k})$ for an arbitrary angle $\varphi$.
Combined with the leading order response matrix, we find
\begin{equation}\begin{aligned}
\Phi_h (\hat{\bf k}) 
&= \frac{1}{2\pi}\int_0^{2\pi} d\varphi~ \Phi_h (R_z(\varphi)\hat{\bf k}) 
= e^{- \i \omega t} D_{(2)}^{mn} \int_0^{2\pi} d\varphi\, \sum_A h^A e_{mn}^A(R_z(\varphi) \hat{\bf k}) + {\cal O}[(\omega L)^3] \\
&= \frac{e^{- \i \omega t}}{2\sqrt{2}}\, h^+  \sin^2 \theta_h\, D^{mn}_{(2)} 
\begin{pmatrix}
-1 & 0 & 0 \\
0 & -1 & 0 \\
0 & 0 & 2
\end{pmatrix}_{mn}
+ {\cal O}[(\omega L)^3].
\label{eq:thm1}
\end{aligned}\end{equation}
By explicit computation, the $h^{\times}$ contribution has vanished, completing the proof.
As a corollary, we note that at leading order an azimuthally symmetric detector can only depend on the incident GW direction through $\sin^2 \theta_h$.

\begin{mdframed}[linewidth=1.5pt, roundcorner=5pt]
\vspace{-8pt}
\paragraph{Selection Rule 2:} For an instrument with azimuthal symmetry, the flux is proportional to either $h^+$ or $h^{\times}$, but not both.
This holds to all orders in $(\omega L)$.\footnotemark
\end{mdframed}
\vspace{-8pt}

\footnotetext{We reiterate that we derive all fluxes from the Biot-Savart law, which is valid only to ${\cal O}[(\omega L)^3]$ (cf. footnote~\ref{footnote:BiotSavart}).}

\textit{Proof:}  Consider a GW of wave vector $\bf k$ incident on the detector.
By the azimuthal symmetry, we can rotate $\bf k$ into the $xz$-plane without loss of generality.
The configuration is now invariant under a $P_y$ reflection, in particular $P_y {\bf k} = {\bf k}$.
If the pickup loop is azimuthally symmetric, the flux will receive a contribution at ${\bf r}^{\prime}$ and $P_y {\bf r}^{\prime}$.
We now compare the contribution at these points, by evaluating the magnetic field transverse to the pickup loop, which has unit normal vector $\hat{\bf n}'({\bf r}^{\prime})$.
\begin{equation}
\hat{\bf n}'(P_y {\bf r}^{\prime}) \cdot {\bf B}_h (P_y {\bf r}^{\prime}, {\bf k}) 
=  [\kappa_y P_y \hat{\bf n}'({\bf r}^{\prime})] \cdot [\sigma \eta_y\,P_y {\bf B}_h ({\bf r}^{\prime}, {\bf k})]
= \sigma \kappa_y \eta_y \, \hat{\bf n}'({\bf r}^{\prime}) \cdot {\bf B}_h ({\bf r}^{\prime}, {\bf k}).
\label{eq:fluxtransform}
\end{equation}
Here we used the transformation properties of ${\bf B}_h$ and $\hat{\bf n}'$ given in Eqs.~\eqref{eq:B_parity} and \eqref{eq:n_parity}.
The various values of $\eta_y$ and $\kappa_y$ are summarised in Tab.~\ref{table:eta}, and recall $h^+$ and $h^{\times}$ transform with $\sigma=\pm 1$.
If $\sigma \kappa_y \eta_y = -1$, then the contributions of the flux at positions ${\bf r}^{\prime}$ and $P_y {\bf r}^{\prime}$ cancel, and hence the total flux vanishes when integrated over an azimuthally symmetric pickup loop.
This uniquely selects the GW polarisation $\sigma$ which can be measured in a detector with azimuthal symmetry, completing the proof.

In Tab.~\ref{table:summary}, we summarise which polarisation survives for which geometry from this argument. 
Together with the first selection rule, this enables us to identify geometries which are potentially suitable to pick up the ${\cal O}[(\omega L)^2]$ component of the induced flux.
Let us work through several explicit examples, each assumed to be azimuthally symmetric, referring to the table for visualisations.
\begin{itemize}
\item[$\circ$] {\it Toroidal magnet with a horizontal pickup loop.} Consider first a configuration with ${\bf B}_0 \propto \hat{\bf e}_{\phi}$ and $\hat{\bf n}^{\prime} \propto \hat{\bf e}_z$, as for instance used by ABRA.
For these choices, the flux in Eq.~\eqref{eq:fluxtransform} transforms with $\sigma \kappa_y \eta_y = - \sigma$.
Accordingly, by selection rule 2 the $h^+$ polarisation ($\sigma=+1$) cannot contribute, only $h^{\times}$ will ($\sigma=-1$).
Concretely, for ABRA this implies that any azimuthally symmetric pickup loop will receive no contribution proportional to $h^+$ for all possible positions $z$ of the pickup loop.
Selection rule 1 further implies that the leading order contribution can only occur at ${\cal O}[(\omega L)^3]$.
Both of these results were observed by explicit calculation in Ref.~\cite{Domcke:2022rgu}.
\item[$\circ$] \textit{Solenoidal magnet with an array of vertical pickup loops.}
Inspired by BASE we next consider a setup with ${\bf B}_0 \propto \hat{\bf e}_z$ and $\hat{\bf n}^{\prime}  \propto \hat{\bf e}_{\phi}$, so that in Eq.~\eqref{eq:fluxtransform}   $\sigma \kappa_y \eta_y = - \sigma$, and only $h^{\times}$ can contribute.
Explicitly, the flux generated by $h^+$ from a pickup loop at $\phi_{\ell}$ will exactly cancel the flux the plus polarisation generates in a pickup loop at $-\phi_{\ell}$.
Thus, even though the geometry has changed significantly, our polarisation selection rule applies identically to the ABRA-type configuration.
This explains the cancellation for azimuthally symmetric solenoidal detectors observed in Sec.~\ref{sec:GWforSolenoid}.

Although the leading order GW flux vanished in the two instances above, using our selection rules we can straightforwardly conceive of azimuthally symmetric geometries where the cancellation does not occur, we simply require $\kappa_y \eta_y = +1$.
One instance is the following example.

\item[$\circ$] \textit{Solenoidal magnet with a horizontal pickup loop.}
Taking ${\bf B}_0 \propto \hat{\bf n}^{\prime} \propto \hat{\bf e}_z$, we have $\sigma \kappa_y \eta_y = + \sigma$ in Eq.~\eqref{eq:fluxtransform}.
As a result, the $h^{\times}$ contributions to the flux vanishes to all orders in $(\omega L)$ whereas the $h^+$ contribution survives, and with it a contribution to the flux of order $(\omega L)^2$.
In this case, one might worry about the feasibility of separating the tiny induced field from the large background magnetic field.
Any detection strategy would exploit the AC nature of the GW flux, as opposed to the (ideally) DC static field, and potentially also the angular dependence of the GW.
(For further discussion, see also Ref.~\cite{Bloch:2023uis}.)
\end{itemize}

The two selection rules derived so far allow us to understand an important consequence for the detection of GWs with axion haloscopes.
In particular, by selection rule 2, the flux in an azimuthally symmetric detector sensitive to the $h^{\times}$ or $\sigma = -1$ component has no dependence on $h^+$.
But then by selection rule 1, such an instrument will not have the optimal sensitivity to the GW, since the ${\cal O}[(\omega L)^2]$ contribution will vanish.
This is an important observation for axion haloscopes, because as shown in App.~\ref{app:scalarEM}, the pseudoscalar axion field transforms with $\sigma=-1$, and therefore to be sensitive to the axion one is forced into a configuration where the leading GW flux vanishes.
The only way of evading this conclusion is to break the azimuthal symmetry.
One could break this maximally by introducing a figure-8 configuration, as done in Ref.~\cite{Domcke:2022rgu}.
This would revive a contribution from $h^+$ at $ {\cal O} [(\omega L)^2] $, however as the axion induced magnetic field ${\bf B}_a$ has no angular dependence, its contribution will vanish.
Hence, to detect \textit{both} axion and GW, one could use a pickup loop with an opening angle smaller than $2 \pi$, which avoids a complete cancellations.
For this purpose, we present the results for $\Phi_h^{(2)}$ with a pickup loop with an arbitrary angle in the next subsection.
Further discussion of these points is provided in App.~\ref{app:scalarEM}, where we also demonstrate that a scalar, $\varphi$, which couples as $\varphi F^2$ transforms with $\sigma=+1$, and so can also be understood through our selection rules.

\begin{mdframed}[linewidth=1.5pt, roundcorner=6pt]
\vspace{-8pt}
\paragraph{Selection Rule 3:} For an instrument with full cylindrical symmetry, $\Phi_h$ will be either an even or odd function of $\omega$.
\end{mdframed}
\vspace{-8pt}

\textit{Proof:} If the instrument has full cylindrical symmetry, then the flux will receive a contribution from ${\bf r}'$ and $P {\bf r}'=-{\bf r}'$, where again $P = P_x P_y P_z$ is the complete parity transformation.
Let us thus consider the property of the induced magnetic flux under ${\bf r}' \rightarrow P {\bf r}'$.
We start by writing the GW in the proper detector frame as power series in $\omega$, $h_{\mu\nu} = \sum_{n = 2}^\infty h^{(n)}_{\mu \nu}$ with $h_{\mu \nu}^{(n)} \propto \omega^n$.
From Eq.~\eqref{eq:metricperturbation}, we can see the components transform as follows,
\begin{equation}\begin{aligned}
&h_{00}^{(n)}(P {\bf r}, {\bf k})  = (-1)^n \, h_{00}^{(n)}({\bf r,k}),
\hspace{0.5cm} 
h^{(n)}_{0i}(P {\bf r}, {\bf k} ) = (-1)^n \, P_{ij}\, h^{(n)}_{0j} ({\bf r,k}), \\
&\hspace{2.4cm}h^{(n)}_{ij}(P {\bf r}, {\bf k}) = (-1)^n \, P_{ik}\, h_{kl}^{(n)}({\bf r,k})\, P_{lj}^T.
\end{aligned}\end{equation}
This is identical to the transformations studied in Eq.~\eqref{eq:h_parity}, but with $\sigma \to (-1)^n$.
Either by proceeding through the same steps as used to derive Eq.~\eqref{eq:B_parity} or by exploiting the above analogy, we find,
\begin{equation}
{\bf B}_h^{(n)}(P {\bf r}',\hat{\bf k}) = \eta (-1)^n P {\bf B}_h^{(n)}({\bf r}',\hat{\bf k}),
\end{equation}
where we have decomposed the induced field as ${\bf B}_h = \sum_{n=2}^{\infty} {\bf B}_h^{(n)}$, with ${\bf B}_h^{(n)} \propto \omega^n$.
Consequently, the flux contribution from the point $P {\bf r}'$ will be determined by
\begin{equation}
{\bf n}'(P {\bf r}^{\prime}) \cdot {\bf B}_h (P {\bf r}^{\prime}, {\bf k})
= \kappa \eta \sum_{n=2}^{\infty} (-1)^n [{\bf n}'({\bf r}^{\prime}) \cdot {\bf B}_h^{(n)} ({\bf r}^{\prime}, \hat{\bf k})].
\label{eq:fluxtransform3}
\end{equation}
The values of $\eta$ and $\kappa$ were given in Tab.~\ref{table:eta}.
Clearly $\kappa \eta = \pm 1$.
If $\kappa \eta = + 1$, then for odd $n$, the flux from $P {\bf r}'$ and ${\bf r}'$ will cancel, whereas for $\kappa \eta = - 1$ a similar conclusion is reached for even $n$, completing the proof.
(Note this result is independent of $\sigma$.)

The combination of our three selection rules show that instruments with full cylindrical symmetry have a highly restricted form of the induced GW flux.
In Tab.~\ref{table:summary}, we apply these selection rules for all possible pickup loop orientations, and for both solenoidal and toroidal external fields, always assuming full cylindrical symmetry.
(Similar results can be derived for a scalar and axion, which we study in the appendix and summarise in Tab.~\ref{table:summary_scalar}.)
In the first line of each cell, we denote the surviving polarisation ($h^+$ or $h^{\times}$), if even or odd powers of $\omega L$ contribute, and what is the leading order contribution.
The second line provides the explicit leading order flux $\Phi_h$.
The flux is computed assuming a parametrically large $H$, and for all cases we took $\rho' \in [0,r]$, with $r \leq R$.
For an array of pickup loops or a continuous pickup surface (which are required for cylindrical symmetry in the case of a vertical pickup loop) we integrate the induced flux over the azimuthal angle $\phi_\ell$.
In the absence of detailed information on the detector setup and resulting inductance, this serves as a proxy to estimate the total flux, and most importantly, will drop out when recasting axion search results (see Sec.~\ref{subsec:GWSensitivity}) since the same factor appears for the axion case.
(See also the discussion in footnote~\ref{footnote:solenoid-8}.)

The result for a horizontal pickup loop in a toroidal magnetic field was previously presented in Ref.~\cite{Domcke:2022rgu}, our new results corrects this expression by a factor $1/3$ which is due taking into account the contribution from the effective surface current previously overlooked.
For the ${\cal O}[(\omega L)^2]$ results in Ref.~\cite{Domcke:2022rgu}, obtained from the use of a figure-8 pickup loop to break the azimuthal symmetry, the surface current contribution vanishes, leaving them unchanged.
Such cases are considered explicitly in the next subsection.

For each case in the table, the selection rules determine the leading order contribution to $\Phi_h$ without any explicit calculation being required, which achieves one of the central goals of this work.
For example, take a solenoidal magnetic field with a radial pickup loop ($\hat{\bf n}' \propto \hat{\bf e}_{\rho}$).
As $\sigma \kappa_y \eta_y = \sigma$ in Eq.~\eqref{eq:fluxtransform}, only $h^+$ can contribute from selection rule 2.
However, as $\kappa \eta = -1$ in Eq.~\eqref{eq:fluxtransform3}, only odd powers of $\omega$ contribute by selection rule 3, and therefore the leading order contribution is at ${\cal O}[(\omega L)^3]$.
If, however, the loop was moved up or down in the vertical direction, breaking the cylindrical symmetry, selection rule 3 would no longer hold, and we would have a contribution at  ${\cal O}[(\omega L)^2]$ as allowed by selection rule 1.
Explicitly, if we place the loop at $z' \in [-H/2,-H/2+l]$, and then expand the flux assuming for simplicity $H \gg l \gg r,a,R$, we find
\begin{equation}
\Phi_h = \frac{e^{-\i \omega t}}{3\sqrt{2}} h^+ \omega^2 B_0 \pi r^2 l^2 \sin^2 {\theta_h},
\label{eq:zshift1}
\end{equation}
so that a leading order contribution has been resurrected.
Consider also the case of a toroidal magnet with a radial pickup loop.
Now $\sigma \kappa_y \eta_y = -\sigma$ and $\kappa \eta = +1$, so that only $h^{\times}$ and even orders of $\omega L$ will contribute.
But by selection rule 1, for such a configuration the leading order contribution cannot occur until ${\cal O}[(\omega L)^4]$, where we already expect corrections from our use of the Biot-Savart law.
Again, placing the loop at the vertical bottom (or top) of the instrument would parametrically enhance the flux to
\begin{equation}
\Phi_h = - \frac{\i e^{-\i \omega t}}{96 \sqrt{2}} h^{\times} \omega^3 B_{\max} \pi r^2 a R  (a+2R) \sin^2 \theta_h.
\label{eq:zshift2}
\end{equation}

\begin{table}[t] \centering \large
\renewcommand{\arraystretch}{1.1}
\begin{tabular}{c|c|c|l}
\cline{2-3}
& \makecell{Solenoid: ${\bf B}_0 \propto \hat{\bf e}_z$ \\[-0.1cm] {\footnotesize ($\eta_y = +1$, $\eta = -1$)}} & \makecell{Toroid: ${\bf B}_0 \propto \hat{\bf e}_{\phi}$ \\[-0.1cm] {\footnotesize ($\eta_y = -1$, $\eta = +1$)}} \\ \cline{1-3}
\multicolumn{1}{ |c| }{\rotatebox[origin=c]{90}{\makecell{$\hat{\bf n}^{\prime} \propto \hat{\bf e}_z$ {\footnotesize ($\kappa_y=+1$, $\kappa=-1$)}}}} 
& \makecell
{\\[-0.4cm] $h^+$, $n$ even $\Rightarrow$ ${\cal O}[(\omega L)^2]$ \\
{\footnotesize$\Phi_h = \frac{e^{-\i\omega t}}{48\sqrt{2}} h^+ \omega^2 B_0 s_{\theta_h}^2 \pi r^2 \left( 11 r^2 + 14 R^2 +16R^2 \ln \frac{R}{H} \right)$} \\[0.1cm]
\includegraphics[height=3cm]{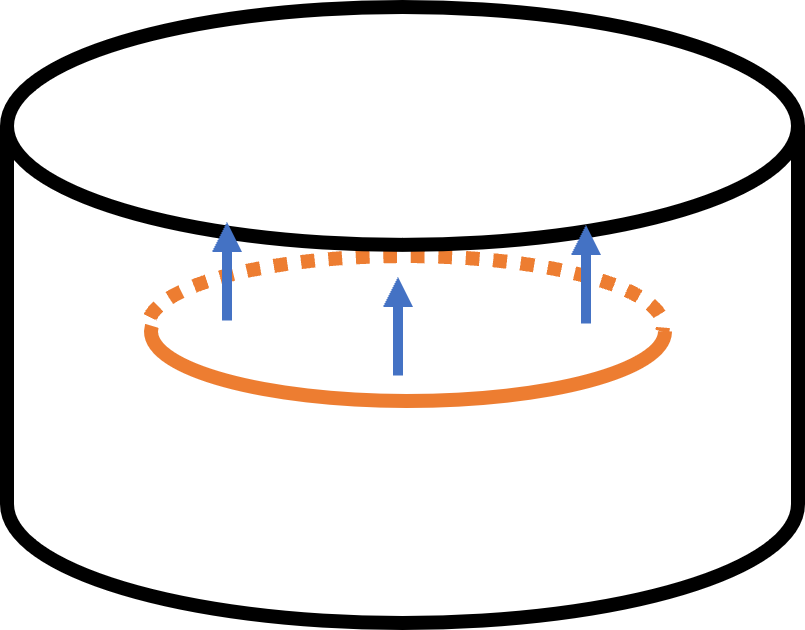}
}
&\makecell
{\\[-0.4cm] $h^{\times}$, $n$ odd $\Rightarrow$ ${\cal O}[(\omega L)^3]$ \\ {\footnotesize $\Phi_h =  \frac{\i e^{-\i\omega t}}{48\sqrt{2}} h^{\times} \omega^3 B_{\max} \pi r^2 a R ( a + 2 R )  s^2_{\theta_h}$} \\[0.1cm]
\includegraphics[height=3cm]{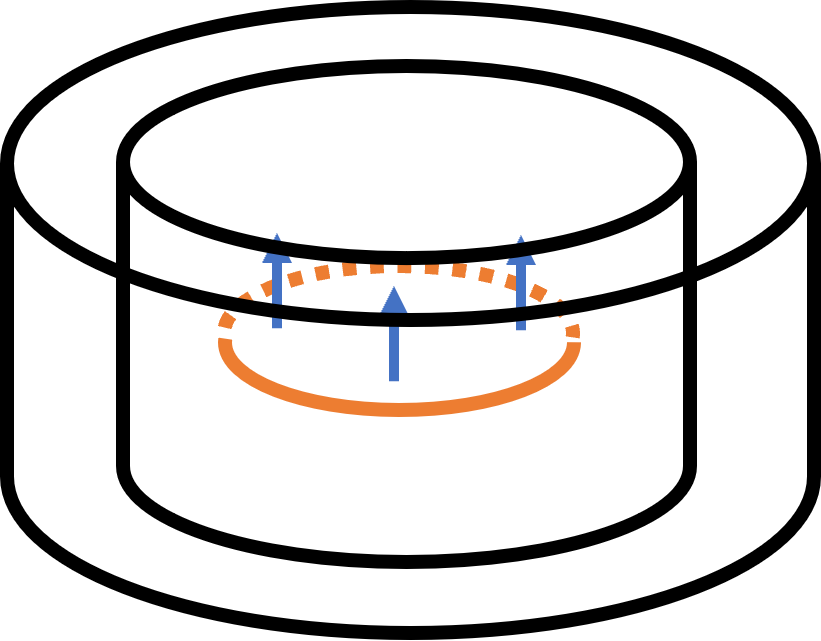}
}
& \\ \cline{1-3}
\multicolumn{1}{ |c| }{\rotatebox[origin=c]{90}{\makecell{$\hat{\bf n}^{\prime} \propto \hat{\bf e}_{\phi}$ {\footnotesize ($\kappa_y=-1$, $\kappa=+1$)}}}}
& \makecell
{\\[-0.4cm] $h^{\times}$, $n$ odd $\Rightarrow$ ${\cal O}[(\omega L)^3]$ \\ {\footnotesize $\Phi_h = \frac{\i e^{-\i\omega t}}{96\sqrt{2}} h^{\times} \omega^3  B_0 \pi r^2 l ( 12 R^2 - 5r^2 ) s^2_{\theta_h}$} \\[0.1cm]
\includegraphics[height=3cm]{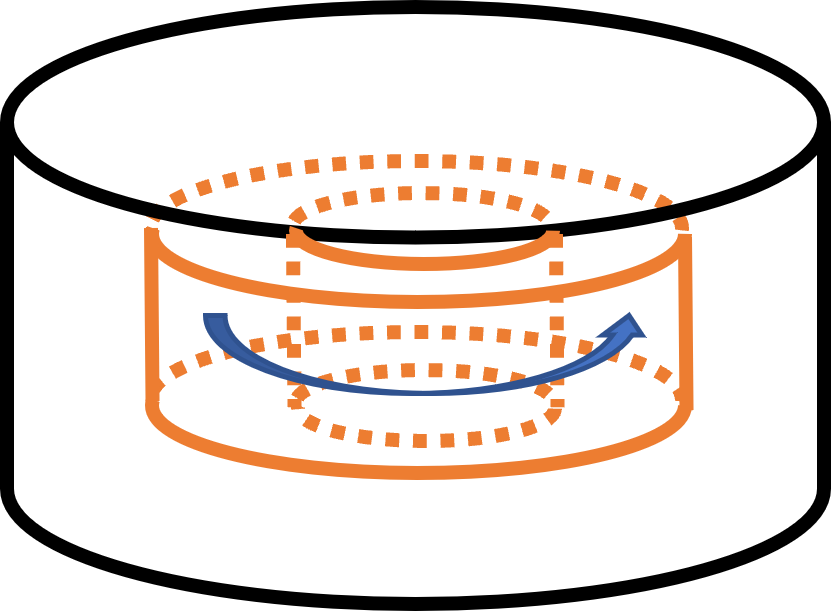} }
& \makecell
{\\[-0.4cm] $h^+$, $n$ even $\Rightarrow$ ${\cal O}[(\omega L)^2]$ \\ 
{\footnotesize $\Phi_h = \frac{3 e^{-i \omega t}}{4\sqrt{2}} h^+ \omega^2 B_{\max} \frac{\pi r^2 a R l (a+2R)}{H^2} s_{\theta_h}^2$} \\[0.1cm]
\includegraphics[height=3cm]{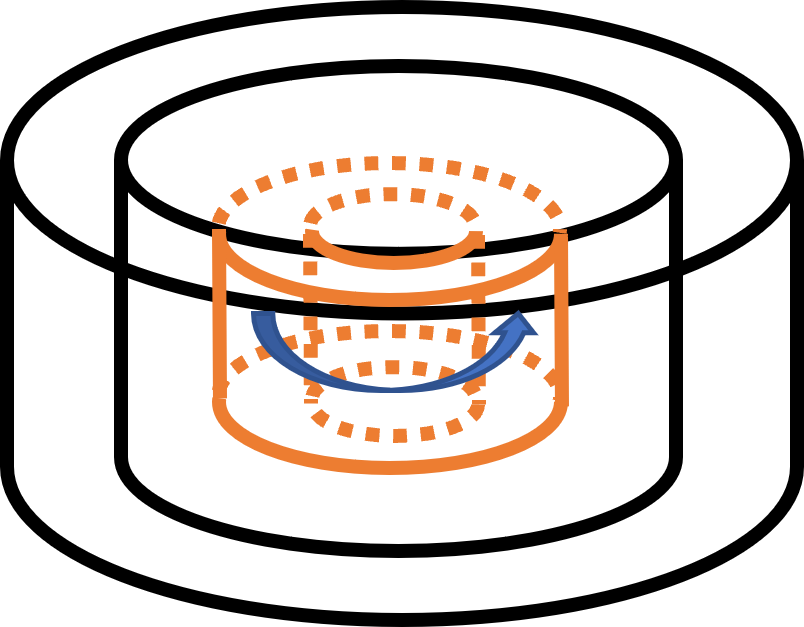} }
&  \\ \cline{1-3}
\multicolumn{1}{ |c| }{\rotatebox[origin=c]{90}{\makecell{$\hat{\bf n}^{\prime} \propto \hat{\bf e}_{\rho}$ {\footnotesize ($\kappa_y=+1, \kappa=+1$)}}}}
& \makecell
{\\[-0.4cm] $h^+$, $n$ odd $\Rightarrow$ ${\cal O}[(\omega L)^3]$  \\
{\footnotesize $\Phi_h = \frac{\i e^{-\i\omega t} }{96\sqrt{2}} h^+ B_0   \omega^3 c_{\theta_h} s_{\theta_h}^2$} \\ {\footnotesize$\times \pi r^2 l \left( 3l^2 - 22(r^2 +2 R^2) - 36 R^2 \ln \frac{R}{H} \right)$} \\[0.1cm]
\includegraphics[height=3cm]{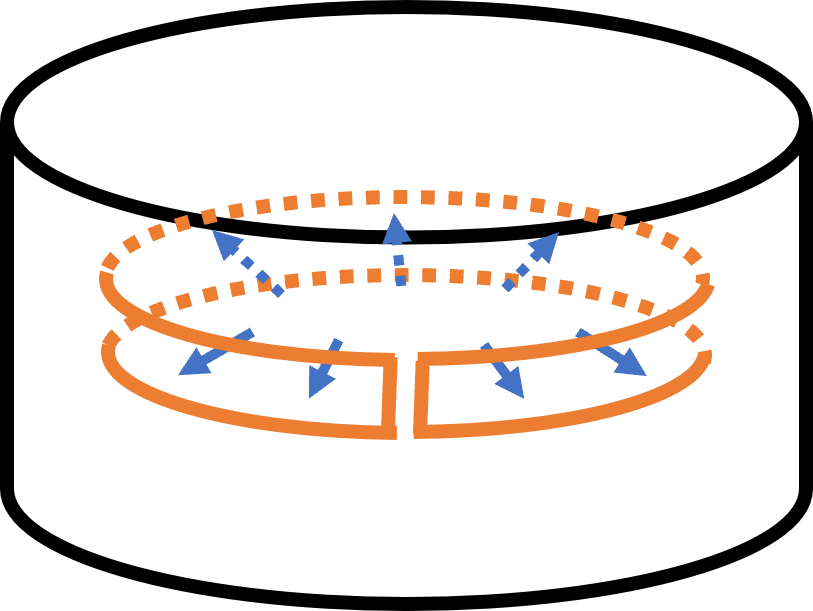}}
& \makecell
{\\[-0.4cm] $h^{\times}$, $n$ even $\Rightarrow$ ${\cal O}[(\omega L)^4]$  \\
{\footnotesize $\Phi_h  = \frac{e^{-\i \omega t}}{32\sqrt{2}} h^{\times} \omega^4 B_{\max} \pi r^2 a R l (a + 2 R)  c_{\theta_h} s_{\theta_h}^2$} \\[0.1cm]
\includegraphics[height=3cm]{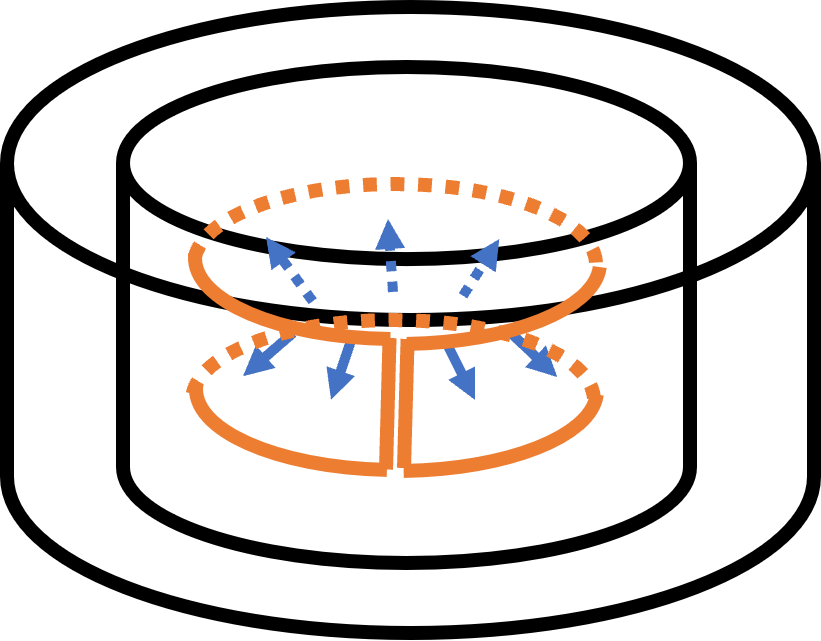} }
&  \\ \cline{1-3}
\end{tabular}
\caption{Summary table of sensitivity to GWs for cylindrical detector geometries depending on the direction of the external magnetic ${\bf B}$ field and the orientation of the pickup loop ${\hat{\bf n}^{\prime}}$.
$\eta_\alpha$ and $\kappa_\beta$ are respectively defined in Eqs.~\eqref{eq:B_parity} and \eqref{eq:n_parity}.
In each cell, the first line indicates the surviving GW polarisation, the selection of odd or even powers of $(\omega L)^n$, and the leading power of the induced flux.
The second line gives the explicit result for the leading order of the induced flux, and finally we provide a cartoon representation of the instrument.
See text for detail.}
\label{table:summary}
\end{table}

\subsection{Increased GW sensitivity for pickup loops that break azimuthal symmetry}

All three selection rules above followed from the full azimuthal symmetry of the detector.
When broken, the restrictions the rules impose are lifted, and in many geometries this allows for a parametric improvement in the GW flux.
While the breaking can occur at the level of the magnetic field or pickup loop, the latter is far more practical.
For instance, as discussed in Ref.~\cite{Domcke:2022rgu}, an instrument could use multiple pickup loops, one for the axion, and another for the GW.

With such a possibility in mind, here we compute the leading ${\cal O}[(\omega L)^2]$ flux for various geometries in the case where the detector has a pickup loop that spans an opening angle $\phi_{\ell} \in [\phi_1, \phi_2]$ for a horizontal or radial readout, or a set of loops that span a fraction $[\phi_1, \phi_2] $ of a toroid in the case of the vertical loop.
As it can be readily restored, we fix $\phi_h = 0$ throughout for ease of notation, and assume $\rho^{\prime} \in [0,r]$ for the radial expanse of the loop.
We again employ the shorthand $c_x = \cos x$ and $s_x = \sin x$.

For a solenoidal magnetic field, the three results determined by the orientation of the pickup loop
$\hat{\bf n}'$ are as follows,
\begin{equation}\begin{aligned}
\hat{\bf e}_z\!:&\hspace{0.25cm} \Phi_h = \frac{e^{-\i\omega t}}{768\sqrt{2}} \omega^2 B_0 r^2 \Bigg[
12 h^{\times} r^2 c_{\theta_h} (c_{2 \phi_2} - c_{2\phi_1})
- 3 h^+ r^2 (3 + c_{2\theta_h}) (s_{2\phi_2} - s_{2\phi_1}) \\
& \hspace{3.85cm} + 8 h^+ (\phi_2 - \phi_1) \left( 11r^2 + 14R^2 + 16 R^2 \ln\frac{R}{H}  \right) s_{\theta_h}^2
\Bigg], \\
\hat{\bf e}_{\phi}\!:&\hspace{0.25cm} \Phi_h = \frac{e^{-\i \omega t}}{144\sqrt{2}}\omega^2 B_0 r l \left(	30R^2  - 13 r^2	\right) s_{\theta_h} \left[	 h^+ c_{\theta_h} ( c_{\phi_2} - c_{\phi_1} )  + h^{\times} ( s_{\phi_2} - s_{\phi_1} )	\right]\!,  \\
\hat{\bf e}_{\rho}\!:&\hspace{0.25cm} \Phi_h = \frac{5 e^{-\i\omega t}}{48\sqrt{2}} \omega^2 B_0 r l \left( 2 R^2 - 3 r^2 \right) s_{\theta_h} \left[ h^+ c_{\theta_h} ( s_{\phi_2} - s_{\phi_1})  - h^{\times} ( c_{\phi_2} - c_{\phi_1} ) \right]\!.
\end{aligned}\end{equation}
In each case we only state the result to ${\cal O}[(\omega L)^2]$ and leading order in $1/H$.
Observe that when $\phi_1=0$ and $\phi_2=2\pi$, only the result for $\hat{\bf n}' \propto \hat{\bf e}_z$ survives, consistent with Tab.~\ref{table:summary}.

For a toroid, the analogous results are as follows (again working only to ${\cal O}[(\omega L)^2]$),
\begin{align}
\hat{\bf e}_z\!:&\hspace{0.25cm} \Phi_h = \frac{e^{-\i \omega t}}{12\sqrt{2}} \omega^2 B_{\max} r^3 R \ln \left( 1 + \frac{a}{R} \right) s_{\theta_h} \left[ h^+ c_{\theta_h} (c_{\phi_2} - c_{\phi_1} ) + h^{\times} (s_{\phi_2} -  s_{\phi_1}) \right]\!, \\
\hat{\bf e}_{\phi}\!:&\hspace{0.25cm}\Phi_h = - \frac{e^{-\i \omega t}}{32\sqrt{2}} \omega^2 B_{\max} r^2 R l \ln \left( 1 + \frac{a}{R} \right) \left[ 4 h^{\times} c_{\theta_h} (c_{2\phi_2} - c_{2\phi_1}) - h^+ (3 + c_{2\theta_h}) (s_{2\phi_2} - s_{2\phi_1})   \right] \nonumber\\
&\hspace{1.06cm} + \frac{3e^{-\i \omega t}}{8\sqrt{2}} h^+ \omega^2 B_{\max} (\phi_2 - \phi_1) \frac{r^2 a R l(a+2R)}{H^2} s_{\theta_h}^2, \nonumber\\
\hat{\bf e}_{\rho}\!:&\hspace{0.25cm} \Phi_h = - \frac{e^{-\i \omega t}}{16\sqrt{2}} \omega^2 B_{\max} r^2 R l  \ln \left( 1 + \frac{a}{R} \right)
\left[ h^+ (3 + c_{2 \theta_h}) (c_{2 \phi_2} - c_{2 \phi_1}) + 4 h^{\times} c_{\theta_h} ( s_{2 \phi_2} - s_{2\phi_1} )  \right]\!. \nonumber
\end{align}
Again taking $\phi_1 = 0$ and $\phi_2 = 2\pi$, there will only be a contribution $\hat{\bf n}' \propto \hat{\bf e}_{\phi}$, reproducing Tab.~\ref{table:summary}.
That contribution only appears at ${\cal O}(1/H^2)$, and this is the only non-leading contribution in $H \to \infty$ we included.

\section{Discussion}

Both axions and GWs induce effective polarisation and magnetisation terms in Maxwell's equations.
While the formalism for exploiting this effect to search for the axion has been in place for four decades~\cite{Sikivie:1983ip}, the GW analogue and its synergies with axion searches remains nascent.
Our work expands our understanding of this latter case, and by focussing on lumped-element circuits for axion detection (such as ABRADACABRA, SHAFT, BASE, ADMX SLIC, WISPLC and the DMRadio program), we estimate their sensitivity to current and future high-frequency GW searches, as shown in Fig.~\ref{fig:result}.

We also expand the theoretical foundations of the interaction of GWs with instruments operating in the magnetoquasistatic regime, extending the earlier results of Ref.~\cite{Domcke:2022rgu} in a number of ways.
Most importantly, we have developed a symmetry based formalism that largely fixes the form of the leading GW signal in various instruments.
We considered external magnetic fields with a cylindrical symmetry -- toroidal or solenoidal fields -- as used in all ongoing and planned axion haloscopes.
We derived selection rules for the signal strength, which, based on symmetry alone, fix the leading power sensitivity in $\omega L$, and hence parametrically determine the GW strain sensitivity without calculation.
This allows one to immediately determine the impact of different geometries for the external magnetic (or electric) field and the pickup loop on the achievable GW strain sensitivity.
As summarised in Tab.~\ref{table:summary}, highly symmetric detectors place strong restrictions on the form of the induced flux as a direct consequence of the tensor nature of the GW.
These arguments can be extended to a scalar or pseudoscalar (axion) coupled to EM, as we show in App.~\ref{app:scalarEM}.
Taken together, we observe that in optimising the sensitivity to axions, existing instruments can often parametrically suppress the GW signal.
Fortunately, however, the observed cancellation can quite easily be remedied by minimally breaking the instruments cylindrical symmetry, for instance by changing the position or shape of the pickup loop.
We demonstrated this for different detector geometries, obtaining a parametric increase for the GW sensitivity.

Our work provides several technical improvements on the formalism and initial studies of Ref.~\cite{Domcke:2022rgu}.
First, we include the contribution from effective surface currents induced by the GW, arising due to the change in effective magnetisation at the boundary of the static magnetic fields.
These effects are generically of the same order as the effects obtained from the interaction of the GW with the magnetic field itself, and for instance modify some of the toroidal results of Ref.~\cite{Domcke:2022rgu}, relevant for ABRACADABRA, SHAFT, and DMRadio-50L.
Second, we give a thorough discussion and prescription of how to compute sensitivities for transient signals, focussing on resonant detectors.
This is particularly important for high-frequency GWs, since the duration of the expected signals can be much shorter than the observation time.
We include the effect of finite coherence time and finite duration of the signal, the scanning strategy and the quality factor of the instrument.
Our prescription is based on bootstrapping the axion search results, allowing an immediate recasting of existing and upcoming axion searches in terms of GW searches.
Third, we introduce linear response matrices describing the detector response to the GW signal.
With this new formalism, we recover the results of Ref.~\cite{Domcke:2022rgu}, but the alternative approach played a key role in revealing the symmetry properties of the detectors, facilitating the derivation of the selection rules mentioned above.
With all this at hand, we provide analytical results for the effective current induced by a GW up to ${\cal O}[(\omega L)^3]$ for a solenoidal magnetic field configuration.

Much work remains.
The symmetry based arguments introduced here can be deployed for the full set of signals axion haloscopes could detect, including, for instance, dark photons.
Such arguments can help determine the full physics reach of the future axion dark-matter program, see Ref.~\cite{Adams:2022pbo}.
A dedicated GW search will require a targeted data analysis strategy as well as a detailed detector simulation.
While this is beyond the scope of the current paper, any such analysis can draw on the tools we have provided here.
A further open question retains to the impact of the mechanical response of the detector to GWs, which may become relevant once the GW frequency lies above the lowest mechanical resonance mode, as discussed in Refs.~\cite{Berlin:2023grv,Bringmann:2023gba}.
We leave this to future work. 
The achievable strain sensitivities (see Fig.~\ref{fig:result}) we obtain by bootstrapping the axion searches still lie above any expected signals from astrophysical or cosmological sources.
Nevertheless, the sensitivities are competitive with other experiments and proposals in this frequency regime.
We aim with this work to join the worldwide effort of paving the way towards high-frequency GW detection.

\acknowledgments

We thank Tael Coren, Jack Devlin, Sebastian Ellis, Joshua Foster, Jai-chan Hwang, Joachim Kopp, Stefan Ulmer, Jonathan Ouellet, Yotam Soreq, Michael Tobar, and Zhongyue Zhang for very helpful discussions, and Francesco Muia for comments on a draft version of this work.
CGC is supported by a Ramón y Cajal contract with Ref.~RYC2020-029248-I.
The work of SML is supported in part by the Hyundai Motor Chung Mong-Koo Foundation Scholarship, and funded by the Korea-CERN Theoretical Physics Collaboration and Developing Young High-Energy Theorists fellowship program (NRF-2012K1A3A2A0105178151).

\paragraph{Note added.}  Following the publication of this and previous work~\cite{Domcke:2022rgu},  Refs.~\cite{Hwang:2023jhs,Hwang:2023cwr,Hwang:2023nqx,Hwang:2024puq} challenged the results based on finding different expressions for the electric and magnetic field when using a covariant definition of these fields introduced in~\cite{Hwang:2023cwr}. This result is, in fact, not surprising as it is well known that the electric and magnetic field are frame dependent in general relativity. The gauge invariant observable in our case is the measured magnetic flux (which is not computed in Refs.~\cite{Hwang:2023jhs,Hwang:2023cwr,Hwang:2023nqx,Hwang:2024puq}). 
As we demonstrate explicitly in App.~\ref{app:new}, one obtains the exact same result as presented here for this observable using the formalism of  Refs.~\cite{Hwang:2023jhs,Hwang:2023cwr,Hwang:2023nqx,Hwang:2024puq}.
The fact that the flux can be computed as in Eq.~\eqref{eq:BioSavartA} without ever resorting to an explicit expression for the magnetic field emphasizes that the effects we describe are independent of the details in how that field is defined.

\appendix

\section{Additional Details of GW Electrodynamics}
\label{app:Maxwell}

In this first appendix, we provide a derivation of the key elements of GW electrodynamics required for the discussion in the main text.
In the first two subsections we will provide a detailed derivation of Eqs.~\eqref{eq:MaxwellGW} and \eqref{eq:effectivecurrent}, which formed the foundation of the physical effect explored in the main text.
After this, we describe the origin of the effective surface current contributions that had been missed in previous analyses, and lastly we explain why GW effects start at ${\cal O}[(\omega L)^2]$ in the proper detector frame.

\subsection{External currents}

The external static fields, $f_{\mu\nu}$, upon which GWs interact to generate electromagnetic effects satisfy Maxwell's equations in flat spacetime 
\begin{equation}
\partial_\nu f^{\mu\nu}  =\left[j^\mu \right]_\text{FLAT},
\hspace{0.5cm}
\partial_{\nu} f_ { \alpha \beta}+\partial_{\alpha} f_{ \beta \nu}+\partial_{\beta} f_{ \nu \alpha} = 0,
\label{eq:covarianthomFLAT}
\end{equation}
where $\left[j^\mu \right]_\text{FLAT}$ is the external electromagnetic current sourcing the  external fields.
For a system of electrons following trajectories described by $x_n^\mu(u)$, with $n$ indexing the various particles, the current is given by~\cite{Weinberg:2008zzc}

\begin{equation}
\left[j^\mu \right]_\text{FLAT} =  \int d u \sum_n e \,\delta^{(4)}\left(x_n(u)-x\right) \frac{d x_n^\mu(u)}{d u}=  \sum_n e\, \delta^{(3)}\left({\bf x}_n(t)-{\bf x}\right) \frac{d x^\mu_n(t)}{d t}.
\label{eq:jFLAT}
\end{equation}
For example, solenoidal and toroidal static configurations in which the electric charges are confined to cylindrical surfaces of constant radius lead to $\left[j^\mu \right]_\text{FLAT}  \propto \delta(\rho-R)$.
For the case of a solenoid infinitely extended in the $z$-direction
\begin{equation}
{\bf j}_\text{FLAT}^\solenoid = B_0  \delta(\rho-R) \, \hat{{\bf e}}_\phi.
\label{eq:jsolenoid}
\end{equation}
Similarly, for the toroidal configurations  in the limit of infinite height
\begin{equation}
{\bf j}^\toroid_\text{FLAT} = B_{\max} \frac{R}{\rho} \left[ \delta\left(\rho-R\right)- \delta\left(\rho-(R+a)\right) \right] \hat{\bf e}_z,
\label{eq:jtoroid}
\end{equation}
because there are two cylindrical surfaces.

Observe also that charge conservation directly follows from  Eq.~\eqref{eq:jFLAT}
\begin{equation}\begin{aligned}
{\bf \nabla} \cdot {\bf j}_\text{FLAT}  
=  &\sum_n e\,{\bf \nabla} \cdot  \left[ \delta^{(3)}\left({\bf x}_n(t)-{\bf x}\right) \right] \frac{d x^\mu_n(t)}{d t} \\
= &- \frac{\partial}{\partial t} \sum_n e\, \delta^{(3)}\left({\bf x}_n(t)-{\bf x}\right) \\
= &- \partial_t j_\text{FLAT}^0.
\label{eq:jFLATC}
\end{aligned}\end{equation}
Let us now discuss how the external currents are modified by the presence of GWs in the proper detector (PD) frame.
A generalisation of Eq.~\eqref{eq:jFLAT} to an arbitrary spacetime can be found by noticing that $\delta^{(4)}\left(x_n(u)-x\right) /\sqrt{-g}$ transforms as a scalar.\footnote{Recall, it is the spacetime volume $d^4 x \sqrt{-g}$ rather than $d^4 x$ that transforms as a scalar.}
The current associated with a set of charges following spacetime trajectories $x_n^\mu(u)$ is thus given by~\cite{Weinberg:2008zzc}
\begin{equation}
j^\mu =   \int d u \sum_n e\,  \frac{1}{\sqrt{-g}}\delta^{(4)}\left(x_n(u)-x\right) \frac{d x_n^\mu(u)}{d u}=  \frac{1}{\sqrt{-g}}\sum_n e\, \delta^{(3)}\left({\bf x}_n(t)-{\bf x}\right) \frac{d x_n^\mu(t)}{d t} .
\label{eq:jingmunu}
\end{equation}
A calculation similar to that in Eq.~\eqref{eq:jFLATC} shows that  $ \partial_\mu  \left(\sqrt{-g} \, j^\mu\right) =0$.
This is equivalent to $\nabla_\mu j^\mu =0$, which can be proven employing the properties of Christoffel symbols.

In the proper detector frame, the effect of a GW on the charged particles is described by a Newtonian force.
As stated above, in this work we assume that the experimental apparatus is rigid, or more precisely, that  such a Newtonian force does not alter the trajectories of the particles.
In particular, this implies that in the proper detector frame we can use the same $x_n^{\mu}(t)$ that led to Eqs.~\eqref{eq:jsolenoid} and \eqref{eq:jtoroid}.
We conclude that
\begin{equation}
\left[\sqrt{-g} \, j^\mu \right]_\text{PD} = \left[j^\mu \right]_\text{FLAT}.
\label{eq:key}
\end{equation}

\subsection{Maxwell's equations in the spacetime of a gravitational wave}

Maxwell's Equations in an arbitrary spacetime read
\begin{equation}
\nabla_\nu \left( g^{\alpha \mu} F_{\alpha\beta} \, g^{\beta \nu} \right) = j^\mu,
\hspace{0.5cm} 
\nabla_{\nu} F_{\alpha \beta}+\nabla_{\alpha} F_{\beta \nu}+\nabla_{\beta} F_{\nu \alpha}=0.
\end{equation}
where the external current is defined by Eq.~\eqref{eq:jingmunu}.
Due to the properties of the Christoffel symbols and the fact the electromagnetic tensor is antisymmetric, these equations can be cast as~\cite{landau1975classical}
\begin{equation} 
\partial_\nu \left( \sqrt{-g} \, g^{\alpha \mu} F_{\alpha\beta} \, g^{\beta \nu} \right) =  \sqrt{-g} \, j^\mu,
\hspace{0.5cm}
\partial_{\nu} F_{\alpha \beta}+\partial_{\alpha} F_{\beta \nu}+\partial_{\beta} F_{\nu \alpha} =0.
\label{eq:covarianthomCUR2} 
\end{equation}

When considering a passing GW in the proper detector frame, the equations are equivalent to Maxwell's equations in a flat spacetime with the GW effects described by an effective current.
To see this, we note that the expression in parenthesis to first order in $h_{\mu\nu} = g_{\mu\nu} -\eta_{\mu\nu}$ is given by
\begin{equation}
\sqrt{-g}\, g^{\alpha \mu} F_{\alpha\beta} \, g^{\beta \nu} = \left(1+\frac{h}{2}\right)   F^{\mu\nu} -F^{\mu \alpha} {h^{\nu}}_{\alpha}
+F^{\nu \alpha} {h^{\mu}}_{\alpha}+{\cal O}(h^2),
\end{equation}
where we employ $\sqrt{-g} = 1+h/2 + {\cal O} (h^2)$.
This motivates us to define the following effective current,
\begin{equation}
j_{\rm eff}^{\mu} \equiv  \partial_\nu 
\left( - \frac{1}{2} h \, F^{\mu\nu}
+ F^{\mu \alpha} {h^{\nu}}_{\alpha}
- F^{\nu \alpha} {h^{\mu}}_{\alpha}
\right)\!,
\end{equation}
which, together with Eqs.~\eqref{eq:covarianthomCUR2} and \eqref{eq:key}, leads to the following form for Maxwell's equations
\begin{equation}
\partial_\nu  F^{\mu\nu} = \left[j^\mu \right]_\text{FLAT} +j_{\rm eff}^{\mu}, 
\hspace{0.5cm}
\partial_{\nu} F_{\alpha \beta}+\partial_{\alpha} F_{\beta \nu}+\partial_{\beta} F_{\nu \alpha} =0.
\label{eq:final}
\end{equation}
This completes our justification of Eqs.~\eqref{eq:MaxwellGW} and \eqref{eq:effectivecurrent}.

We combine Maxwell's equations in the presence of a GW with their flat space analogues in Eq.~\eqref{eq:covarianthomFLAT} to isolate the induced fields, defined by
\begin{equation}
F^h_{ \,\mu\nu} \equiv - f_{\mu\nu}  + F_{\mu\nu}.
\end{equation}
In particular, these fields satisfy the following equations
\begin{equation}
\partial_\nu F^{h \, \mu\nu}  = j^\mu_{\rm eff},
\hspace{0.5cm}
\partial_{\nu} F^h_{\alpha \beta}+\partial_{\alpha} F^h_{\beta \nu}+\partial_{\beta}  F^h_{\nu \alpha} = 0.
\label{eq:MW_app}
\end{equation}
Alternatively, one may also write these equations in terms of a contravariant electromagnetic field strength tensor, $g^{\alpha \mu} F_{\alpha\beta} \, g^{\beta \nu}- f^{\mu\nu}$, in which case the second set of equations acquire a source term, as shown in Refs.~\cite{Sokolov:2022dej,Hwang:2023jhs}.
In the absence of external currents, this leads to an ambiguity in the definition of the electromagnetic field.
This is the well-known duality of Maxwell's equations in the absence of charges and currents.
However, in contrast to the claims of those references, such an ambiguity does not arise here because of the proper detector external current.
Furthermore, the derivation presented here shows that it is not necessary to define the electric and magnetic field vectors in curved spacetime in order to describe the effect of GWs propagating in external electromagnetic fields, as one can always work at the level of $F^{\mu \nu}$.
For simplicity, in the main text and in the rest of the manuscript, we drop the FLAT subscript from the current in Eq.~\eqref{eq:final}.

\subsection{Covariant definition of EM fields}
\label{app:new}

Here we address the claims in Refs.~\cite{Hwang:2023jhs,Hwang:2023cwr,Hwang:2023nqx,Hwang:2024puq}, that different results to what we have presented are obtained using a covariant definition of the electric and magnetic fields. As already noted, focusing on the definition of the electromagnetic fields can be misleading, since the gauge invariant observable (the magnetic flux $\Phi_B$) can be expressed without resorting to explicit expressions for the magnetic fields or even the explicit use of co- or contra variant Lorentz indices,
\begin{align}
\Phi_B = \int_{\partial S} A = \int_S F,
\label{eq:forms}
\end{align}
with $A$ denoting the 1-form vector potential, $F$ the two-form field strength tensor, $S$ the area of the pickup loop, and $\partial S$ its boundary. $A$ and $F$ can be obtained directly as solutions of the Maxwell equations in curved space time, derived from the kinetic gauge field term in the action, and all quantities are well defined under Lorentz transformations. This demonstrates that there can be no ambiguities of the type discussed in Refs.~\cite{Hwang:2023jhs,Hwang:2023cwr,Hwang:2023nqx,Hwang:2024puq} that would impact the physical observable $\Phi_B$.

Nevertheless, let us illustrate the point further by explicitly rederiving Eq.~\eqref{eq:flux} using the formalism proposed in Ref.~\cite{Hwang:2023cwr}.
We consider a static external magnetic field in the proper detector frame, ${ \mathbf{ \overline B}} = B_0 \, \hat{\mathbf{e}}_z$ and a rigid observer,\footnote{As in the main text, we stress that this is an assumption which we expect to hold only in the low frequency limit, but which may be violated in realistic experimental setups as the GW frequency approaches the mechanical resonance frequencies of the apparatus.} whose 4-velocity is given by
\begin{align}
 u^\mu = \frac{1}{\sqrt{-g_{00}}} (1, 0,0,0) = \left(1 + \frac{1}{2} h_{00} \right) (1,0,0,0),
\end{align}
which follows directly from $g_{\mu \nu} u^\mu u^\nu = -1$.
We can now define co- and contra variant magnetic fields as
\begin{align}
B^\mu &=  \frac{1}{2} \eta^{\mu \nu \rho} F_{\nu \rho} = - \frac{1}{2 \sqrt{-g}} \epsilon^{\mu \nu \rho \sigma} u_\sigma F_{\nu \rho} \nonumber  \\
& =  \frac{1}{2} \epsilon^{\mu \nu \rho 0} \bar F_{\nu \rho} + \frac{1}{2} \epsilon^{\mu \nu \rho 0}  F^h_{\nu \rho}
{- \frac{h}{4} \epsilon^{\mu \nu \rho 0} \bar F_{\nu \rho} - \frac{1}{4} h_{00} \epsilon^{\mu \nu \rho 0} \bar F_{\nu \rho} - \frac{1}{2} \epsilon^{\mu\nu\rho i} h_{i0} \bar F_{\nu \rho}} \nonumber \\
& = \bar B^\mu + B_h^\mu - \frac{1}{2}(h+h_{00}) \bar{B}^{\mu} - \frac{1}{2} \epsilon^{\mu\nu\rho i} h_{i0} \bar{F}_{\nu\rho} \equiv \bar B^\mu + \delta B^\mu, \label{eq:B1} \\
B_\mu &= g_{\mu \nu} B^\mu =  \eta_{\mu \nu} \bar B^\nu + \eta_{\mu \nu} \delta B^\nu {+ h_{\mu \nu} \bar B^\nu },
\label{eq:B2}
\end{align}
where we have introduced
\begin{align}
    \eta_{\mu\nu\rho} = \eta_{\mu\nu\rho\sigma} u^{\sigma}, &&
    \eta_{\mu\nu\rho\sigma} = - \sqrt{\vert g \vert} \epsilon_{\mu\nu\rho\sigma}, && \eta^{\mu\nu\rho\sigma} = - \frac{1}{ \sqrt{\vert g \vert} } \epsilon^{\mu\nu\rho\sigma},
\end{align}
with $\epsilon^{\mu\nu\rho\sigma}$ being the totally antisymmetric Levi-Civita symbol, and we have used $\sqrt{-g} = 1 + h/2$, $u_0 = g_{00} u^0 = - 1 + 1/2 \, h_{00}$, and $u_i = h_{i0}$.
Here, $\bar{B}$ denotes the background magnetic field, and $\delta B$ the correction induced by the GW. The latter consists of two parts: The part labelled $B_h$ is the contribution discussed in the main text as sourced by the effective current, obtained as a solution of Eq.~\eqref{eq:MW_app}.
The second part is of the structure $h \bar B$ and describes a ${\cal O}(h)$ oscillation of the background magnetic field due to the impact of the GW on the observer and on the metric. Ref.~\cite{Hwang:2023nqx} claims that this second part was overlooked in the analysis in the main text and could change the results.\footnote
{
Ref.~\cite{Hwang:2023jhs} chooses $\delta u_{i} = 0$ instead of  $\delta u^{i} = 0$. Inserting this into the expressions above,  Eq.~\eqref{eq:B2} becomes equivalent to Eqs.~(15,18) in~\cite{Hwang:2023jhs}. Here and in the main text, we follow the usual textbook convention of denoting position vectors with raised indices. This can of course be done differently (see also comment below Eq.~\eqref{eq:MW_app}), but then needs to be done consistently throughout the entire computation and cannot impact any physical observables.}
(As an aside, we note that for setups such as ABRACADABRA where the pickup loop is placed in a region with approximately vanishing background field, these terms trivially vanish.)

We can now compute the flux as
\begin{equation}\begin{aligned}
\Phi_{B}  & =  \frac{1}{2} \int_S F_{\mu\nu} ~ dx^{\mu} \wedge dx^{\nu} \\
& = - \frac{1}{2} \int \eta_{\mu\nu\rho\sigma} u^{\rho} B^{\sigma} ~ dx^{\mu} \wedge dx^{\nu} = \frac{1}{2} \int \epsilon_{\mu\nu\rho\sigma} \sqrt{-g} u^{\rho} B^{\sigma} ~ dx^{\mu} \wedge dx^{\nu}. 
\end{aligned}\end{equation}
Inserting  Eq.~\eqref{eq:B1} yields
\begin{equation}\begin{aligned}
  \Phi_{B} & = \frac{1}{2} \int \epsilon_{\mu\nu\rho\sigma} \left( 1 + \frac{h}{2} \right) (\bar u^\rho + \delta u^\rho) (\bar B^\sigma + \delta B^\sigma)  ~ dx^{\mu} \wedge dx^{\nu} \\
  & = \bar \Phi_B + \frac{1}{2} \int \epsilon_{\mu\nu\rho\sigma} \bar{u}^{\rho} \left[ \delta B^{\sigma} + \frac{1}{2} (h + h_{00})  \bar{B}^{\sigma}    \right] dx^{\mu} \wedge dx^{\nu} \\
  & = \bar \Phi_B + \frac{1}{2} \int \epsilon_{\mu\nu\rho\sigma} \bar u^\rho B_h^\sigma ~ dx^{\mu} \wedge dx^{\nu} \\
   & = \bar \Phi_B + \frac{1}{2} \int F^h_{\mu\nu} ~ dx^{\mu} \wedge dx^{\nu},
   \label{eq:PhiF}
 \end{aligned}\end{equation}
where we have used
\begin{align}
 \epsilon_{\mu\nu\rho\sigma} \bar u^\rho \epsilon^{\sigma \alpha \beta i}\bar F_{\alpha \beta} = \epsilon_{\mu\nu\rho\sigma} \delta^{\rho 0} \epsilon^{\sigma \alpha \beta i} \bar F_{\alpha \beta} = 0,
\end{align}
for $\bar F_{\alpha \beta}$ containing only external magnetic and not electric fields. 

Equation~\eqref{eq:PhiF} is identical to Eq.~\eqref{eq:flux} in the main text. We conclude that the magnetic flux as measured by the observer is fully accounted for by taking $ F^h_{\mu \nu}$, defined as the solution to Eq.~\eqref{eq:MW_app} (i.e.\ sourced by the effective current) and inserting it into the last line of Eq.~\eqref{eq:PhiF}. The additional terms found in Eq.~\eqref{eq:B1} drop out, as expected from the general argument around Eq.~\eqref{eq:forms}.

\subsection{Effective surface currents}
\label{app:surface_currents}

Comparing the magnetic fields in Eqs.~\eqref{eq:toro_mag_field} and \eqref{eq:sol_mag_field} with the currents in Eqs.~\eqref{eq:jsolenoid} and \eqref{eq:jtoroid}, one can note that the currents close to the surface $\rho=R$ are given by
${\bf j} =  \delta(\rho -R)\, \hat{\bf n} \times {\bf B}$, with  $\hat{\bf n} = - \hat{\bf e}_\rho$ for the solenoid and  $\hat{\bf n} = \hat{\bf e}_\rho$  for the toroid.
This can also be written as
\begin{equation}
{\bf j} =  \delta(\rho -R)\, {\bf K},
\hspace{0.5cm}\text{with} \hspace{0.5cm}
{\bf K} =  \hat{\bf n} \times {\bf B},
\label{eq:jboundary}
\end{equation}
where $\bf K$ is often called the surface current density.
The result in Eq.~\eqref{eq:jboundary} is a particular example of a phenomenon that takes place whenever there is discontinuity in the magnetic field  across a surface.
It is possible to prove (see, for instance, Ref.~\cite{jackson_classical_1999}) that at the interface of two bodies with different values of ${\bf H} = {\bf B} - \bf M_{\rm eff}$, Maxwell's equations predict the existence of a surface current density,
\begin{equation}
{\bf K} = \hat{\bf n} \times\left({\bf H}_2-{\bf H}_1\right)\!,
\label{eq:KdeltaH}
\end{equation}
where $\hat{\bf n}$ is the unit vector normal to the surface from medium 1 to 2.
Note that Eq.~\eqref{eq:jboundary} is a particular case of Eq.~\eqref{eq:KdeltaH} when the magnetic field vanishes on one side of the surface  $\rho=R$,  and $ {\bf M}_{\rm eff}=0$ everywhere,  i.e. there is no GW or (pseudo-)scalar field.

In the presence of GWs or a (pseudo-)scalar field, the magnetisation does not vanish.
For the toroid and solenoid cases of Eqs.~\eqref{eq:toro_mag_field} and \eqref{eq:sol_mag_field}, in addition to Eq.~\eqref{eq:jboundary}, there is an effective current on the surface given by
\begin{equation}
{\bf j}_{S,{\rm eff}} =  \delta(\rho -R)\, \hat{\bf n} \times {\bf M}_{\rm eff} \big|_{\rho=R}.
\label{eq:jeffboundary}
\end{equation} 
For instance, as we show in App.~\ref{app:scalarEM}, a scalar coupled to electromagnetism will generate ${\bf M}_{\rm eff} =-\gp \varphi {\bf B}_0 $, and therefore ${\bf j}_{S,{\rm eff}} = -\gp \varphi\, {\bf j}$.
For axions, however, ${\bf M}_{\rm eff} = \gagg a {\bf E}$ and ${\bf j}_{S,{\rm eff}} = 0$, as the tangential component of the electric field must be continuous across any boundary~\cite{jackson_classical_1999}.
This will be found by an explicit calculation in App.~\ref{app:scalarEM}.
To obtain the effective surface for GWs, Eq.~\eqref{eq:jeffboundary} can be cast as $\hat{\bf u} \cdot {\bf j}_{S,{\rm eff}}  =  \pm \delta(\rho -R)\, (\hat{\bf u} \times \hat{\bf e}_\rho) \cdot {\bf M}_{\rm eff} \big|_{\rho=R} $, with $\hat{\bf u}$ an arbitrary unit vector, or in components
\begin{eqnarray}
j_{S,{\rm eff},\phi} &=&  \mp  \,\delta(\rho -R) (\hat{\bf e}_z )_i  \left(- h_{ij} B_j - \frac{1}{2} h B_i + h_{jj} B_i \right)\!, \\
j_{S,{\rm eff},z} &=&  \pm \,\delta(\rho -R) (\hat{\bf e}_\phi )_i  \left(- h_{ij} B_j - \frac{1}{2} h B_i + h_{jj} B_i \right)\!,
\end{eqnarray}
where we assume the system is interacting purely with a magnetic field.

\subsection{$(\omega L)$ power counting in the proper detector frame}
\label{app:PDframe}

The proper detector frame is the closest analogue to the inertial reference frame of the laboratory, and therefore allows for a simple description of the experimentally generated electromagnetic fields.
As Eq.~\eqref{eq:metricperturbation} demonstrates, in the proper detector frame the leading order frequency contribution to $h$ occurs at ${\cal O}[(\omega L)^2]$.
Accordingly, when the GW interacts with static electromagnetic fields, the leading order gauge invariant contribution to the induced magnetic field and measurable magnetic flux will also be $(\omega L)^2$, as seen explicitly in, for example, Refs.~\cite{Berlin:2021txa,Domcke:2022rgu}.
In this appendix we explain the physical origin of this scaling, and in particular justify the absence of any contribution at ${\cal O}(\omega L)$.

The starting point is that the proper detector frame corresponds to Fermi normal coordinates~\cite{FortiniGualdi,Marzlin:1994ia,Rakhmanov:2014noa}, which are freely falling locally inertial coordinates defined along a geodesic, $x_0$.
In effect, Fermi normal coordinates are the extension of Riemmann normal coordinates to an entire worldline.
In terms of the specified geodesic, we can evaluate the metric at an arbitrary spacetime point as follows,
\begin{equation}
g_{\mu \nu}(x) = g_{\mu \nu}(x_0) + (x-x_0)^{\alpha} \partial_{\alpha} g_{\mu \nu}(x_0) + (x-x_0)^{\alpha} (x-x_0)^{\beta} \partial_{\alpha} \partial_{\beta} g_{\mu \nu}(x_0) + \ldots.
\label{eq:gexp}
\end{equation}
To wit, we can take $x_0$ to represent the worldline of the center of our detector, and then $x$ could represent an arbitrary point in the detector where we wish to evaluate the impact of the GW.
This implies that parametrically $(x-x_0) \sim L$, where as throughout the main text $L$ is a characteristic length scale of the instrument.
Now, Fermi normal coordinates are locally flat, and therefore we have $g_{\mu \nu}(x_0) = \eta_{\mu \nu}$.
Performing the usual linear decomposition of the metric for a GW propagating in flat space, $g_{\mu \nu} = \eta_{\mu \nu} + h_{\mu \nu}$, we can therefore identify the contribution from the GW with the derivative terms in Eq.~\eqref{eq:gexp}.
Assuming there is no backreaction from the instrument on the GW, and in addition that the detector can be treated as rigid, if we have a monochromatic incident source of frequency $\omega$, then $\partial^n g_{\mu \nu}(x_0) \sim \omega^n g_{\mu \nu}(x_0)$.
In this language, the absence of an ${\cal O}(\omega)$ term in the description of the GW is reduced to explaining why the single derivative contribution to Eq.~\eqref{eq:gexp} must vanish.
This is straightforward: Fermi normal coordinates are a locally inertial reference frame, so that all Christoffel symbols vanish along $x_0$, and therefore $\partial_{\alpha} g_{\mu \nu}(x_0) = 0$.
Accordingly, the GW in Fermi normal coordinates has a leading contribution at ${\cal O}(\omega^2)$.
The final ingredient is to transform from a freely falling frame to the non-inertial frame of a laboratory on the surface of the Earth, which define the proper detector frame.
However, it can be shown that this transformation introduces contributions only at significantly lower frequencies than we consider here~\cite{Ni:1978zz}, and therefore does not impact the above argument.

\section{Scalar and Axion Electrodynamics}
\label{app:scalarEM}

In this appendix, we expand on our discussion of scalar and axion electrodynamics. 
In particular, we will compare the well studied pseudoscalar axion interaction $-\tfrac{1}{4} \gagg a F^{\mu \nu} \tilde{F}_{\mu \nu}$,\footnote{We define the dual field strength tensor as
$\tilde{F}^{\mu\nu} \equiv \frac{1}{2} \epsilon^{\mu \nu\alpha \beta} F_{\alpha\beta}$, where  $\epsilon^{\mu \nu\rho \sigma}$ is the ordinary totally antisymmetric symbol (with $\epsilon^{0123}=1$).} to the scalar equivalent\footnote{We remain agnostic as to the UV details of this scenario, for a discussion, see e.g. Refs.~\cite{Arvanitaki:2014faa,Banerjee:2022sqg}.
Our focus is simply to study how such a coupling would differ from the conventional axion interaction.}
\begin{equation}
{\cal L} \supset - \frac{1}{4} \gp \varphi F^{\mu \nu} F_{\mu \nu}.
\label{eq:scalarL}
\end{equation}
In the main text, we used this as a motivating example, as it involved several of the features we explored for GW signals, explicitly the importance of the symmetry of the detector and further the additional contributions we receive from the boundaries of the detector.
Here we will provide a more complete discussion of the specific differences between axion and scalar electrodynamics, and then determine the flux scalar dark matter could induce in various lumped-element circuit instruments.

In direct analogy to the axion, a scalar field that couples as in Eq.~\eqref{eq:scalarL} will modify Maxwell's equations.
If we work perturbatively in the coupling $g$, writing $F^{\mu \nu} = F^{\mu \nu}_0 + F^{\mu \nu}_{a/\varphi} + {\cal O}(g^2)$, we have the equations of motion for the induced fields,
\begin{equation}\begin{aligned}
\partial_{\nu} F^{\mu \nu}_{\varphi} &= \partial_{\nu} (\gp \varphi F_0^{\nu \mu}) = \gp (\partial_{\nu} \varphi) F_0^{\nu \mu} - \gp \varphi j^{\mu}, \\
\partial_{\nu} F^{\mu \nu}_a &= \partial_{\nu} (\gagg a \tilde{F}_0^{\nu \mu}) = \gagg (\partial_{\nu} a) \tilde{F}_0^{\nu \mu},
\label{eq:sa-eom}
\end{aligned}\end{equation}
The final results follow from $\partial_{\nu} \tilde{F}^{\mu \nu} = 0$ for the axion, and $\partial_{\nu} F_0^{\mu \nu} = j^{\mu}$ for the scalar.
The presence of a coupling to the current, which also must be included for the GW, is a novelty that does not arise for the axion.
For instance, this interaction will give rise to an oscillating contribution to the fields generated by $j^{\mu}$, as explored in Ref.~\cite{Bloch:2023uis}.
(There have also been discussions of using axion haloscope inspired instruments to detect $\gp$, see Refs.~\cite{Flambaum:2022zuq,McAllister:2022ibe}.)

For both cases, Eq.~\eqref{eq:sa-eom} demonstrates that we can define an effective magnetisation and polarisation tensor as in Ref.~\cite{Domcke:2022rgu}.
In particular, we have ${\cal M}_{\varphi}^{\mu \nu} = \gp \varphi F^{\mu \nu}$ and ${\cal M}_a^{\mu \nu} = \gagg a \tilde{F}^{\mu \nu}$, which yields explicit polarisation and magnetisation vectors,
\begin{equation}
{\bf P}_{\varphi} = \gp \varphi {\bf E},\hspace{0.5cm}
{\bf M}_{\varphi} = -\gp \varphi {\bf B},\hspace{0.5cm}
{\bf P}_a = \gagg a {\bf B},\hspace{0.5cm}
{\bf M}_a = \gagg a {\bf E},
\end{equation}
and, therefore, the following inhomogeneous equations for the induced fields,
\begin{equation}\begin{aligned}
\nabla \cdot {\bf E}_{\varphi} &= - \gp {\bf E} \cdot \nabla \varphi - \gp \varphi \rho, \\
\nabla \cdot {\bf E}_a &= - \gagg {\bf B} \cdot \nabla a, \\
\nabla \times {\bf B}_{\varphi} &= \partial_t {\bf E}_{\varphi} - \gp (\nabla \varphi) \times {\bf B} + \gp (\partial_t \varphi) {\bf E} - \gp \varphi {\bf j}, \\
\nabla \times {\bf B}_a &= \partial_t {\bf E}_a \hspace{0.05cm}+ \gagg (\nabla a) \times {\bf E} \hspace{0.1cm}+ \gagg (\partial_t a) {\bf B}.
\label{eq:sa-maxwell}
\end{aligned}\end{equation}
From these, we can determine the induced fields a scalar or axion would generate for various laboratory field configurations.

We now specialise to the situation where the background electric field vanishes, as is commonly employed in experiments.
The effective currents that will then source magnetic fields are determined from Eq.~\eqref{eq:sa-maxwell} as ${\bf j}^{a}_{\rm eff} = \gagg (\partial_t a) {\bf B}$ and ${\bf j}^{\varphi}_{\rm eff} = -\gp \nabla \times (\varphi {\bf B})$.
From here, in analogy to Eq.~\eqref{eq:j_parity}, we find the following transformation properties for these currents
\begin{equation}
{\bf j}_{\rm eff}^{a}(P_\alpha {\bf r},P_\alpha {\bf k}) = \eta_{\alpha} P_{\alpha}\, {\bf j}_{\rm eff}^{a}({\bf r},{\bf k}),\hspace{0.5cm}
{\bf j}_{\rm eff}^{\varphi}(P_\alpha {\bf r},P_\alpha {\bf k}) = - \eta_{\alpha} P_{\alpha}\, {\bf j}_{\rm eff}^{\varphi}({\bf r},{\bf k}). 
\label{eq:sa-jtransform}
\end{equation}
In the language we introduced for gravitational waves, we see that the axion and scalar transform with $\sigma=-1$ and $+1$ respectively, being the spin-0 counterparts of $h^{\times}$ and $h^+$.
According to selection rule 2 derived in Sec.~\ref{sec:othergeom}, for a detector with cylindrical symmetry, there is only sensitivity to either $h^{\times}$ and or $h^+$.\footnote{We emphasise once more that this statement assumes the polarisations being defined as in Eq.~\eqref{eq:polarisation}.}
The proof of the selection rule only used Eq.~\eqref{eq:B_parity} which followed from Eq.~\eqref{eq:j_parity}, that is directly analogous to the transformations in Eq.~\eqref{eq:sa-jtransform}.
Accordingly axions and scalars must also obey selection rule 2, demonstrating that when full azimuthal symmetry is in place an instrument can only be sensitive to one of the two scalar waves.
This is shown explicitly in Tab.~\ref{table:summary_scalar}, which summarises our symmetry based results for those geometries that are sensitive to an axion and scalar.
This is the spin-0 analogue to Tab.~\ref{table:summary}.
Next we will expand upon these claims by presenting the explicit results for several geometries (these same three cases were considered for the GW in Sec.~\ref{sec:othergeom}).

\begin{table}[t] \centering \large
\renewcommand{\arraystretch}{1.2}
\begin{tabular}{c|c|c|}\cline{2-3}
& Solenoid: ${\bf B}_0 \propto \hat{\bf e}_z$ & Toroid: ${\bf B}_0 \propto \hat{\bf e}_{\phi}$ \\ \cline{1-3}
\multicolumn{1}{|c|}{$\hat{\bf n}' \propto \hat{\bf e}_z$}
& \makecell{scalar \\ $\Phi_a \equiv 0$, $\Phi_{\varphi} \neq 0$}
& \makecell{axion (ABRA) \\ $\Phi_a \neq 0$, $\Phi_{\varphi} \equiv 0$}
\\ \cline{1-3}
\multicolumn{1}{|c|}{$\hat{\bf n}' \propto \hat{\bf e}_{\phi}$}
& \makecell{axion (BASE) \\ $\Phi_a \neq 0$, $\Phi_{\varphi} = 0$}
& \makecell{scalar \\ $\Phi_a \equiv 0$, $\Phi_{\varphi} \neq 0$} 
\\ \cline{1-3}
\multicolumn{1}{|c|}{$\hat{\bf n}' \propto \hat{\bf e}_{\rho}$}
& \makecell{scalar \\ $\Phi_a \equiv 0$, $\Phi_{\varphi} = 0$}
& \makecell{axion \\ $\Phi_a = 0$, $\Phi_{\varphi} = 0$} 
\\ \cline{1-3}
\end{tabular}
\caption{A summary of the sensitivity of various detector geometries to an axion or scalar coupled to electromagnetism.
Columns specify the laboratory magnetic field configuration and rows specify the component of the induced magnetic field which is measured.
As manifest in Eq.~\eqref{eq:sa-jtransform} there is a direct analogy to these results and those for the GW given in Tab.~\ref{table:summary}.
The magnetic fluxes stated are derived under the assumption of full cylindrical symmetry.
Cases where we write $\Phi=0$ implies that the cylindrical symmetry has killed these results, as we had either $B^{a,\varphi} \propto z'$, or for the scalar something of the form $B^{a,\varphi} \propto \sin(\phi'-\phi_{\varphi})$.
On the other hand, where we write $\Phi \equiv 0$ sensitivity can only be recovered by modifying ${\bf B}$, as the vanishing flux originates arises from $B^{a,\varphi} = 0$.}
\label{table:summary_scalar}
\end{table}

\paragraph{Toroidal magnet with a horizontal pickup loop.}
We first study an instrument with ${\bf B}_0 \propto \hat{\bf e}_{\phi}$ and where $\hat{\bf n}^{\prime} \propto \hat{\bf e}_z$.
An explicit example of such a geometry is ABRACADABRA.
The magnetic field in this case was already provided in Eq.~\eqref{eq:toro_mag_field}, and again the toroid has inner and outer radii given by $R$ and $R+a$, and a height $H$ which we take to be parametrically larger than both.
The induced field in the $z$ direction -- the one the pickup loop will measure -- for each case can be computed as,
\begin{equation}\begin{aligned}
B^{\varphi}_z({\bf r}') &= 0, \\
B^{a}_z({\bf r}') &= \gagg (\partial_t a) B_0 R \left[ 
\ln \left( 1+ \frac{a}{R} \right) - \frac{a(a+2R)}{H^2}
\right]\!.
\end{aligned}\end{equation}
As in the main text, ${\bf r}' = (\rho',\,\phi',\,z')$ is the cylindrical coordinate system where the field is measured and integrated over by the pickup loop.
The axion contribution here is only stated to ${\cal O}(H^{-2})$, whereas the scalar result is exact: there is no scalar induced magnetic field in the $z$ direction.
This results from the azimuthal symmetry of the toroid, a toroid with a wedge in $\phi$ removed will have a non-zero $B^{\varphi}_z({\bf r}')$.
Regardless, the existing and planned toroidal axion instruments operating in this range, such as ABRACADABRA, SHAFT, or DMRadio-50L would have an exactly vanishing sensitivity to a scalar dark-matter signal, as they would for the $h^+$ polarisation of a GW, both consistent with selection rule 2.

\paragraph{Solenoidal magnet with an array of vertical pickup loops.}
Next we consider the primary configuration studied in the main text, where ${\bf B}_0 \propto \hat{\bf e}_z$ and $\hat{\bf n}^{\prime} \propto \hat{\bf e}_{\phi}$, as pursued by, for example, the BASE collaboration.
We will compute the azimuthal component of the magnetic field, as measured by a vertical pickup loop, and again assume that the height of the solenoid is parametrically larger than the other scales.
Adopting the magnetic field in Eq.~\eqref{eq:sol_mag_field}, to ${\cal O}(H^{-2})$ we have\footnote{To facilitate comparison with the axion, in the scalar case we used the fact that the phase of the dark-matter wave will be $\sim m(t-{\bf v} \cdot {\bf x})$ to write $\nabla \varphi = - {\bf v} (\partial_t \varphi)$.}
\begin{equation}\begin{aligned}
B^{\varphi}_{\phi}({\bf r}') &= 2 \gp (\partial_t \varphi) v B_0 z' \frac{R^2}{H^2} \sin \theta_{\varphi} \sin (\phi'-\phi_{\varphi}), \\
B^{a}_{\phi}({\bf r}') &= \frac{1}{2} \gagg (\partial_t a) B_0 \rho'  \left[ 1 - \frac{2R^2}{H^2} \right]\!,
\label{eq:sa-solenoid}
\end{aligned}\end{equation}
with $(\theta_{\varphi},\,\phi_{\varphi})$ the coordinates of the scalar field's velocity on the celestial sphere.
If we compare the two results, we see that if the experiment measures the magnetic flux within a pickup loop symmetric in $z'$ -- as done in ADMX SLIC or BASE -- then the axion flux will grow proportional to the height of the loop, whereas the scalar flux will exactly vanish as $B^{\varphi}_{\phi}({\bf r}') \propto z'$.
Further, for any loop in a plane of constant $\phi'$, if we wrap it in a full circle in $\phi'$ as DMRadio-m$^3$ plans, then again while the axion flux increases the scalar flux will vanish, consistent with selection rule 2.
It is straightforward to confirm that these results persist to all orders in $H$.
In other words, solenoidal axion instruments will generically have exactly vanishing sensitivity to a scalar dark-matter signal, even though $B^{\varphi}_{\phi}({\bf r}') \neq 0$.

As discussed in the main body, the key difference between these scalar and pseudoscalar axion results can be understood on basic symmetry grounds.
In particular, consistent with Tab.~\ref{table:eta}, the azimuthal component of the magnetic field will flip sign under parity.
For the axion interaction, this sign flip is produced by the pseudoscalar field itself, whereas for the scalar, the flip is generated by the $z'$ in Eq.~\eqref{eq:sa-solenoid}, as $P z' = - z'$.
(In the axion case, the equivalent dimensions were made up by $\rho'$, which is invariant under parity.)
Nevertheless, in spite of these differences, there are pickup loops that can be designed which would have sensitivity to both the scalar and axion.
This is not the case for a background toroidal field.
Similar arguments can be used to understand the remaining results in Tab.~\ref{table:summary_scalar}.

\paragraph{Solenoidal magnet with a horizontal pickup loop.}
Finally, we consider a configuration which is optimally sensitive to a scalar and minimally sensitive to the axion.
From selection rule 2, we simply need a configuration sensitive to the $\sigma=+1$ contribution, or the $h^+$ component for a GW.
From Tab.~\ref{table:summary}, one possibility is a solenoidal magnetic field ${\bf B}_0 \propto \hat{\bf e}_z$, with a pickup loop that reads out the $z$ component of the induced field, $\hat{\bf n}^{\prime} \propto \hat{\bf e}_z$.
For such a configuration, we have
\begin{equation}\begin{aligned}
B^{\varphi}_z({\bf r}') &= - \gp \varphi B_0 + \frac{1}{2} \gp (\partial_t \varphi) v B_0 \rho' \sin \theta_{\varphi} \cos (\phi'-\phi_{\varphi}), \\
B^{a}_z({\bf r}') &= 0.
\label{eq:B-solenoid-horizontal}
\end{aligned}\end{equation}
The scalar result is stated to leading order in $1/H$, whereas the axion result exactly vanishes.

Observe that in Eq.~\eqref{eq:B-solenoid-horizontal} the first term for the scalar is proportional to $\varphi$ rather than a derivative of the field, and arises from the unique current term for the scalar already visible in Eq.~\eqref{eq:sa-eom}.
As discussed, similar terms arise for the GW.
In the scalar case, these contributions are particularly simple.
For instance, as $-\gp \varphi {\bf j} \propto {\bf j}$, this contribution must be directly proportional to the background magnetic field, which is also produced by ${\bf j}$, or ${\bf B}^{\varphi} \propto {\bf B}_0$.
This implies that in Tab.~\ref{table:summary_scalar} such a contribution can only be measured when $\hat{\bf n}' \propto {\bf B}_0$, which only occurs for two of the six cases in the table.
This leaves one case where a scalar contribution is expected by selection rule 2, but where there is no contribution from the surface current: a solenoidal magnet with $\hat{\bf n}' \propto \hat{\bf e}_{\rho}$ for the pickup loop.
In this case, the magnetic field is generated purely from $(\nabla \varphi) \times {\bf B}_0$, and therefore must be proportional to $\sin(\phi'-\phi_{\varphi})$ or $\cos(\phi'-\phi_{\varphi})$.
When the pickup loop has azimuthal symmetry, however, such terms vanish, explaining why we write $\Phi_{\varphi}=0$ in the table, although it could be recovered by using a pickup loop that violates the rotational symmetry.

\section{Parity Properties of External Magnetic Fields}
\label{app:Bdecomposition}

In this appendix, we will demonstrate that it is possible to  decompose static cylindrically symmetric magnetic fields -- those invariant under azimuthal and $z$-reflection symmetries -- as a sum of a toroidal and a solenoidal piece, which as we will show take the form ${\bf B}^\toroid \propto \hat{\bf e}_{\phi}$ and ${\bf B}^\solenoid \propto \hat{\bf e}_z$, respectively.
In the idealised treatment of static laboratory magnetic fields, we usually imagine them dropping sharply to zero beyond the boundary of a  well-defined region.
According to the discussion in App.~\ref{app:surface_currents}, a current flows on the surface of such a boundary, which sources the magnetic field inside.
Azimuthal symmetry dictates that the current takes the form ${\bf j}\propto \delta(\rho-\rho_S(z))$.
Together with ${\bf \nabla}\cdot {\bf j} = 0$, this can be used to show that current can always be cast as ${\bf j} = {\bf j}^\toroid+{\bf j}^\solenoid$, with 
\begin{equation}\begin{aligned}
{\bf j}^\toroid &=\frac{{\cal C}^\toroid}{\rho}   \delta(\rho-\rho_S(z)) \left( [\partial_z\rho_S(z)]\hat{\bf e}_\rho+  \hat{\bf e}_z  \right)\!, \\
{\bf j}^\solenoid &=  {\cal C}^\solenoid(z)  \delta(\rho-\rho_S(z))  \,\hat{\bf e}_\phi,
\label{eq:currentfz}
\end{aligned}\end{equation}
where ${\cal C}^\solenoid (z) $ is a function of $z$ while ${\cal C}^\toroid$ is a constant.
In addition, $z$-reflection symmetry implies that
\begin{equation}
{\cal C}^\solenoid (-z)  = {\cal C}^\solenoid (z),\hspace{0.5cm}
\rho_S (-z) =\rho_S (z).
\label{eq:Csym}
\end{equation}
Observe that the two terms in Eq.~\eqref{eq:currentfz} are conserved separately.
As a result, we can similarly decompose the magnetic field sourced by ${\bf j}$ as ${\bf B} = {\bf B}^\toroid+{\bf B}^\solenoid$, with ${\bf \nabla} \times {\bf B}^\toroid = {\bf j}^\toroid$ and ${\bf \nabla} \times {\bf B}^\solenoid = {\bf j}^\solenoid$.
We will further discuss each case below, and justify that ${\bf B}^\toroid \propto \hat{\bf e}_{\phi}$ and ${\bf B}^\solenoid \propto \hat{\bf e}_z$.
From this we conclude that toroidal and solenoidal magnetic fields exhaust all the realistic configurations invariant under cylindrical and $z$-reflection symmetries, and further this justifies the symmetry transformations encoded in ${\bf B} (P_\alpha {\bf r}) = \eta_{\alpha} P_\alpha {\bf B} ({\bf r})$ and Tab.~\ref{table:eta}.

\paragraph{Toroidal fields.}
As we now show, the magnetic field arising from the toroidal current takes a simple form, regardless of the function $\rho_S(z)$.
Due to the cylindrical symmetry, the magnetic field sourced by ${\bf j}^\toroid$ points in the azimuthal direction, ${\bf B}^\toroid = B(\rho,z) \hat{\bf e}_\phi$.
For a fixed $z$, the circulation of the magnetic field along a circular path is $2\pi \rho B(\rho,z)$.
According to Amp\`ere's law, this must equal $ \int d\rho \int d\phi\, \rho \, j^\toroid_z = 2\pi {\cal C}^\toroid$.
Hence
\begin{equation}
{\bf B}^\toroid (\rho,z) = \frac{{\cal C}^\toroid}{\rho} \Theta(\rho-\rho_S (z)) \, \hat{\bf e}_\phi.
\label{eq:BtoroidGeneral}
\end{equation}
The $1/\rho$ dependence of the magnetic field can be alternatively derived by considering an arbitrary field of the form ${\bf B} = B(\rho,z)  \, \hat{\bf e}_\phi$, and then demanding that ${\bf \nabla} \cdot {\bf B} = {\bf \nabla} \times {\bf B} = 0$, which holds sufficiently far from the surface.

For the realistic toroidal fields produced in the lab, the current flows up and then down again, so the total field is the sum of two pieces.
Furthermore, we can usually neglect the $z$-dependence of $\rho_S(z)$.
In that case, the general result in Eq.~\eqref{eq:BtoroidGeneral} reduces to Eq.~\eqref{eq:toro_mag_field}.

\paragraph{Solenoidal fields.}
The solenoidal configuration does not lead to an expression as simple as Eq.~\eqref{eq:BtoroidGeneral}.
Nonetheless, Eq.~\eqref{eq:Csym} shows that ${\cal C}^{\solenoid} (z)$ and $\rho_S(z)$ are even, and hence $\partial_z {\cal C}^{\solenoid}(0) = \partial_z\rho_S(0)=0$.
Since axion experiments often measure the induced magnetic flux away from the $z$-boundaries of the external magnetic field, to a reasonable approximation we can take $\rho_S$ and $\partial_z {\cal C}^{\solenoid}$ as constants.
In that case ${\bf j}^\solenoid$ reduces to Eq.~\eqref{eq:jsolenoid} and the magnetic field takes the form given in Eq.~\eqref{eq:sol_mag_field}.
In general, however, this need not hold, and for instance ${\bf B}^{\solenoid}$ can develop a contribution $\propto \hat{\bf e}_{\rho}$.

\section{Response Matrix} 
\label{app:responsematrix}

In this appendix, we derive the response matrix introduced in Eq.~\eqref{eq:flux_Dmn}, and provide an explicit example.
Let us first note  that the effective current is a linear functional of the GW, which implies that there exists a tensor ${\cal J}^i_{m n}({\bf r}, {\bf k})$ such that
\begin{equation}
j_{\rm eff}^i =e^{- \i \omega t} {\cal J}^i_{m n}({\bf r}, {\bf k}) \sum_A h^A e^A_{m n}(\hat{\bf k}).
\end{equation}
Following Eq.~\eqref{eq:BioSavartA}, this gives Eq.~\eqref{eq:flux_Dmn}
\begin{equation}
\Phi_h=e^{- \i \omega t } D^{m n}({\bf k}) \sum_A h^A e^A_{m n}(\hat{{\bf k}}),
\hspace{0.5cm}
\text{with}
\hspace{0.5cm}
D^{m n}({\bf k})=\int_{\ell} d r^{\prime}_i \int_{V_{B}} \frac{d^3 {\bf r}}{4 \pi} \frac{{\cal J}^i_{m n}}{\left|{\bf r}-{\bf r}^{\,\prime}\right|}.	
\end{equation}

As an explicit example, for the solenoidal magnetic field in Eq.~\eqref{eq:sol_mag_field}, we find
\begin{equation}\begin{aligned}
{\cal J}^{\rho} & =  
\frac{1}{12}  e^{- \i \omega t} \omega^2 B_0 \Theta(R - \rho)
\begin{pmatrix}
\rho s_{2\phi} & - \rho c_{2 \phi} & 4z s_{\phi} \\
- \rho c_{2\phi} & - \rho s_{2\phi} & - 4z c_{\phi} \\
4z s_{\phi} & - 4z c_{\phi} & 0
\end{pmatrix}\!, \\
{\cal J}^{\phi} &  = 
\frac{1}{12}  e^{- \i \omega t} \omega^2 B_0  \Theta(R - \rho)
\begin{pmatrix}
\rho (3 + c_{2\phi}) & \rho s_{2 \phi} & 4z c_{\phi} \\
\rho s_{2 \phi} & \rho( 3 - c_{2 \phi} ) &  4z s_{\phi} \\
4z c_{\phi} & 4z s_{\phi} & - 4 \rho
\end{pmatrix} \\ 
& \;\; -\frac{1}{12}  e^{- \i \omega t} \omega^2 B_0  \delta(R - \rho)
\begin{pmatrix}
z^2 + 2 \rho^2 + \rho^2 c_{2\phi} 	&  \rho^2 s_{2\phi} & 4z \rho c_{\phi} \\
\rho^2 s_{2\phi} & z^2 + 2 \rho^2 - \rho^2 c_{2\phi}  & 4z \rho s_{\phi} \\
4z \rho c_{\phi}	&  4z \rho s_{\phi} & 5z^2 - \rho^2
\end{pmatrix}\!, \\
{\cal J}^{z} & = 
\frac{1}{4} e^{- \i \omega t} \omega^2 B_0  \Theta(R- \rho)
\begin{pmatrix}
0 & 0 & - \rho s_{\phi} \\
0 & 0 &  \rho c_{\phi} \\
- \rho s_{\phi} & \rho c_{\phi} & 0
\end{pmatrix} \\
& \;\; - \frac{1}{12}  e^{- \i \omega t} \omega^2 B_0 \delta(R- \rho)
\begin{pmatrix}
z \rho s_{2\phi} & - z \rho c_{2 \phi } & - \rho^2 s_{\phi} \\
- z \rho c_{2 \phi } &  - 2 z \rho c_{\phi} s_{\phi} &  \rho^2 c_{\phi} \\
- \rho^2 s_{\phi} & \rho^2 c_{\phi} & 0
\end{pmatrix}\!,
\end{aligned}\end{equation}
at ${\cal O}[(\omega L)^2]$.
From here, considering the specific loop geometry described in Fig.~\ref{fig:solenoid} with $r_2=r$ and $r_1=0$, the response matrix is
\begin{equation}
D^{mn}({\bf k}) =\frac{e^{- \i \omega t}}{288} \omega^2 B_0 r l \left(	13 r^2 - 30 R^2	\right)  
\begin{pmatrix}
0 & 0 & - s_{\phi_{\ell}} \\
0 & 0 & c_{\phi_{\ell}} \\
- s_{\phi_{\ell}} & c_{\phi_{\ell}} & 0
\end{pmatrix}\!.
\end{equation}
Note that as shown in Eq.~\eqref{eq:thm1}, for a cylindrical symmetric setup, only the diagonal part of the response matrix matters, and therefore the zeros in the response function are consistent with the fact that we do not have any contribution at ${\cal O}[(\omega L)^2]$ for the vertical loop.

\section{Explicit Expressions for the Effective Current}
\label{app:explicitcurrent}

In this appendix, we provide the analytic expressions for the current induced by a GW to $ {\cal O}[(\omega L)^3] $ for both a solenoidal and toroidal external magnetic field as in Eq.~\eqref{eq:sol_mag_field} and Eq.~\eqref{eq:toro_mag_field}.
Together with Eq.~\eqref{eq:BiotSavart} or Eq.~\eqref{eq:BioSavartA}, these are used to calculate the induced magnetic field and flux which can then be measured by a pickup loop.
To slightly simplify the expressions that follow, throughout this appendix we have taken $\phi_h=0$, but this can be restored immediately by sending $\phi \to \phi-\phi_h$.

\paragraph{Solenoidal magnet.}
The three components of the current ${\bf j}=(j_{\rho},\,j_{\phi},\,j_z)$ are given as follows, where in each case organise the results by polarisation and power in $\omega$,
\begin{align}
j_{\rho} e^{\i \omega t}  =\, & B_0 \Theta(R - \rho) \label{eq:current_full_rho} \\
\times &\Bigg[
\frac{1}{24\sqrt{2}} h^+ \omega^2
\left\{ \rho (3 + c_{2 \theta_h} ) s_{2\phi} - 8 z s_{2 \theta_h} s_{\phi} \right\} -\frac{1}{6\sqrt{2}} h^{\times} \omega^2
\left\{ \rho c_{\theta_h } c_{2\phi} - 4 z c_{\phi} s_{\theta_h} \right\} \nonumber  \\
&\hspace{-0.25cm} +\frac{\i}{192\sqrt{2}} h^+ \omega^3
\Big\{ \left[  (45 \rho^2 + 5 \rho^2 c_{2\phi} -58z^2) s_{\theta_h} + (\rho^2 + \rho^2 c_{2\phi}-18z^2) s_{3 \theta_h} 
\right] s_{\phi} \nonumber \\
&\hspace{2.5cm} + 2 z\rho (19 c_{\theta_h} + 5 c_{3 \theta_h } ) c_{2\phi} \Big\}    \nonumber \\
&\hspace{-0.25cm} - \frac{\i}{48\sqrt{2}}  h^{\times} \omega^3
\left\{ 4 z\rho (1 + 2 c_{2 \theta_h} c_{2\phi} - 6 s_{\theta_h}^2 ) + (5 \rho^2 + \rho^2 c_{2\phi} -14z^2) c_{\phi} s_{2 \theta_h} \right\}
\Bigg], \nonumber
\end{align}
\begin{align}
j_{\phi} e^{\i \omega t}  =\, & B_0 \Theta(R-\rho) \label{eq:current_full_phi} \\
\times &\Bigg[ \frac{1}{24\sqrt{2}} h^+ \omega^2  
\left\{ \rho ( 3 + c_{2 \theta_h} ) c_{2\phi} + 14 \rho s_{\theta_h}^2 - 8 z c_{\phi} s_{2 \theta_h} \right\} + \frac{1}{3\sqrt{2}} h^{\times} \omega^2  
(\rho c_{\theta_h}  c_{\phi} - 2 z s_{\theta_h} ) s_{\phi} \nonumber \\
&\hspace{-0.25cm} - \frac{\i}{192\sqrt{2}}  h^+ \omega^3 
\left\{ -10 z \rho c_{3 \theta_h }  c_{2\phi}
- 2 z \rho c_{\theta_h}  ( -26 + 26 c_{2 \theta_h} + 19 c_{2\phi} )  \right. \nonumber \\
& \hspace{2.6cm} \left. + c_{\phi }
\left[  ( 58z^2 + 51 \rho^2 - 5 \rho^2 c_{2\phi} ) s_{\theta_h} 
+ ( 18z^2 - 17 \rho^2 - \rho^2 c_{2\phi} ) s_{3 \theta_h}
\right] \right\}  \nonumber \\
&\hspace{-0.25cm} + \frac{\i}{48\sqrt{2}} h^{\times} \omega^3
\left\{ 4 z\rho   (1 + 2 c_{2 \theta_h } ) s_{2\phi}  
+ (-14^2 + \rho^2 + \rho^2 c_{2\phi)}  s_{2 \theta_h } s_{\phi} \right\}  \Bigg] \nonumber \\
+\,& B_0 \delta(R-\rho) \nonumber \\
\times&  \Bigg[ \frac{1}{24\sqrt{2}}  h^+ \omega^2  
\left\{ - 8z^2 s_{\theta_h}^2 + \rho(  - \rho c_{2\theta_h} (3 + c_{2\phi}) + 8 z c_{\phi} s_{2 \theta_h} + 6 \rho s_{\phi}^2  )
\right\} \nonumber \\
&\hspace{-0.25cm} - \frac{1}{3\sqrt{2}}  h^{\times} \omega^2  \rho 
( \rho c_{\theta_h}  c_{\phi} - 2 z s_{\theta_h} ) s_{\phi} \nonumber  \\
&\hspace{-0.25cm} - \frac{\i}{192\sqrt{2}} h^+ \omega^3 
\left\{ z c_{\theta_h} \left( \rho^2 (-22 + 22 c_{2 \theta_h} + c_{2\phi} ) + 24z^2 s_{\theta_h}^2  \right) \right. \nonumber \\
&\hspace{2.5cm} + \rho \left(   c_{\phi} (   ( 6z^2 - 15\rho^2 + 5\rho^2 c_{2\phi} ) s_{\theta_h} + (- 18z^2 + 5\rho^2 + \rho^2 c_{2\phi} ) s_{3\theta_h} ) \right. \nonumber \\ 
&\hspace{2.5cm} \left. \left. + 7 z \rho c_{3 \theta_h} c_{2\phi} \right) \right\} \nonumber \\
&\hspace{-0.25cm} + \frac{\i}{48\sqrt{2}}  h^{\times} \omega^3 
\left\{ 2 z\rho   (-1 + 2 c_{2 \theta_h} ) s_{2\phi}  
+ (-6 z^2 + \rho^2 + \rho^2 c_{2\phi})  s_{2 \theta_h} s_{\phi} \right\} \Bigg], \nonumber
\end{align}
and
\begin{align}
j_z e^{\i \omega t} =\, & B_0 \Theta(R - \rho) \label{eq:current_full_z} \\
\times  &\Bigg[ \frac{1}{4\sqrt{2}}  h^+ \omega^2  \rho s_{2 \theta_h} s_{\phi} - \frac{1}{2\sqrt{2}} h^{\times} \omega^2 \rho c_{\phi} s_{\theta_h} \nonumber  \\
&\left. \hspace{-0.25cm} + \frac{\i}{3\sqrt{2}} h^+ \omega^3 \rho c_{\theta_h}  s_{\theta_h} (z c_{\theta_h} + \rho c_{\phi} s_{\theta_h} ) s_{\phi} - \frac{\i}{3\sqrt{2}} h^{\times} \omega^3 \rho c_{\phi} s_{\theta_h } (z c_{\theta_h} + \rho c_{\phi } s_{\theta_h} ) \right]
\nonumber \\
+\, &\rho B_0 \delta(R- \rho) \nonumber \\
\times &\Bigg[  \frac{-1}{24\sqrt{2}}  h^+ \omega^2 \left\{ 2\rho s_{2\theta_h} s_{\phi} + z(3+c_{2\theta_h}) s_{2\phi} \right\}  + \frac{1}{6\sqrt{2}} h^{\times} \omega^2 (  z c_{\theta_h} c_{2\phi} + \rho c_{\phi} s_{\theta_h} ) \nonumber  \\
&\hspace{-0.25cm} - \frac{\i}{48\sqrt{2}} h^+ \omega^3 \left\{ z \cos \theta_h + \rho c_{\phi} s_{\theta_h}  (2 \rho s_{2\theta_h} s_{\phi} + z (3+c_{2\theta_h}) s_{2\phi} \right\} \nonumber \\
&\hspace{-0.25cm}  + \frac{\i}{12\sqrt{2}} h^{\times} \omega^3 
(z c_{\theta_h} + \rho c_{\phi } s_{\theta_h} ) (z c_{\theta_h} c_{2\phi} + \rho c_{\phi} s_{\theta_h} ) \Bigg]. \nonumber
\end{align}
In these expressions we employ the coordinate system introduced in Sec.~\ref{sec:framework}.
Observe that for the $\phi$ and $z$ components, the current can be divided into two parts: one proportional to $\Theta(R-\rho)$ and the other proportional to $\delta(R-\rho)$.
The latter can be understood as a contribution coming from the change in the magnetisation at the edge of magnetic field, see Eq.~\eqref{eq:jeffboundary}.

\paragraph{Toroidal magnet.} Equivalent results can be derived for a toroidal magnet.
Again, the results are divided into two parts: one proportional to $\Theta(R + a - \rho) - \Theta(R-\rho)$ and the other to $\delta(R + a - \rho) - \delta(R-\rho)$.
\begin{align}
j_{\rho} e^{\i \omega t} =\, &\frac{B_{\max} R}{\rho} \left[  \Theta(R+a - \rho) - \Theta(R-\rho) \right] \\
\times &\Bigg[ \frac{1}{12\sqrt{2}} h^+ \omega^2 
\left\{ 2z (3 + c_{2\theta_h}) c_{2\phi} + 6 z s_{\theta_h}^2 - \rho c_{\phi} s_{2\theta_h}  \right\} + \frac{1}{6\sqrt{2}} h^{\times} \omega^2 \left( 8 z c_{\theta_h}c_{\phi} - \rho s_{\theta_h}\right) s_{\phi} \nonumber \\ 
&\hspace{-0.25cm}+ \frac{\i}{192\sqrt{2}} h^+ \omega^3 
\Big\{ z \rho \left( - 9 c_{\phi} (-11 s_{\theta_h}  +  s_{3\theta_h} ) + 5 c_{3\phi} (5 s_{\theta_h} + s_{3 \theta_h}) \right) \nonumber \\
&\hspace{2.5cm} + 2 c_{\theta_h} \left( (30z^2 - 21 \rho^2 + (10z^2 + \rho^2) c_{2\theta_h}  ) c_{2\phi} + (24 z^2 - 2\rho^2) s_{\theta_h}^2    \right) \Big\} \nonumber \\
&\hspace{-0.25cm} + \frac{\i}{48\sqrt{2}}h^{\times} \omega^3 \left\{ 
2 \left( 5z^2 - 4 \rho^2 + (5z^2 - \rho^2) c_{2\theta_h} \right) s_{2\phi} + z \rho s_{2\theta_h} (9 s_{\phi} + 5 s_{3\phi} ) \right\} \Bigg], \nonumber
\end{align}
\begin{align}
j_{\phi} e^{\i \omega t} =\, & \frac{B_{\max} R}{\rho} \left[  \Theta(R+a - \rho) - \Theta(R-\rho) \right] \\
\times &\Bigg[ \frac{-1}{6\sqrt{2}} h^+ \omega^2  \left\{ z (3 + c_{2\theta_h}) c_{\phi} + \rho s_{2\theta_h} \right\}  s_{\phi} +  \frac{1}{3\sqrt{2}} h^{\times} \omega^2 \left( z c_{\theta_h} c_{2\phi} + \rho c_{\phi} s_{\theta_h}\right) \nonumber \\ 
& \hspace{-0.25cm}- \frac{\i}{64\sqrt{2}} h^+ \omega^3 \Big\{ 4c_{\theta_h} \left( 3z^2 + \rho^2 + (z^2 - \rho^2 ) c_{2\theta_h} \right)   s_{2\phi} + z \rho s_{3\theta_h} ( 5s_{\phi} + s_{3\phi} )   \nonumber \\ 
&\hspace{2.5cm} + z\rho s_{\theta_h} ( 9 s_{\phi} + 5 s_{3\phi} )  
 \Big\} \nonumber \\
& \hspace{-0.25cm} + \frac{\i}{16\sqrt{2}}h^{\times} \omega^3 \left\{ 
( 2 z^2 + \rho^2 ) c_{2\phi} - c_{2\theta_h} \left( \rho^2 + (- 2z^2 + \rho^2) c_{2\phi} \right)
+ \rho (\rho + 4z c_{\phi}^3 s_{2\theta_h}) \right\} \Bigg] \nonumber \\
+\, &B_{\max} R \left[  \delta(R+a - \rho) - \delta(R-\rho) \right] \nonumber \\
\times &\Bigg[ \frac{-1}{12\sqrt{2}} h^+ \omega^2  \left\{ z (3 + c_{2\theta_h}) c_{\phi} + \rho s_{2\theta_h} \right\}  s_{\phi}
-  \frac{1}{6\sqrt{2}} h^{\times} \omega^2 \left( z c_{\theta_h} c_{2\phi} + \rho c_{\phi} s_{\theta_h}\right)  \nonumber \\ 
& \hspace{-0.25cm}- \frac{\i}{48\sqrt{2}} h^+ \omega^3  \left( z c_{\theta_h} + \rho c_{\phi} s_{\theta_h} \right) \left( 2 \rho s_{2\theta_h} s_{\phi} + z (3 + c_{2\theta_h}) s_{2\phi} \right) \nonumber \\
&\hspace{-0.25cm} - \frac{\i}{12\sqrt{2}}h^{\times} \omega^3 \left( z c_{\theta_h} + \rho c_{\phi} s_{\theta_h} \right) \left( z c_{\theta_h} c_{2\phi} + \rho c_{\phi} s_{\theta_h} \right) \Bigg], \nonumber
\end{align}
and
\begin{align}
j_z e^{\i \omega t} =\, &\frac{B_{\max} R}{\rho^2} \left[  \Theta(R+a - \rho) - \Theta(R-\rho) \right] \\
\times &\Bigg[ \frac{1}{12\sqrt{2}} h^+ \omega^2  
\left\{ (z^2 - 2 \rho^2) (3 + c_{2\theta_h}) c_{2\phi} + 3z \rho c_{\phi} s_{2\theta_h} \right\} \nonumber \\ 
&\hspace{-0.25cm}+  \frac{1}{6\sqrt{2}} h^{\times} \omega^2 \left\{ 4(z^2 - 2 \rho^2) c_{\theta_h} c_{\phi} + 3z \rho s_{\theta_h} \right\}  s_{\phi} \nonumber \\ 
&\hspace{-0.25cm}- \frac{\i}{192\sqrt{2}} h^+ \omega^3 \Big\{ 2z(2z^2 - 7\rho^2) (7 c_{\theta_h} + c_{3\theta_h}) c_{2\phi} + 8 z \rho^2 c_{\theta_h} s_{\theta_h}^2 \nonumber \\ 
&\hspace{2.5cm}  + \rho c_{\phi} \big[ (50 z^2 - 9 \rho^2 + 5 (4z^2 - 9 \rho^2) c_{2\phi} ) s_{\theta_h} \nonumber \\
&\hspace{4.2cm}
+ ( 10 z^2 + 3\rho^2 + ( 4 z^2 - 9 \rho^2 ) c_{2\phi} ) s_{3\theta_h} \big] \Big\} \nonumber \\
&\hspace{-0.25cm} + \frac{\i}{48\sqrt{2}}h^{\times} \omega^3 \left\{ 
\rho \left(14z^2 - 9 \rho^2 + (4z^2 - 9 \rho^2)c_{2\phi} \right) s_{2\theta_h} s_{\phi} + 4z(2z^2 - 7\rho^2) c_{\theta_h}^2 s_{2\phi} \right\} \Bigg]  \nonumber \\
+\, &\frac{B_{\max} R}{\rho} \left[  \delta(R+a - \rho) - \delta(R-\rho) \right] \nonumber \\
\times &\Bigg[ \frac{1}{24\sqrt{2}} h^+ \omega^2 
\left\{ (z^2 + 2\rho^2) (3+c_{2\theta_h}) c_{2\phi}  + 2z s_{\theta_h} (- 4\rho c_{\theta_h} c_{\phi} + 3z s_{\theta_h}) \right\} \nonumber \\
&\hspace{-0.25cm}+ \frac{1}{3\sqrt{2}} h^{\times} \omega^2  \left[ (z^2 + 2\rho^2) c_{\theta_h} c_{\phi} - z \rho s_{\theta_h} \right] s_{\phi}  \nonumber \\
&\hspace{-0.25cm}- \frac{\i}{192\sqrt{2}} 
 h^+ \omega^3 
\Big\{ \rho c_{\phi} \left[ (8z^2 + 3 \rho^2 + 5(2z^2 + 3\rho^2)c_{2\phi} ) s_{\theta_h} \right. \nonumber \\ 
& \left. \hspace{3.8cm} + ( - 8 z^2 - \rho^2 + (2z^2 + 3 \rho^2) c_{2\phi} ) s_{3\theta_h} \right] \nonumber \\
& \hspace{2.5cm} + 2 z c_{\theta_h} \left( (6z^2 + 7 \rho^2 + (2z^2 + 5\rho^2)c_{2\theta_h} ) c_{2\phi}  + (8z^2 - 2 \rho^2) s_{\theta_h}^2 \right) \Big\}\nonumber \\
&\hspace{-0.25cm} - \frac{\i}{48\sqrt{2}} h^{\times} \omega^3 
\left\{ \rho(3 \rho^2 + (2z^2 + 3\rho^2) c_{2\phi} ) s_{2\theta_h} s_{\phi} + 2z (z^2 + \rho^2 + (z^2 + 2\rho^2)c_{2\theta_h})s_{2\phi} \right\} \Bigg]. \nonumber
\end{align}

\section{Recasting Dark Matter Sensitivity to GW Strain Sensitivity}
\label{app:CoherenceRatio}

Using the techniques described so far, the magnetic flux induced by a GW passing through various lumped-element detectors can be computed.
To determine the sensitivity to these signals, that information then needs to be combined with various other properties of the signal, such as its duration and frequency profile, as well as the backgrounds characteristic of the individual detector.
In this appendix, we present an alternative way to estimate the parametric sensitivity of instruments to a GW signal, which involves bootstrapping the known sensitivity to axion dark matter.
The starting point is a calculation of the magnetic flux for axion dark matter and the GW.
However, we cannot simply equate these fluxes, as there are two other properties of the signal that will determine their detectability: the signal duration, and the signal coherence.
Roughly, the more coherent and longer a signal, the more straightforward it is to detect.
Our approach will be to correct the detectability of the two fluxes with a \textit{coherence ratio}, ${\cal R}_c$, determined such that
\begin{equation}
\Phi_h = {\cal R}_c\, \Phi_a.
\end{equation}
As $\Phi_h \propto h$ (and $\Phi_a \propto \gagg$) this then determines the GW strain sensitivity.
In most cases, the highly coherent nature of dark matter will lead to ${\cal R}_c > 1$, and therefore suppress our sensitivity, but as we outline, there are cases where the coherence ratio can be less than unity.
We emphasise that our approach should be viewed as a heuristic, a shortcut to exploit known axion sensitivities to determine the parametric sensitivity to a GW.
For an individual experiment, the correct approach is always to determine the full sensitivity to the instrument.

In what follows, we will firstly detail the relevant time scales that need to be combined to compute ${\cal R}_c$, and following this we will explain how that computation is performed.
The results can depend on the scan strategy adopted by the instrument, and in order to account for this we will then introduce a simple model for the DMRadio scan strategy.
We will then outline how two specific examples, black hole superradiance and PBH mergers can be described using the formalism we have introduced.

\subsection{The relevant times scales for GW and dark matter signals}

With the motivations and caveats detailed, let us expand on the physics of the problem.
We will consider a GW signal that lasts a time $T_h$ and has a finite bandwidth, and therefore coherence time $\tau_h$, although we will have $\tau_h \leq T_h$.
The exact values of these parameters will depend on specific models -- we will consider the cases of black hole superradiance and PBH mergers later in this appendix -- but for now we will keep the discussion general.
For instance, while we keep $T_h$ arbitrary, by making it longer than the experimental run time $T_\textrm{exp}$ we can account for a persistent signal.
The GW signal will then be compared against axion DM, which is a persistent signal that is highly coherent, with a coherence time $\tau_a = 2 \pi\, Q_a/m_a \simeq 4~\textrm{s}\, (1~\textrm{neV}/m_a)$, specified in terms of an effective quality factor for the signal, which for DM is given by $Q_a = 10^6$.
If the mean frequency of the GW is $\omega_h$, we can analogously define a quality factor as $2 \pi\, Q_h = \tau_h\, \omega_h \leq T_h\, \omega_h$.

Thus far, we have introduced the scales associated with the two signals we wish to compare, $\{\tau_h,\,T_h,\,\tau_a\}$, however we must also account for the relevant time scales associated with the experiment we are searching for the signal in.
In this work, we focus solely on instruments that pursue a resonant scan strategy: the instrument is tuned to resonantly enhance an angular frequency $\omega_m$, interrogating this frequency for a time $T_m$, before moving onto the next frequency.
The total run time for the instrument is then $T_\textrm{exp} = \sum_m T_m$, where the choice of $\{\omega_m,\,T_m\}$ defines an experimental scan strategy.
The final relevant quantity for a resonant instrument is the quality of the resonant response at the frequency $\omega_m$, which defines a timescale $\tau_r = 2\pi \,Q_r/\omega_m$.
Note that $\tau_r$ and $Q_r$ can vary as we change the resonant mass considered; cf. the resonant response of a microwave cavity to that in nuclear magnetic resonance (NMR).

\subsection{Exponential statistics and the coherence ratio}

We now need to combine the previously discussed timescales into a single coherence ratio.
As mentioned at the outset, we cannot simply match the fluxes because certain signals are easier to detect than others.
The formal way of capturing this is with statistical significance.
As demonstrated in Ref.~\cite{Foster:2017hbq} (see also Ref.~\cite{Foster:2020fln}), for axion dark matter searches, the signal and background are expected to be exponentially distributed, and based on similar arguments, we assume the GW signals discussed here are also.
An estimate for the signal sensitivity for an exponentially distributed quantity is given by the signal to noise ratio, $\textrm{SNR} \sim P_s/P_b$, where $P_s$ and $P_b$ are the power associated with the signal (with $s = a,h$) and background.
If we were instead performing a counting experiment, the analogous expression would be $S/\sqrt{B}$ -- a consequence of the Poisson likelihood -- where $S$ and $B$ are the number of signal and background counts.
For $N$ independent measurements, the significance in both cases grows as $\sqrt{N}$.
By matching the significance for the GW and dark matter signals, we will quantify the notion of detectability, and then as $P_s \propto \langle \Phi_s^2 \rangle$, we will determine ${\cal R}_c$.

The goal then is to compute $P_s$ and $P_b$ for each signal, and where relevant, the number of independent bins.
As we do so, we need only keep track of factors specific to each signal, as any common factors will cancel when we compute the coherence ratio.
Consider first the power associated with a general signal, with flux $\Phi_s$, coherence time $\tau_s$ and duration $T_s > \tau_s$.
For a given resonant bandwidth, the longest time we can effectively interrogate the signal is given by $T_{m,s} = \min[T_m,\,T_s]$.
In principle, this need not be longer than $\tau_r$, in which case we would not fully ring up the resonant cavity.
Given this and that such resonant systems can be effectively reduced to the study of the simple harmonic oscillator, the recent analysis for NMR based instruments in Ref.~\cite{Dror:2022xpi} can be deployed.
Using this, the differential signal power at the resonant frequency, $\omega_0$ depends on the hierarchy of scales as follows,\footnote{Note that in the event $T_{m,s} = T_s > \tau_s$, the first and third hierarchies here are unphysical, and only the remaining two need be considered, with the relevant comparison being between $T_s$ and $\tau_r$.
This holds for all of the expressions that follow.}
\begin{equation}\begin{aligned}
\frac{dP_s}{d\omega}(\omega_0) \propto \langle \Phi_s^2 \rangle\left\{ \begin{array}{lcc}
T_{m,s}^3 && T_{m,s} \ll \tau_s, \tau_r, \\
T_{m,s}^2 \tau_s && \tau_s \ll T_{m,s} \ll \tau_r, \\
T_{m,s} \tau_r^2 & & \tau_r \ll T_{m,s} \ll \tau_s, \\
\tau_s \tau_r^2 & & \tau_s,\tau_r \ll T_{m,s}.
\end{array} \right.
\label{eq:Psomega0}
\end{aligned}\end{equation}
What enters the significance is the integrated power.
For the first three cases in Eq.~\eqref{eq:Psomega0}, where $T_{m,s}$, the signal will not be resolved in a single bin in the analysis; the frequency resolution in the discrete Fourier transform is set by $\Delta \omega = 2 \pi/T_{m,s}$, which must be narrower than the product of the signal and the instrument transfer function before the signal is resolved (for further details, see Ref.~\cite{Dror:2022xpi}).
Accordingly, for the first three cases the width is simply $2 \pi/T_{m,s}$.
In the final case the signal becomes resolved, and therefore the integration range is set by the minimum of the signal and resonator widths, or $\textrm{min}[\omega_m/Q_s,\,\omega_m/Q_r] = 2\pi\,\textrm{min}[1/\tau_s,\,1/\tau_r]$.
Dropping the common factor of $2\pi$, the total signal power is then,
\begin{equation}\begin{aligned}
P_s \propto \langle \Phi_s^2 \rangle\left\{ \begin{array}{lcc}
T_{m,s}^2 && T_{m,s} \ll \tau_s, \tau_r, \\
T_{m,s} \tau_s && \tau_s \ll T_{m,s} \ll \tau_r, \\
\tau_r^2 & & \tau_r \ll T_{m,s} \ll \tau_s, \\
\tau_r\, \textrm{min}[\tau_s,\,\tau_r] & & \tau_s,\tau_r \ll T_{m,s}.
\end{array} \right.
\label{eq:Pstotal}
\end{aligned}\end{equation}
Note that $\textrm{min}[\tau_s,\,\tau_r] \propto \textrm{min}[Q_s,\,Q_r]$, so in the resolved scenario the signal power can only be rung up to the minimum of the instrumental and signal $Q$-factors.

The consideration of the background will be more straightforward.
If we assume it is flat in frequency, then the total background power is just controlled by the width over which the signal is distributed, which we have already discussed.
Lastly, for the case where the signal is resolved into multiple frequency bins, we will receive a $\sqrt{N}$ enhancement to the SNR as discussed above.
In particular, the number of bins is given by $T_{m,s}/\textrm{max}[\tau_s,\,\tau_r]$, so we arrive at
\begin{equation}\begin{aligned}
\textrm{SNR}/\langle \Phi_s^2 \rangle \propto {\cal T}_s \equiv \left\{ \begin{array}{lcc}
T_{m,s}^3 && T_{m,s} \ll \tau_s, \tau_r, \\
T_{m,s}^2 \tau_s && \tau_s \ll T_{m,s} \ll \tau_r, \\
T_{m,s} \tau_r^2 & & \tau_r \ll T_{m,s} \ll \tau_s, \\
\tau_s \tau_r^2 \sqrt{T_{m,s}/\textrm{max}[\tau_s,\,\tau_r]} & & \tau_s, \tau_r \ll T_{m,s}.
\end{array} \right.
\label{eq:SNR}
\end{aligned}\end{equation}
For axion DM, $\Phi_s \propto \gagg$ and $T_{m,s} = T_m$, so that from this result for the four regimes in Eq.~\eqref{eq:SNR}, our sensitivity would scale as $\gagg \propto \{T_m^{-3/2},\,T_m^{-1},\,T_m^{-1/2},\,T_m^{-1/4} \}$ for the four cases considered, as claimed in Ref.~\cite{Dror:2022xpi}.

For a given resonant bandwidth, Eq.~\eqref{eq:SNR} allows us to determine the relative SNR for an axion dark matter and GW signal, and therefore the appropriate coherence ratio, as ${\cal R}_c = \sqrt{{\cal T}_a/{\cal T}_h}$.
There is, however, one final factor that must be included.
As the instrument executes its scan strategy in search of dark matter -- scanning each $\omega_m$ for a time $T_m$ -- if the GW wave signal is sufficiently long and incoherent, then the signal will persist across multiple resonant bandwidths, and if there are $M_h$ of these, the GW SNR receives a $\sqrt{M_h}$ enhancement.
Determining $M_h$ requires knowledge of the exact scan strategy executed by the instrument, and a number of considerations enter into the determination of the optimal scan strategy, as discussed in, for example, Refs.~\cite{Chaudhuri:2018rqn,Chaudhuri:2019ntz}.
Instead, we adopt a simplified approach.
We assume the scan strategy is determined by a choice not to scan any putative dark matter mass more than once, so that $\omega_{m+1}-\omega_m = \textrm{max}[\omega_m/Q_a,\,\omega_m/Q_r]$.
For a persistent GW signal, the width is approximately $\omega_m/Q_h$, so that $M_h \sim \textrm{max}[1,\,\textrm{min}[Q_a,\,Q_r]/Q_h]$.
For a transient signal, $M_h$ can be reduced from this value, as the signal may not persist as the various bins are scanned, and that reduction must be accounted for.
As a specific example, in the limit where $\tau_h = T_h$, we will always have $M_h=1$.

In summary then, we have
\begin{equation}
{\cal R}_c = \frac{1}{M_h^{1/4}}\sqrt{\frac{{\cal T}_a}{{\cal T}_h}},
\end{equation}
with $M_h$ the number of resonant bandwidths the GW signal appears in as discussed in the previous paragraph, and ${\cal T}_a$ and ${\cal T}_h$ are determined by Eq.~\eqref{eq:SNR}.
To build some intuition for this result, let us determine ${\cal R}_c$ explicitly in several cases.
The complexity of the expressions above largely originate from the many scales and the possible hierarchies between them.
As we will see, once several are fixed, the results simplify.
After this, we will show how the formalism can be deployed for the specific examples of superradiance and PBH mergers.

\paragraph{Persistent signal and a long interrogation time.} For axion DM, the signal duration is effectively infinite ($T_a = \infty$), and often we take $T_m \gg \tau_a,\,\tau_r$.
If we assume a similar hierarchy holds for the GW ($T_h \geq T_\textrm{exp}$ and $T_m \gg \tau_h,\,\tau_r$), then both signals are descsribed by the final line of Eq.~\eqref{eq:SNR}, and accounting for the additional factor of $M$, we have
\begin{equation}\begin{aligned}
{\cal R}_c = &\sqrt{\frac{Q_a}{Q_h}} \left( \frac{\textrm{max}[Q_h,\,Q_r]}{\textrm{max}[Q_a,\,Q_r]} \right)^{1/4} \left(\frac{1}{\textrm{max}[1,\,\textrm{min}[Q_a,\,Q_r]/Q_h]} \right)^{1/4} \\
= &\left\{ \begin{array}{lcc}
(Q_a/Q_h)^{1/2} & & Q_a < Q_h < Q_r, \\
(Q_a^2/Q_r Q_h)^{1/4} & & Q_a < Q_r < Q_h, \\
(Q_a/Q_h)^{1/4} & & \textrm{otherwise}.
\end{array}\right.
\end{aligned}\end{equation}
Given the highly coherent nature of the dark matter signal, in most cases we will be in the regime where $Q_h < Q_a$, and then we find ${\cal R}_c = (Q_a/Q_h)^{1/4}$.
This is exactly the scaling argued for in Ref.~\cite{Domcke:2022rgu}, although we can now see that this result only holds for the particular set of assumptions we invoked in this paragraph.\footnote{In Ref.~\cite{Dror:2021nyr}, it was argued that a persistent signal of relativistic axions has a coherence ratio analogous to our ${\cal R}_c = (Q_a/Q_h)^{1/4}$, which was claimed to hold for both resonant and broadband strategies.
Even though that result essentially matched the more accurate experimental sensitivity determined in Ref.~\cite{ADMX:2023rsk}, we emphasise that for a resonant experiment a deviation from the $(Q_a/Q_h)^{1/4}$ scaling should be expected whenever $T_{m,s}$ is not the longest timescale, or if dark matter is the less coherent signal.}

\paragraph{Transient signal of equal duration and coherence time.} Next we consider a transient GW signal, but taking $\tau_h = T_h$ for simplicity.
From Eq.~\eqref{eq:SNR} we have (using $T_{m,h} = \textrm{min}[T_m,\,T_h]$)
\begin{equation}\begin{aligned}
{\cal T}_h =& \left\{ \begin{array}{lcc}
T_{m,h}^3 && T_{m,h} \ll \tau_r, \\
T_{m,h} \tau_r^2 & & \tau_r \ll T_{m,h}.
\end{array} \right. \\
= &T_{m,h}\, \textrm{min}[T_{m,h}^2,\,\tau_r^2].
\end{aligned}\end{equation}
We then need to compare this to DM.
Let us again assume $T_m$ is such that dark matter is well resolved (the more general case is straightforward), so that we can immediately read of ${\cal T}_a$ from the final line of Eq.~\eqref{eq:SNR}, and obtain
\begin{equation}
{\cal R}_c = \sqrt{\frac{\tau_a}{T_{m,h}}} \frac{\tau_r}{\, \textrm{min}[T_{m,h},\,\tau_r] } \left( \frac{T_{m,a}}{\textrm{max}[\tau_a,\,\tau_r]} \right)^{1/4}\!,
\label{eq:Rctransient}
\end{equation}
as in this case we will always have $M_h=1$.
For $T_h < T_m$, this result is then exactly Eq.~\eqref{eq:Rexplicit} from the main text.

\subsection{A model for the DMRadio scan strategy}

Equation~\eqref{eq:Rctransient} demonstrates explicitly that the coherence ratio can depend on the scan strategy of a resonant detector, which again is defined by the choice of $\{\omega_m,\,T_m\}$.
Therefore an explicit strategy is required if we are to compute the gravitational wave sensitivity in this limit.
Here we detail an explicit, albeit simplified, strategy for the most sensitive proposed instrument we consider in our frequency range: DMRadio-GUT, as described in Ref.~\cite{DMRadio:2022jfv}.

Considerable effort has been put into determining the optimal scan strategy for resonant instruments, for instance, see Refs.~\cite{Chaudhuri:2018rqn,Chaudhuri:2019ntz}.
Here, however, we adopt a simpler approach.
Working with the parameters forecast for DMRadio-GUT, the instrument will have $Q_r = 2 \times 10^7 > Q_a$, so that the instrumental bandwidth is narrower than that expected for the axion.
Therefore, a simple strategy that ensures no axion mass is overlooked, is to adjust the resonant frequency by the larger axion bandwidth, i.e. $\omega_{m+1} = \omega_m (1+1/Q_a)$.
For the frequency range DMRadio-GUT will cover, $m_a \in [0.4,120]\,\textrm{neV}$ this then defines the set of $\omega_m$.

Next we specify the integration time spent at each frequency, or $T_m$.
To begin with, for the nominal scan rate for the instrument is~\cite{DMRadio:2022jfv}
\begin{equation}
\frac{d\nu}{dt} = 41~\frac{\rm kHz}{\rm year} \left( \frac{\gagg}{10^{-19}~{\rm GeV}^{-1}} \right)^4 \left( \frac{\nu}{100~{\rm kHz}} \right)\!.
\label{eq:DMscan}
\end{equation}
The broad target for axion dark-matter instruments is the QCD axion line, where $\gagg$ varies as a function of mass.
In particular, two common targets are the KSVZ~\cite{Kim:1979if,Shifman:1979if} and DFSZ~\cite{Dine:1981rt,Zhitnitsky:1980tq} axion models, where
\begin{equation}
|\gagg| \simeq 1.62 \times 10^{-19}~{\rm GeV}^{-1} \left( \frac{m_a}{100~{\rm kHz}} \right) \times \left\{ \begin{array}{llc} 
1 & & \textrm{KSVZ}, \\
0.389 & & \textrm{DFSZ}.
\end{array} \right.
\end{equation}
Over the full mass range, Eq.~\eqref{eq:DMscan} implies that it would take DMRadio-GUT roughly 37 days to reach the KSVZ prediction, and 4.4 years for the smaller DFSZ.
Focusing on DFSZ, which is the target for the instrument, if we combine this result with the strategy for choosing $\omega_m$ already described, we find
\begin{equation}
T_m \simeq 14.3\,{\rm s}~ \left( \frac{1~{\rm neV}}{m_a} \right)^4\!.
\end{equation}
So in this simplified strategy very little time is spent at any particular axion mass, but this can be accounted for using the details of Eq.~\eqref{eq:SNR} being utilised.

\subsection{Examples for high-frequency GW signals}

\paragraph{Superradiance.}
One possible nearly persistent, highly coherent GW source is axion superradiance (for a comprehensive review, see Ref.~\cite{Brito:2015oca} and references therein) with the expected strain
\begin{equation}
h \sim 10^{-18} \left( \frac{\alpha}{l} \right) \epsilon \left(\frac{1\,\text{kpc}}{d} \right) \left( \frac{m_{\rm BH}}{2M_{\odot}} \right)\!,
\end{equation}
where $\alpha = G m_{\rm BH} m_a$ with $m_a$ being the axion mass, $l$ is the orbital angular momentum of the decaying axions from the hosting black hole, $d$ is the distance between the observer the black hole, and $\epsilon$ is the fraction of black hole mass accumulated in the axion cloud~\cite{Arvanitaki:2012cn,Aggarwal:2020olq}.
For a definite illustration, we choose $d = 1~\text{kpc}$, $\alpha / l = 0.5$, $\alpha=0.1$ and $\epsilon = 10^{-3}$ for our discussion.
The frequency of the signal is determined by the mass of the axion as $f = m_a / \pi$.

Superradiance can generate highly coherent signals, with $\dot{f} \sim 10^{-20} \left( \alpha / 0.1 \right)^{17} f^2$~\cite{Baryakhtar:2020gao,Berlin:2023grv}, which corresponds to very long coherence times $\tau_h$ which here indicate the time scale over which the GW frequency changes by a factor ${\cal O}(1)$.
The signal quality factor $Q_h = \tau_h f$ is correspondingly large, $Q_h \sim \int df f / \dot f \sim f^2 / \dot f \sim 10^{20} (0.1/\alpha)^{17}$.
Note that for resonant detectors, the number of observable cycles is limited by the detector bandwidth $\Delta f$ to $\sim f/\dot f \, \Delta f$.
In accordance with our definitions above, we denote also in this case the intrinsic coherence of the GW signal with $Q_h$, whereas the detector bandwidth is accounted for by the detector parameters introduced above.
The expected strain $h$ as well as effective signals $h/{\cal R}_c$ for different values of experimental run time $T_m$ and the quality factor of the detector $Q_r$ are depicted in Fig.~\ref{fig:superradiance}.
Because of its highly coherence nature, in principle we can have ${\cal R}_c \ll 1$ especially for high frequency regime.

\begin{figure}[!t]
\centering
\includegraphics[width=0.55\textwidth]{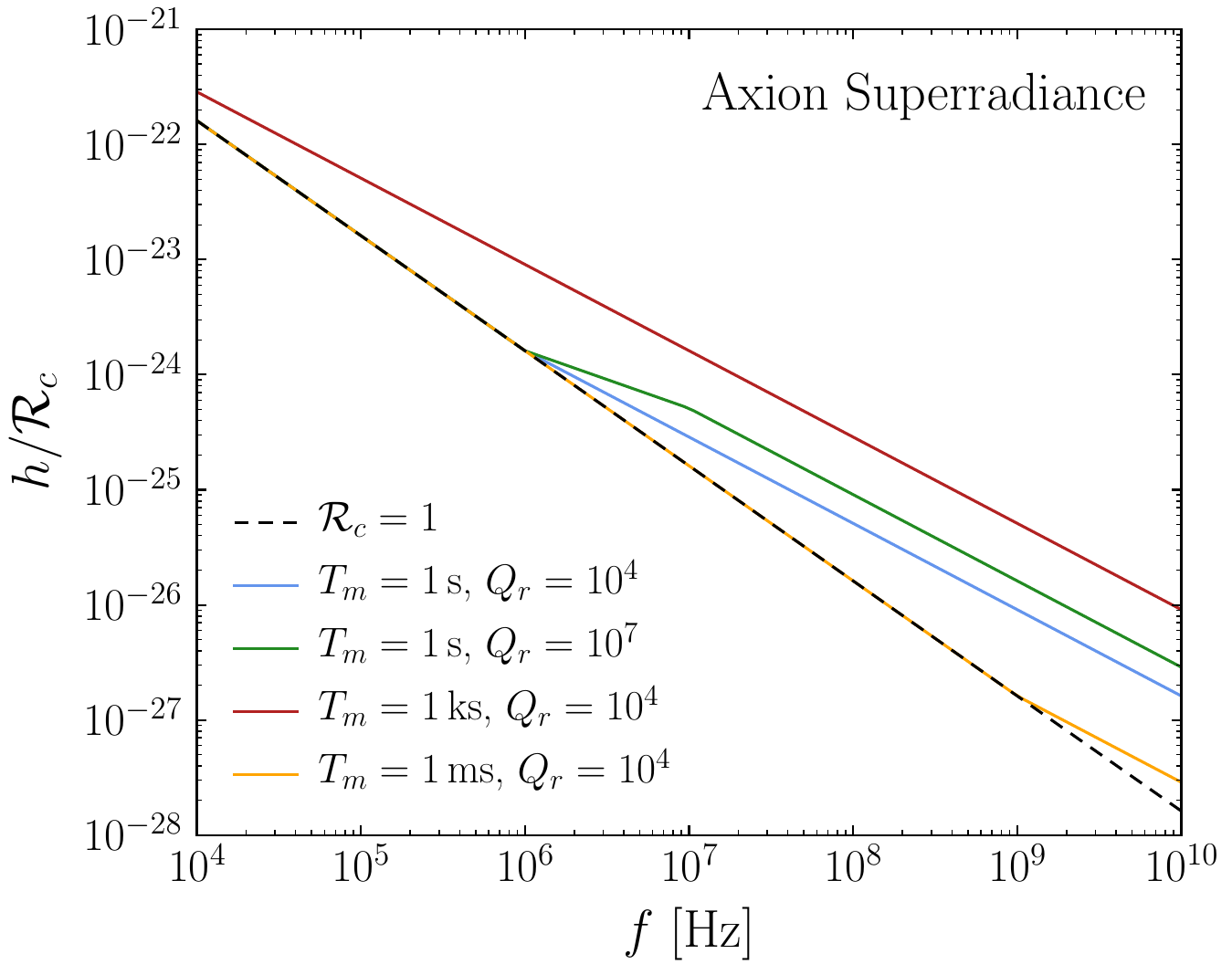}
\caption{Expected strain from the superradiance (blue line, ${\cal R}_c = 1$) and effective signals for various values of $(T_m, Q_r)$.
See the text for details and the parameters chosen.}
\label{fig:superradiance}
\end{figure}

\paragraph{PBH mergers.}
As another benchmark signal, we consider PBH binary mergers. For simplicity, we will take both black holes to have the same mass, $m_{\rm PBH}$.
In this case, we can determine the various time scales of the GW signal as a function of $m_{\rm PBH}$ and the GW frequency $f$.

The rotation frequency of the PBH binary is initially given by Kepler's law with the distance between the two BHs $R$ and the total mass $2m_{\rm PBH}$.
Here, we take the rotation frequency of the binary as a free parameter determined by this initial condition.
Due to the emission of GWs, the binary loses energy which leads to a reduction in $R$ and a corresponding increase in frequency,
\begin{equation}
\dot{f} = \frac{48 \cdot 2^{2/3}}{5}\pi^{8/3} \left( \frac{G m_{\rm PBH}}{c^3} \right)^{5/3} f^{11/3}.
\end{equation}
The radius continues to decrease until the innermost circular orbit (ISCO) is reached, at which point the radius and rotation frequency are given by
\begin{equation}
r_{\rm ISCO} = \frac{12 G m_{\rm PBH}}{c^2},\hspace{0.5cm} f_{\rm ISCO} = \frac{c^3}{24\sqrt{6} \pi G m_{\rm PBH}} \simeq 1.1\,{\rm kHz} \left( \frac{M_{\odot}}{m_{\rm PBH}} \right)\!.
\end{equation}
This sets the maximum frequency of the GW signal as $ \sim 2 f_{\rm ISCO}$~\cite{Maggiore:2007ulw}.

This increase  in the frequency of the merger  signal severely limits the coherence time $\tau_h$ of the signal.
As above, the corresponding quality factor can be obtained as  $Q_h \sim \int df\, f/\dot f \sim f^2/\dot f$, and since the integral is dominated by its lower boundary, we have $T_h \simeq \tau_h$.
In addition to setting the coherence time, $Q_h$ is also the number of cycles the orbit will undertake until merger, explaining why the coherence decreases as the ISCO is approached.

Following the same procedure described in the supplementary material of Ref.~\cite{Domcke:2022rgu}, we can determine expected strain sensitivity $h$ from PBH binary systems assuming an event per year.
In Fig.~\ref{fig:pbh signal}, we plotted the effective GW signal, $h/{\cal R}_c$ accounting for the suppression factor depending on the run time and the quality factor of the instrument.
In the figure, the grey region is excluded as it corresponds to $f > 2 f_{\rm ISCO}$.

\begin{figure}[!t] \centering
\centering
\includegraphics[width=0.45\textwidth]{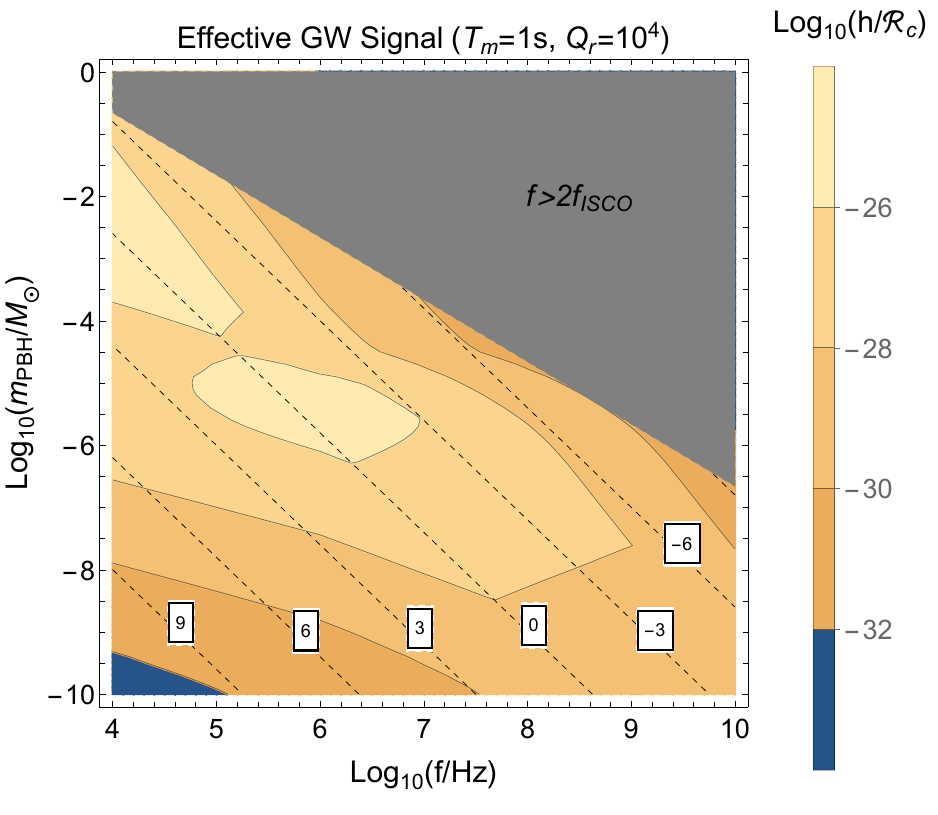} 
\includegraphics[width=0.45\textwidth]{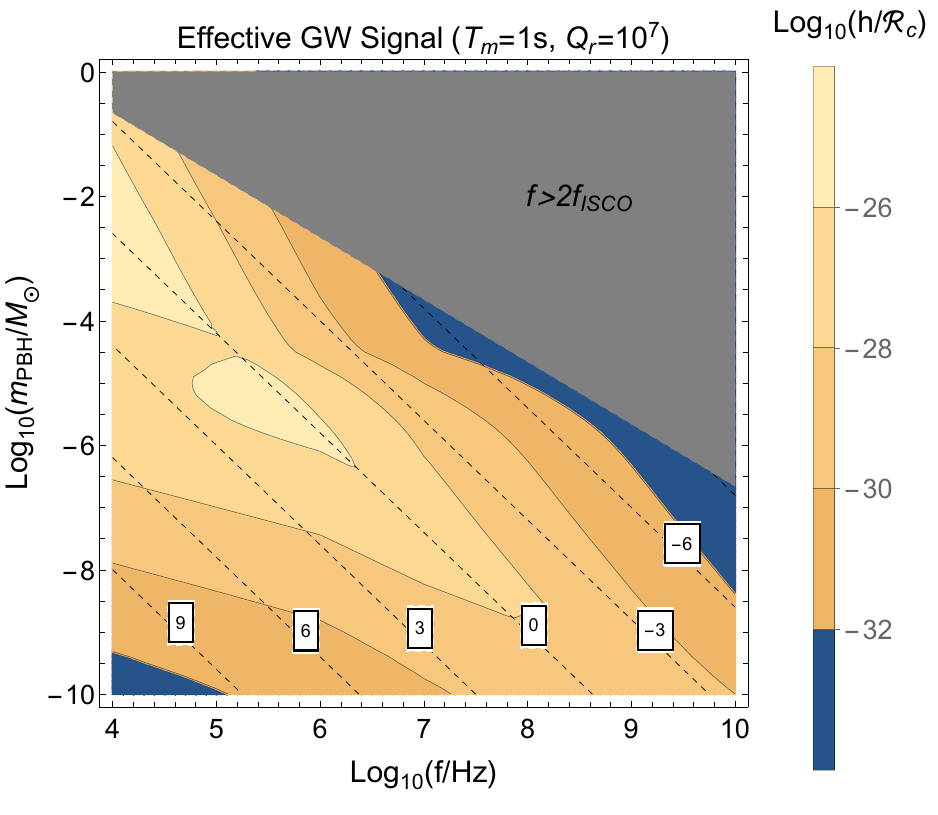} 
\includegraphics[width=0.45\textwidth]{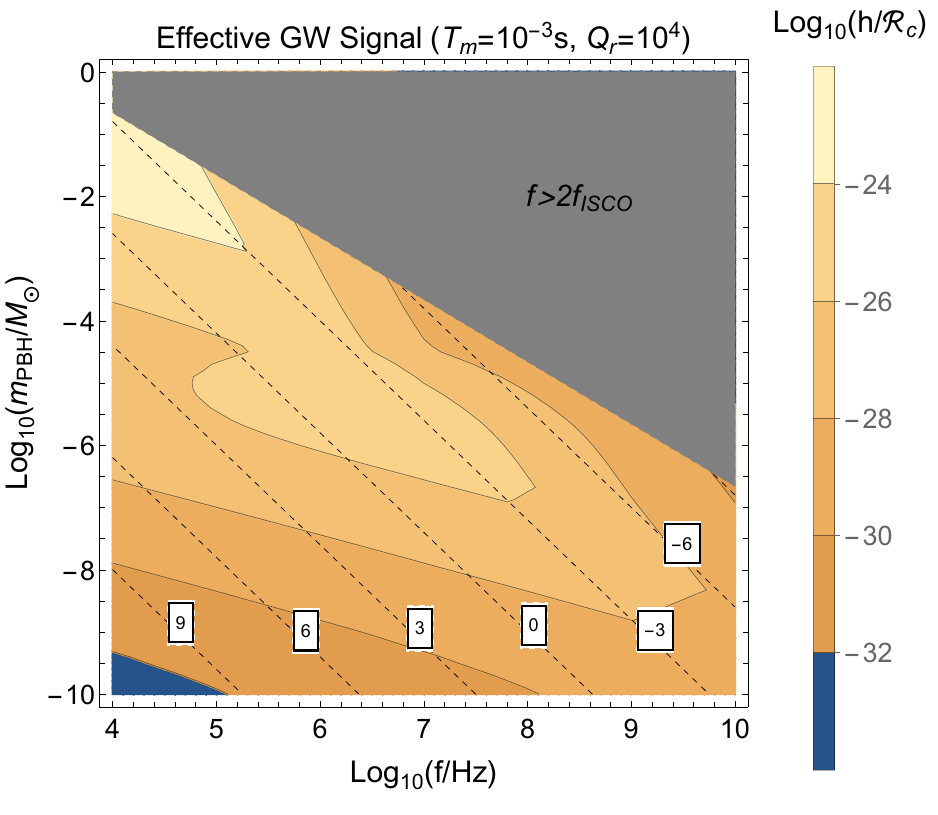}  
\includegraphics[width=0.45\textwidth]{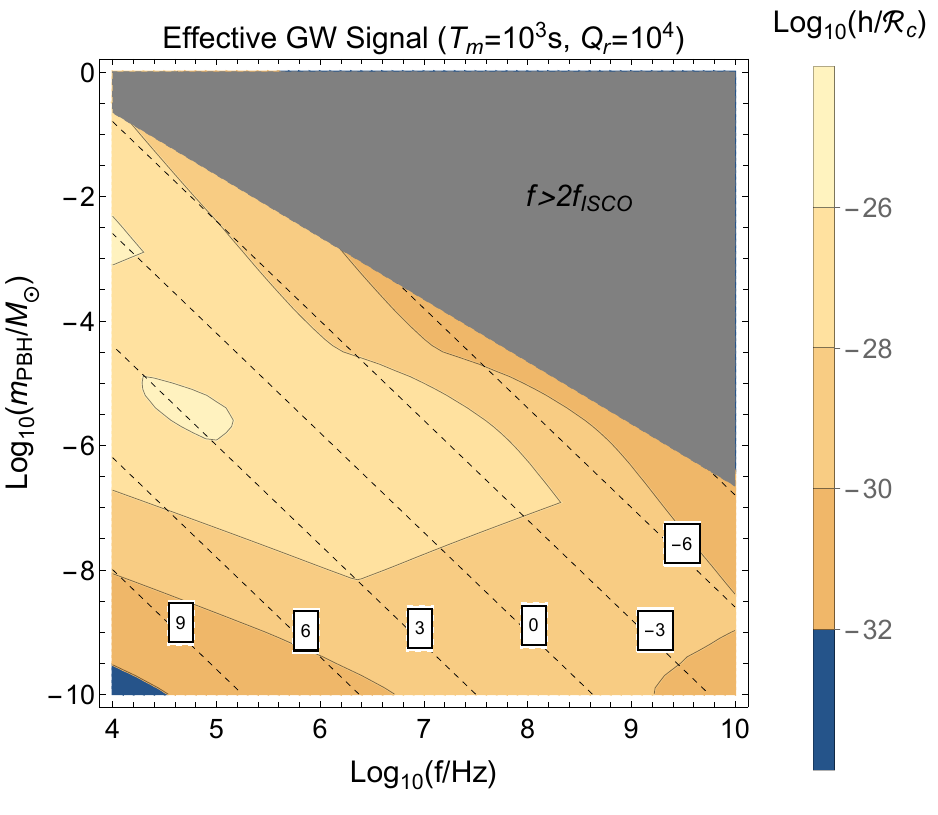} 
\caption{Effective GW signal for four benchmarks of $(T_m, Q_r)$ as a function of the PBH binary frequency $f$ and the mass $m_{\rm PBH}$.
The gray region is unphysical with $f > 2 f_{\rm ISCO}$.
The dotted black contour denotes the duration of GW signal, given in terms of $\log_{10}(T_h/1\,{\rm s})$.}
\label{fig:pbh signal}
\end{figure}

\section{GW Detection with an Electric Field}
\label{app:Efield}

As briefly discussed in the main text, in principle one can also search for the electromagnetic response when a GW passes through an electric field.
For axion dark matter, any such effect will be suppressed by dark matter's non-relativistic velocity, as the coupling arises from $\nabla a$ rather than $\partial_t a$.
No similar suppression occurs for the GW, although it remains true that for a given volume, the largest laboratory magnetic fields will have an enhanced energy density compared to the largest electric fields.
Nevertheless, for completeness we here demonstrate how the symmetry arguments apply in the case of an experiment with an electric field, providing parametric estimates for a single configuration.
The example we consider is an instrument with a solenoidal electric field, ${\bf E} = E_0 \hat{\bf e}_z$.
The exact details of the experimental electric field (such as the form of the electric field at the boundary) we will not consider.
We will take all length scales in the problem to be $L$, and simply study the angular dependence and $\omega$ scaling of the results.

For such a configuration, the leading ${\cal O}[(\omega L)^2]$ contribution is given as 
\begin{equation}
\Phi_h^{(2)} \sim e^{-\i \omega t} \omega^2  E_0 L^4 s_{\theta_h} \left( h^+ c_{\phi_h} - h^{\times} c_{\theta_h} s_{\phi_h - \phi_{\ell}} \right)\!.
\end{equation}
If we impose azimuthal symmetry on the pickup loop configuration (either as in BASE or DMRadio-m$^3$), the ${\cal O}[(\omega L)^2]$ order vanishes as in the magnetic field case, and the ${\cal O}[(\omega L)^3]$ order has only a $h^+$ contribution,
\begin{equation}
\Phi_h^{(3)} \sim e^{-\i \omega t} h^+ \omega^3 E_0 L^5 s^2_{\theta_h}.
\end{equation}
Compared to the equivalent magnetic field result studied in Sec.~\ref{sec:GWforSolenoid}, the form is similar except for the appearance of $h^+$ rather than $h^{\times}$.
More generally, for an electric external field, it can be shown that the same arguments and selection rules given in Sec.~\ref{sec:othergeom} apply with the exchange of $h^{\times} \leftrightarrow h^+$, as expected given that the electric field is a vector while the magnetic field is a pseudovector.
Hence, changing from a magnetic to electric field will leave the leading power conclusions in Tab.~\ref{table:summary} unchanged.

\bibliographystyle{JHEP}
\bibliography{detector.bib}

\end{document}